\documentclass[format=acmsmall, review=false, screen=true]{acmart}

\usepackage{booktabs} 
\usepackage{multirow}
\usepackage{amsmath}
\usepackage{amssymb}

\usepackage[ruled]{algorithm2e} 

\SetAlFnt{\small}
\SetAlCapFnt{\small}
\SetAlCapNameFnt{\small}
\SetAlCapHSkip{0pt}
\IncMargin{-\parindent}
\acmJournal{CSUR}
\acmVolume{0}
\acmNumber{0}
\acmArticle{0}
\acmYear{0}
\acmMonth{0}
\copyrightyear{2017}

\setcopyright{rightsretained}
\acmPrice{0.00}

\acmDOI{0000001.0000001}

\received{XXXXX}
\received[revised]{XXXXX}
\received[accepted]{XXXXX}

\newcommand{\massa}{MASS}
\newcommand{\mass}{MASS}

\begin{document}
\title[Music in digital audio]{Musical elements in the discrete-time representation of sound}  
\author{Renato Fabbri}
\affiliation{%
  \institution{University of S\~ao Paulo}
  \department{Institute of Mathematics and Computer Sciences}
  \streetaddress{Avenida Trabalhador S\~ao Carlense, 400 - Centro}
  \city{S\~ao Carlos}
  \state{SP}
  \postcode{13566-590}
  \country{Brazil}}
\author{Vilson Vieira da Silva Junior}
\affiliation{%
  \institution{Cod.ai}
  \city{Berlin}
  \state{BE}
  \postcode{???}
  \country{DE}
}
\author{Ant\^onio Carlos Silvano Pessotti}
\affiliation{%
  \institution{Universidade Metodista de Piracicaba}
  \department{??}
  \city{Piracicaba}
  \state{SP}
  \postcode{???}
  \country{Brazil}}
\author{D\'ebora Cristina Corr\^ea}
\affiliation{%
  \institution{University of Western Australia}
  \department{??}
  \city{Piracicaba}
  \state{SP}
  \postcode{???}
  \country{AU}
  }
\author{Osvaldo N. Oliveira Jr.}
\affiliation{%
  \institution{University of S\~ao Paulo}
  \department{S\~ao Carlos Institute of Physics}
  \streetaddress{Avenida Trabalhador S\~ao Carlense, 400 - Centro}
  \city{S\~ao Carlos}
  \state{SP}
  \postcode{13566-590}
  \country{Brazil}
  }

\begin{abstract}
The representation of basic elements of music in terms of discrete audio
signals is often used in software for musical creation and design.
Nevertheless, there is no unified approach that relates these elements to the discrete samples of digitized sound.
In this article, each musical element is related by equations and algorithms to the discrete-time samples of sounds,
and each of these relations are implemented in scripts within a software toolbox,
referred to as MASS (Music and Audio in Sample Sequences).
The fundamental element, the musical note with duration, volume, pitch and timbre,
is related quantitatively to characteristics of the digital signal.
Internal variations of a note, such as tremolos, vibratos and spectral fluctuations,
are also considered, which enables the synthesis of notes inspired by real instruments and new sonorities.
With this representation of notes, resources are provided for the generation of higher scale musical structures,
such as rhythmic meter, pitch intervals and cycles.
This framework enables precise and trustful scientific experiments, data sonification and is useful for education and art.
The efficacy of MASS is confirmed by the synthesis of small musical pieces using basic notes,
elaborated notes and notes in music, which reflects the organization of the toolbox and thus of this article.
It is possible to synthesize whole albums through collage of the scripts and settings specified by the user.
With the open source paradigm, the toolbox can be promptly scrutinized,
expanded in co-authorship processes and used with freedom by musicians, engineers and other interested parties.
In fact, MASS has already been employed for diverse purposes which include music production,
artistic presentations, psychoacoustic experiments and computer language diffusion where
the appeal of audiovisual artifacts is exploited for education.
\end{abstract}

%
%


%
%


\thanks{This work is supported by FAPESP and CNPq.}

\maketitle

\renewcommand{\shortauthors}{R. Fabbri et al.}

\section{Introduction}\label{sec:level1}
Music is usually defined as the art whose medium is sound.
The definition might also state that the medium includes silences
and temporal organization of structures, or that music
is also a cultural activity or product.
In physics and in this document, sounds are longitudinal waves of mechanical pressure.
The human auditory system perceives sounds in the frequency bandwidth between $20Hz$ and $20kHz$,
with the actual boundaries depending on the person,
climate conditions and the sonic characteristics themselves.
Since the speed of sound is $\approx 343.2 m/s$,
such frequency limits corresponds to  wavelengths of $\frac{343.2}{20} \approx 17.16\,m$ and
$\frac{343.2}{20000} \approx 17.16\,mm$.
Hearing involves stimuli in bones, stomach, ears,
transfer functions of head and torso,
and processing by the nervous system.
The ear is a dedicated organ for the appreciation of these waves,
which decomposes them into their sinusoidal spectra and delivers to the nervous system.
The sinusoidal components are crucial to musical phenomena,
as one can recognize in the constitution of sounds of musical interest
(such as harmonic sounds and noises, discussed in Sections~\ref{sec:discNote} and~\ref{sec:internalVar}),
and higher level musical structures (such as tunings, scales and chords, in Section~\ref{sec:notesMusic}).~\cite{Roederer}

\begin{figure*}[!h]
    \centering
        \includegraphics[width=.7\textwidth]{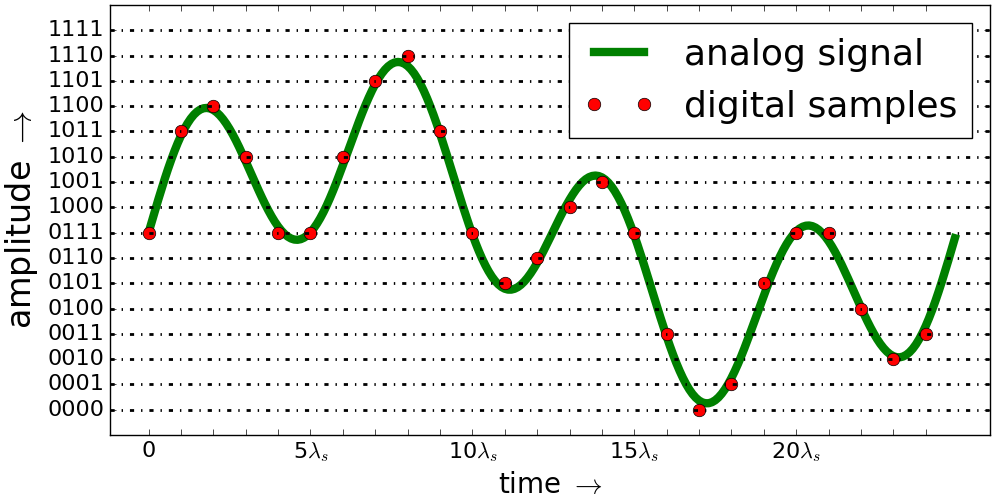}
	\caption{Example of PCM audio:
	a sound wave is represented by 25 samples equally spaced in time where each sample has an amplitude specified with 4 bits.}
        \label{fig:PCM}
\end{figure*}

The representation of sound can take many forms,
from musical scores and texts in a phonetic language to electric analog signals and binary data.
It includes sets of features such as wavelet or sinusoidal components.
Although the terms 'audio' and 'sound' are often used without distinction
and 'audio' has many definitions which depend on the context and the author,
audio most often means a representation of the amplitude through time.
In this sense,
audio expresses sonic waves yield by synthesis or input by microphones,
although these sources are not always neatly distinguishable
e.g. as captured sounds are processed to generate new sonorities.
Digital audio protocols often imply in quality loss (to achieve smaller files, ease storage and transfer)
and are called \emph{lossy}~\cite{loss}.
This is the case e.g. of MP3 and Ogg Vorbis.
Non-lossy representations of digital audio, called \emph{lossless} protocols or formats,
on the other hand, assures perfect reconstruction of the analog wave within any convenient precision.
The standard paradigm of lossless audio consists of
representing the sound with samples equally spaced
by a duration $\delta_s$, and specifying the amplitude of each sample by a fixed number of bits.
This is the linear Pulse Code Modulation (LPCM) representation of sound,
herein referred to as PCM.
A PCM audio format has two essential attributes:
a sampling frequency $f_s=\frac{1}{\delta_s}$ (also called e.g. sampling rate or sample rate),
which is the number of samples used for representing a second of sound;
and a bit depth, which is the number of bits used for specifying the amplitude of each sample.
Figure~\ref{fig:PCM} shows $25$ samples of a PCM audio with a bit depth of $4$,
which yields $2^4=16$ possible values for the amplitude of each sample
and a total of $4 \times 25= 100$ bits for representing the whole sound.

The fixed sampling frequency and bit depth
yield the quantization error or quantization noise.
This noise diminishes as the bit depth increases
while greater sampling frequency allows higher frequencies to be represented.
The Nyquist theorem asserts that the sampling frequency
is twice the maximum frequency that the represented signal can contain~\cite{Openheim}.
Thus, for general musical purposes, it is suitable to use a sample rate of
at least twice the highest frequency heard by humans,
that is, $f_s \geq 2\times 20kHz = 40kHz$.
This is the basic reason for the adoption of sampling frequencies
such as $44.1kHz$ and $48kHz$,
which are standards in Compact Disks (CD)
and broadcast systems (radio and television), respectively.

Within this framework for representing sounds,
musical notes can be characterized.
The note often stands as the 'fundamental unit' of musical structures
(such as atoms in matter or cells in macroscopic organisms) and,
in practice, it can unfold into sounds that uphold other approaches to music.
This is of capital importance because science and scholastic artists
widened the traditional comprehension of music in the twentieth century
to encompass discourse without explicit rhythm, melody or harmony.
This is evident e.g. in the concrete, electronic, electroacoustic,
and spectral musical styles.
In the 1990s, it became evident that popular
(commercial) music had also incorporated
sound amalgams and abstract discursive arcs\footnote{There
are well known incidences of such characteristics in ethnic music,
such as in Pygmy music, but western theory assimilated them
only in the last century~\cite{Wisnick}.}.
Notes are also convenient for another reason:
the average listener -- and a considerable part of the specialists --
presupposes rhythmic and pitch organization 
(made explicit in Section~\ref{sec:notesMusic})
as fundamental musical properties,
and these are developed in traditional musical theory in terms of notes.
Thereafter, in this article we describe musical notes in PCM audio through equations
and then indicate mechanisms for deriving higher level musical structures.
We understand that this is not the unique approach to mathematically express music in
digital audio, but musical theory and practice suggest that
this is a proper framework for understanding and making computer music,
as should become patent in the reminder of this text and is verifiable
by usage of the \massa\ toolbox.
Hopefully, the interested reader or programmer will be able to use
this framework to synthesize music beyond traditional conceptualizations when intended.

This document provides a
fundamental description of musical structures
in discrete-time audio.
The results include mathematical relations,
usually in terms of musical characteristics and PCM samples,
concise musical theory considerations,
and their implementation as software routines both as
very raw and straightforward algorithms and in the context of rendering musical pieces.
Despite the general interests involved,
there are only a few books and computer implementations that tackle the subject directly.
These mainly focus on computer implementations and ways to mimic traditional instruments,
with scattered mathematical formalisms for the basic notions.
Articles on the topic appear to be lacking, to the best of our knowledge,
although advanced and specialized developments are often reported.
A compilation of such works and their contributions is in the Appendix G of~\cite{dissertacao}.
Although current music software uses the analytical descriptions presented here,
there is no concise mathematical description of them, and it is far from trivial
to achieve the equations by analyzing the available software implementations.

Accordingly, the objectives of this paper are:
\begin{enumerate}
	\item Present a concise set of mathematical and algorithmic relations between basic musical elements and sequences of PCM audio samples.
	\item Introduce a framework for sound and musical synthesis with control at sample level which entails potential uses in psychoacoustic experiments, data sonification and synthesis with extreme precision (recap in Section~\ref{cap:conclusao}).
	\item Provide a powerful theoretical framework which can be used to synthesize musical pieces and albums.
	\item Provide approachability to the developed framework\footnote{All
		the analytic relations presented in this article are implemented as small scripts in public domain.
		They constitute the \massa\ toolbox, available in an open source Git repository~\cite{gitBook}.
		These routines are written in Python and make use of Numpy,
		which performs numerical routines efficiently (e.g. through LAPACK),
		but the language and packages are by no means mandatory.
		Part of the scripts has been ported to JavaScript
		(which favors their use in Web browsers such as Firefox and Chromium)
		and native Python~\cite{numpy, tutpython, python}.
		These are all open technologies, published using licenses 
		that grant permission for copying,
		distributing, modifying and usage in research, development, art and education.
		Hence, the work presented here aims at being compliant with recommended
		practices for availability and validation
		and should ease co-authorship processes~\cite{Raymond,Lessig}.}.
	\item Provide a didactic presentation of the content, which is highly multidisciplinary,
		involving signal processing, music, psychoacoustics and programming.
\end{enumerate}

The reminder of this article is organized as follows:
Section~\ref{sec:discNote} characterizes the basic musical note;
Section~\ref{sec:internalVar} develops internal dynamics of musical notes;
Section~\ref{sec:notesMusic} tackles the organization of musical notes into
higher level musical structures~\cite{Wisnick,Webern,Lerdahl,Cook,Lacerda,Zamacois,Schoenberg,microsound}.
As these descriptions require knowledge on topics such as psychoacoustics, cultural traditions,
and mathematical formalisms, the text points to external complements as needed
and presents methods, results and discussions altogether.
Section~\ref{cap:conclusao} is dedicated to final considerations and further work.

\subsection{Additional material}
One Supporting Information document~\cite{massListings}
holds commented listings of all the equations, figures, tables and sections in this document
and the scripts in the \massa\ toolbox.
Another Supporting Information document~\cite{massCode} is a PDF version of
the code that implements the equations and concepts in each section\footnote{
	The toolbox contains a collection of Python scripts which:
\begin{itemize}
    \item implement each of the equations;
    \item render music and illustrate the concepts;
    \item render each of the figures used in this article.
\end{itemize}
The documentation of the toolbox consists of this article, the Supporting Information documents and the scripts themselves.}.
The Git repository~\cite{MASSA} holds all the PDF documents and Python scripts.
The rendered musical pieces are referenced when convenient and linked directly through URLs,
and constitute another component of the framework.
They are not very traditional, which facilitates the understanding of specific techniques
and the extrapolation of the note concept.
There are \mass -based software packages~\cite{music,figgus} and
further musical pieces that are linked in the Git repository.

\subsection{Synonymy, polysemy and theoretical frames (disclaimer)}\label{sec:disc}
Given that the main topic of this article (the expression of musical elements
in PCM audio) is multidisciplinary and involves art,
the reader should be aware that much of the vocabulary admits different choices of terms and definitions.
More specifically, it is often the case where many words can express the same concept
and where one word can carry different meanings.
This is a very deep issue which might receive a dedicated manuscript.
The reader might need to read the rest of this document to understand this
small selection of synonymy and polysemy in the literature,
but it is important to illustrate the point before the more dense sections:
\begin{itemize}
	\item a ``note'' can mean a pitch or an abstract construct with pitch and duration or a sound emitted from a musical instrument or a specific note in a score or a music.
	\item The sampling rate (discussed above) is also called the sampling frequency or sample rate.
	\item A harmonic in a sound is most often a sinusoidal component which is in the harmonic series of the fundamental frequency. Many times, however, the terms harmonic and component are not distinguished. A harmonic can also be a note performed in an instrument by preventing certain overtones (components).
	\item Harmony can refer to chords or to note sets related to chords or even to ``harmony'' in a more general sense, as a kind of balance and consistency.
	\item A ``tremolo'' can mean different things: e.g. in a piano score, a tremolo is a fast alternation of two notes (pitches) while in computer music theory it is (most often) an oscillation of loudness.
\end{itemize}

We strived to avoid nomenclature clashes and the use of more terms than needed.
Also, there are many theoretical standpoints for understanding musical phenomena,
which is an evidence that most often there is not a single way to express or characterize musical structures.
Therefore, in this article, adjectives such as "often", "commonly" and "frequently" are abundant and they would probably be even more numerous if we wanted to be pedantically precise.
Some of these issues are exposed when the context is convenient, such as in the first considerations of timbre.

\section{Characterization of the musical note in discrete-time audio} \label{sec:discNote}\label{sec:notaDisc}
In diverse artistic and theoretical contexts, music is conceived as constituted by fundamental units referred to as notes, ``atoms'' that constitute music itself~\cite{Wisnick, Lovelock, Webern}.
In a cognitive perspective, notes are understood as discernible elements that facilitate and enrich the transmission of information through music~\cite{Roederer, Lacerda}.
Canonically, the basic characteristics of a musical note are duration, loudness, pitch and timbre~\cite{Lacerda}.
All relations described in this section are implemented in the file \texttt{src/sections/eqs2.1.py}.
The musical pieces \emph{5 sonic portraits} and \emph{reduced-fi} are also available online to corroborate and illustrate the concepts.

\subsection{Duration}
The sample frequency $f_s$ is defined as the number of samples in each second of the discrete-time signal. Let $T=\{t_i\}$ be an ordered set of real samples separated by $\delta_s=1/f_s$ seconds ($f_s=44.1kHz \Rightarrow \delta_s=1/44100\approx 0.023ms$).
A musical note of duration $\Delta$ seconds can be expressed as a sequence $T^{\Delta}$ with $\Lambda = \lfloor \Delta . f_s \rfloor$ samples.
That is, the integer part of the multiplication is considered, and an error of at most $\delta_s$ missing seconds is admitted, which is usually fine for musical purposes. Thus:

\begin{equation}\label{eq:dur}
T^{\Delta}={\{t_i\}}_{i=0}^{\lfloor \Delta . f_s \rfloor -1} = \{t_i\}_0^{\Lambda-1}
\end{equation}

\subsection{Loudness}\label{subsec:volume}
Loudness\footnote{Loudness and ``volume'' are often used indistinctly.
In technical contexts, loudness is used for the subjective perception of sound intensity
while volume might be used for some measurement of loudness or to a change
in the intensity of the signal by equipment.
Accordingly, one can perceive a sound as loud or soft and change the volume
by turning a knob.
We will use the term loudness and avoid the more ambiguous term volume.} 
is a perception of sonic intensity that depends on reverberation,
spectrum and other characteristics described in Section~\ref{sec:varInternas}~\cite{Chowning}.
One can achieve loudness variations through the power of the wave~\cite{Chowning}:

\begin{equation}\label{eq:potencia}
pow(T)=\frac{\sum_{i=0}^{\Lambda -1} t_i^2}{\Lambda}
\end{equation} 

The final loudness is dependent on the amplification of the signal by the speakers.
Thus, what matters is the relative power of a note in relation to the others around it,
or the power of a musical section in relation to the rest. Differences in loudness are the result of complex psychophysical phenomena but can often be reasoned about in terms of decibels, calculated directly from the amplitudes through energy or power:

\begin{equation}\label{decibels}
V_{dB}=10log_{10}\frac{pow(T^{'})}{pow(T)}
\end{equation}

The quantity $V_{dB}$ has the decibel unit ($dB$). 
By standard, a ``doubled loudness'' is associated to a gain of $10dB$ (10 violins yield double the loudness of a violin).
A handy reference is $10dB$ for each step in the musical intensity scale: \emph{pianissimo}, \emph{piano}, \emph{mezzoforte}, \emph{forte} and \emph{fortissimo}. Other useful references are $dB$ values related to double amplitude or power:

\begin{align}
t_i^{'}=2 t_i \Rightarrow pow(T^{'})=4 pow(T) \Rightarrow V^{'}_{dB}=10log_{10} 4 \approx 6 dB \label{eq:ampVol}\\
t_i^{'}=\sqrt{2} t_i \Rightarrow pow(T^{'})=2 pow(T) \Rightarrow V^{'}_{dB}=10log_{10} 2 \approx 3 dB\label{eq:potVol}
\end{align}

\noindent and the amplitude gain for a sequence whose loudness has been doubled ($10dB$):

\begin{equation}\label{eq:dobraVol}
\begin{split}
10log_{10}\frac{pot(T^{'})}{pot(T)} = 10 \quad \Rightarrow \\ \Rightarrow \quad \sum_{i=0}^{\lfloor \Delta.f_s \rfloor -1}t^{'2}_i=10\sum_{i=0}^{\Lambda-1}t_i^2=\sum_{i=0}^{\Lambda-1}(\sqrt{10}.t_i)^2 \\
\therefore \quad t^{'}_i=\sqrt{10}t_i \quad \Rightarrow \quad t^{'}_i \approx 3.16t_i
\end{split}
\end{equation}

Thus, an amplitude increase by a factor slightly above 3 is required for achieving a doubled loudness.
These values are guides for increasing or decreasing the absolute values in sample sequences.
The conversion from decibels to amplitude gain (or attenuation) is straightforward:

\begin{equation}\label{ampDec}
A = 10^{\frac{V_{dB}}{20}}
\end{equation}

\noindent where $A$ is the multiplicative factor that relates the amplitudes before and after amplification.

\subsection{Pitch}
The perception sounds as 'higher' or 'lower' is usually thought in terms of pitch.
An exponential progression of frequency ($f_i = f . X^i, \forall \;X > 0, i \geq 1$) yields a linear variation of the pitch,
a fact that will be further exploited in Sections~\ref{sec:internalVar} and~\ref{sec:notesMusic}.
Accordingly, a pitch is specified by a (fundamental) frequency $f$ whose cycle has duration $\delta=1/f$.
This duration, multiplied by the sampling frequency $f_s$, yields the number of samples per cycle:
$\lambda=f_s . \delta =f_s/f$.
For didactic reasons, let $f$ divide $f_s$ and result $\lambda$ integer.
If $T^f$ is a sonic sequence with fundamental frequency $f$, then:

\begin{equation}\label{periodicidade}
     T^f=\left\{ t_i^f \right\}=\left\{ t^f_{i+\lambda}  \right\}= \left\{ t^f_{i+\frac{f_s}{f}} \right\}
\end{equation}

In the next section, frequencies $f$ that do not divide $f_s$ will be considered. This restriction does not imply in loss of generality of this current section's content.

\subsection{Timbre}\label{sec:timbre}
A spectrum is said harmonic if all the (sinusoidal) frequencies $f_n$ it contains are (whole number) multiples of a fundamental frequency $f_0$ (lowest frequency): $f_n=(n+1)f_0$.
From a musical perspective, it is critical to internalize that energy in a component with frequency $f$
is a sinusoidal oscillation in the constitution of the sound in that frequency $f$.
This energy, specifically concentrated on $f$,
is separated from other frequencies by the ear for further cognitive processes 
(this separation is performed by diverse living organisms by mechanisms similar to the human cochlea).
The sinusoidal components are responsible for timbre\footnote{The 
timbre of a sound is a subjective and complex characteristic.
The timbre can be considered by the temporal evolution of energy in the spectral components that 
are harmonic or noisy (and by deviations of the harmonics from the ideal harmonic spectrum).
In addition, the word timbre is used to designate different things:
one same note can have (be produced with) different timbres,
an instrument has different timbres, two instruments of the same family have,
at the same time, the same timbre that blends them into the same family,
and different timbres as they are different instruments.
Timbre is not only about spectrum: culture and context alter our perception of timbre.~\cite{Roederer}} qualities (including pitch).
If their frequencies do not relate by small integers, the sound is perceived as noisy or dissonant,
in opposition to sonorities with an unequivocally established fundamental.
Accordingly, the perception of absolute pitch relies on the similarity of the spectrum to the harmonic series.~\cite{Roederer}

A sound with a harmonic spectrum has a wave period (wave cycle duration) which corresponds to the inverse of the fundamental frequency.
The trajectory of the wave inside the period is the \emph{waveform} and implies a specific combination of amplitudes and phases of the harmonic spectrum.
Sonic spectra with minimal differences can result in timbres with crucial differences and,
consequently, distinct timbres can be produced using different waveforms.

\begin{figure*}
    \centering
        \includegraphics[width=.7\textwidth]{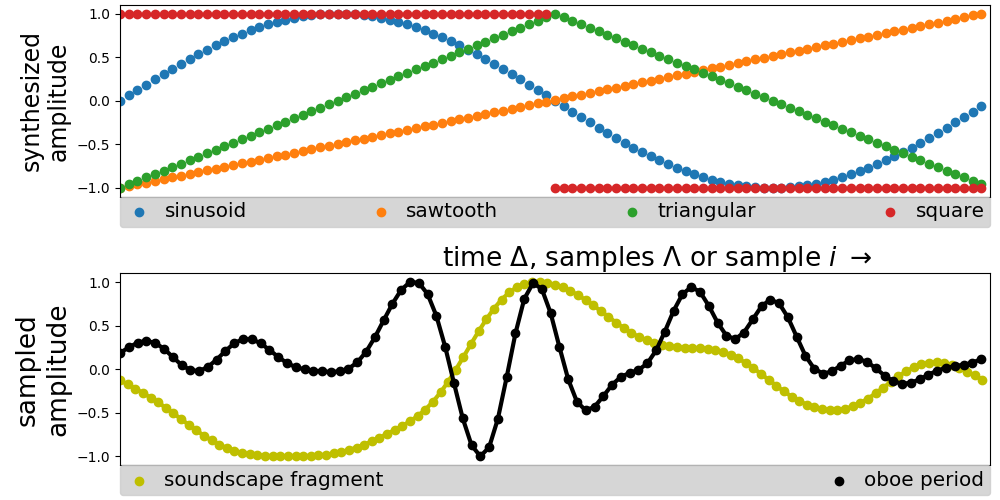}
    \caption{Basic musical waveforms: (a) the basic synthetic  waveforms
	given by the Equations~\ref{sinusoid},~\ref{sawTooth},~\ref{triangular} and~\ref{square};
	(b) real waveforms.
	Because of the period with $\approx 100$ samples ($\lambda_f\approx 100$),
	if $f_s=44.1kHz$
	the basic and oboe waves have a fundamental frequency of $f=\frac{f_s}{\lambda_f}\approx \frac{44100}{100} = 441 \; Hz$,
	whatever the waveform is.}
        \label{fig:formasDeOnda}
\end{figure*}

High curvatures in the waveform are clues of energy in the high frequencies.
Figure~\ref{fig:formasDeOnda} depicts a wave, labeled as ``soundscape fragment''.
The same figure also displays a sampled period from an oboe note.
One can notice from the curvatures: the oboe's rich spectrum at high frequencies and the greater contribution of the lower frequencies in spectrum of the soundscape fragment.

The sequence $R=\{ r_i \}_0^{\lambda_f-1}$ of samples in a real sound (e.g. of Figure~\ref{fig:formasDeOnda}) can be taken as a basis for a sound $T^f$ in the following way:

\begin{equation}\label{sampleandoFormaDeOnda}
     T^f=\{ t_i^f \}=\Bigl\{ r_{(i\,\%\lambda_{f})} \Bigr\}
\end{equation}

\begin{figure*}
    \centering
        \includegraphics[width=.7\textwidth]{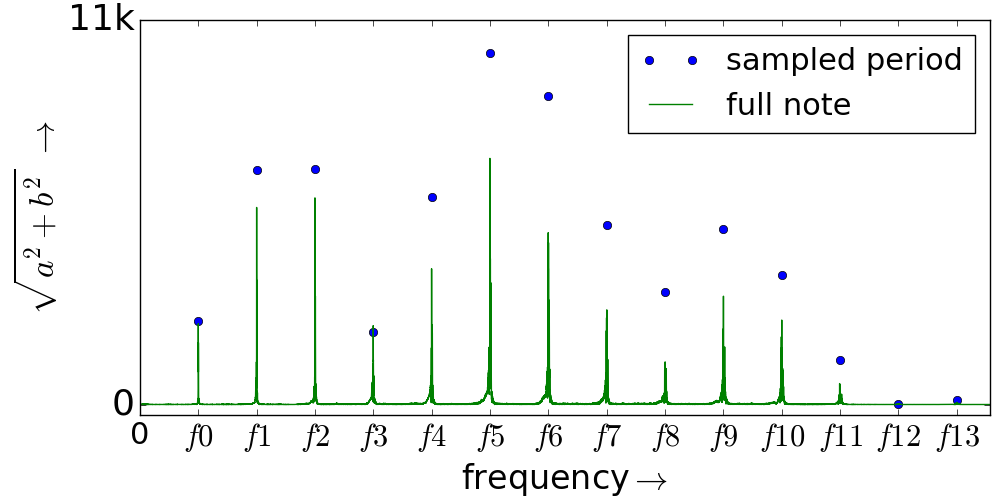}
	\caption{Spectra of the sonic waves of a natural oboe note and obtained through a sampled period. The natural sound has fluctuations in the harmonics and in its noise, while the sampled period note has a perfectly harmonic (and static) spectrum.}
        \label{fig:espectroOboe}
\end{figure*}

The resulting sound has the spectrum of the original waveform.
As a consequence of the identical repetitions,
the spectrum is perfectly harmonic,
without noise or variations of the components
which are typical of natural phenomena.
This can be observed in Figure~\ref{fig:espectroOboe},
which shows the spectrum of the original oboe note and a note with the same duration,
whose samples consist of the repetition of the cycle on Figure~\ref{fig:formasDeOnda}.

The simplest case is the spectrum with only one frequency, which is a sinusoid, often regarded as a ``pure'' oscillation (e.g. in terms of the \emph{simple harmonic motion}).
Let $S^f$ be a sequence whose samples $s_i^f$ describe a sinusoid with frequency $f$:

\begin{equation}\label{sinusoid}
     S^f=\{ s^f_i \}=\Bigl\{ \sin\bigl(2\pi \frac{i}{\lambda_f} \bigr)  \Bigr\} = \Bigl\{ \sin\bigl(2\pi f \frac{i}{f_s}\bigr)  \Bigr\} 
\end{equation}

\noindent where $\lambda_f=\frac{f_s}{f}=\frac{\delta_f}{\delta_s}$ is the number of samples in the period.

Other artificial waveforms are used in music for their spectral qualities and simplicity.
While the sinusoid is an isolated node in the spectrum, any other waveform presents a succession of harmonic components (harmonics).
The most basic waveforms are specified by Equations~\ref{sinusoid},~\ref{sawTooth},~\ref{triangular} and~\ref{square}, and are illustrated in Figure~\ref{fig:formasDeOnda}.
These artificial waveforms are traditionally used in music for synthesis and oscillatory control of variables. They are also useful outside musical contexts~\cite{Openheim}.

The sawtooth presents all the harmonics with a decreasing energy of $-6dB/octave$\footnote{In 
musical jargon, an ``octave'' means a frequency and twice such frequency ($f$ and $2f$), or the bandwidth $[f,2f]$.}.
The sequence of temporal samples can be described as:

\begin{equation}\label{sawTooth}
	D^f=\left\{ d^f_i \right\}=\left\{ 2\frac{i\,\%(\lambda_f+1)}{\lambda_f} -1 \right\}
\end{equation}

The triangular waveform has only odd harmonics falling with $-12dB/octave$:
\begin{equation}\label{triangular}
     T^f=\left\{ t^f_i \right\}=\left\{1- \left| 2 - 4\frac{i\,\%\lambda_f}{\lambda_f} \right| \right\}
\end{equation}

The square wave also has only odd harmonics but falling at $-6dB/octave$:
\begin{equation}\label{square}
     Q^f=\left\{ q^f_i \right\}= \left\{
         \begin{array}{l l}
              1 & \quad \text{for } \; \; (i\,\%\lambda_f)   <  \lambda_f /2  \\
              -1 & \quad \text{otherwise}\\
         \end{array} \right.
\end{equation}

\begin{figure*}
    \centering
        \includegraphics[width=.7\textwidth]{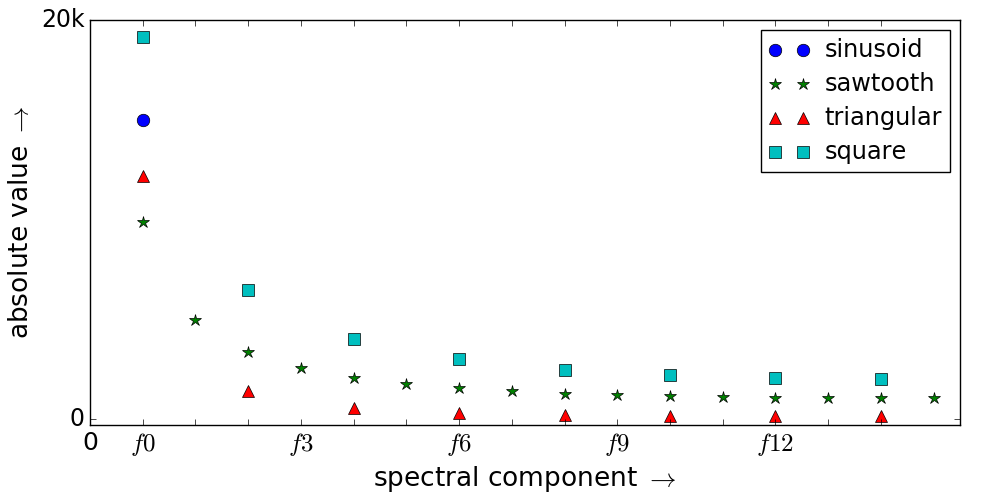}
    \caption{Spectra of basic artificial waveforms.
The isolated and exactly harmonic components of the spectra is a consequence of the fixed period.
	The figure exhibits the spectra described in Section~\ref{sec:timbre}:
	the sawtooth is the only waveform with a complete harmonic series (odd and even components);
	triangular and square waves have the same components (odd harmonics), decaying at $-12dB/octave$ and $-6dB/octave$, respectively;
	the sinusoid consists of a unique node in the spectrum.
	}
        \label{fig:espectroDeOndas}
\end{figure*}

The square wave can be used in a subtractive synthesis with the purpose of mimicking a clarinet.
This instrument has only the odd harmonics and the square wave is convenient with its abundant energy at high frequencies.
The sawtooth is a common starting point for subtractive synthesis,
because it has both odd and even harmonics with high energy.
In general, these waveforms are appreciated as excessively rich in sharp harmonics,
and attenuation by filtering on treble and middle parts of the spectrum is 
especially useful for achieving a more natural and pleasant sound. 
The relatively attenuated harmonics of the triangle wave makes it the more functional 
- among the listed possibilities - to be used in the synthesis of musical notes without any further processing.
The sinusoid is often a nice choice, but a problematic one.
While pleasant if not loud in a very high pitch (above $\approx 500Hz$ it requires careful dosage),
the pitch of a pure sinusoid is not accurately detected by the human auditory system,
particularly at low frequencies.
Also, it requires a great amplitude gain for an increase in loudness of a sinusoid
if compared to other waveforms. Both particularities are understood in the scientific
literature as a consequence of the nonexistence of pure sinusoidal sounds in nature~\cite{Roederer}.
The spectra of each basic waveform is illustrated in Figure~\ref{fig:espectroDeOndas}.

\subsection{Spectra of sampled sounds}
The sinusoidal components in the discretized sound have some particularities.
Considering a signal $T$ and its corresponding Fourier decomposition 
$\mathcal{F}\langle T\rangle=C=\{c_k\}_0^{\Lambda-1}=\left\{ \sum_{i=0}^{\Lambda-1}t_ie^{-j i (k \frac{2 \pi}{\Lambda})} \right\}_-1^{\Lambda-1}$,
the recomposition is the sum of the frequency components to yield the temporal samples\footnote{The 
factor $\frac{1}{\Lambda}$ can be distributed among the Fourier transform and its reconstruction, as preferred.
Note that $j$ here is the imaginary unit $j^2=-1$.}:

\begin{equation}\label{recomposicaoFourier}
\begin{split}
t_i = & \frac{1}{\Lambda}\sum_{k=0}^{\Lambda-1}c_ke^{j \frac{2\pi k}{\Lambda} i } \\ 
    = & \frac{1}{\Lambda}\sum_{k=0}^{\Lambda-1}(a_k+ j . b_k)\left[cos(w_k i)   +j . sen(w_k i)\right]
\end{split}
\end{equation}

\noindent where $c_k = a_k + j . b_k$ defines the amplitude and phase of each frequency:
$w_k=\frac{2\pi}{\Lambda}k$ in radians or $f_k=w_k\frac{f_s}{2\pi}=\frac{f_s}{\Lambda}k$ in Hertz,
and are limited by $w_k\leq\pi$ and $f_k\leq\frac{f_s}{2}$ as given by the Nyquist Theorem.

For a sonic signal, samples $t_i$ are real and are given by the real part of Equation~\ref{recomposicaoFourier}:

\begin{equation}\label{moduloEfase}
\begin{split}
t_i& = \frac{1}{\Lambda}\sum_{k=0}^{\Lambda-1}\left[a_k cos(w_k i) -b_k sen(w_k i)\right] \\
	& = \frac{1}{\Lambda}\sum_{k=0}^{\Lambda-1}\sqrt{a_k^2 + b_k^2} \; cos\left[w_k i - arctan(b_k, a_k)\right]
\end{split}
\end{equation}

\noindent where $arctan(x, y) \in [0, 2\pi]$ is the inverse tangent with the right choice of the quadrant
in the imaginary plane.

$\Lambda$ real samples $t_i$ result in $\Lambda$ complex coefficients $c_k=a_k+j.b_k$.
The coefficients $c_k$ are equivalent two by two,
corresponding to the same frequencies and with the same contribution to its reconstruction.
They are complex conjugates: $a_{k1}=a_{k2}$ and $b_{k1}=-b_{k2}$ and, as a consequence, the modules are equal and phases have opposite signs.
Recalling that $f_k = k\frac{f_s}{\Lambda}, \; k \in \left\{0, ..., \left \lfloor \frac{\Lambda}{2} \right \rfloor \right\} $.
When $k > \frac{\Lambda}{2}$, the frequency $f_k$ is mirrored through $\frac{f_s}{2}$ in this way: $f_k=\frac{f_s}{2} - (f_k-\frac{f_s}{2})=f_s-f_k=f_s - k\frac{f_s}{\Lambda}=(\Lambda-k)\frac{f_s}{\Lambda} \;\;\;\; \Rightarrow \;\;\;\; f_k\equiv f_{\Lambda-k} \; ,\;\; \forall \;\; k<\Lambda$.

The same applies to $w_k=f_k\frac{2\pi}{f_s}$ and the periodicity $2\pi$:
it follows that $w_k=-w_{\Lambda-k} \; ,\;\; \forall \;\; k<\Lambda$.
Given the cosine (an even function) and the inverse tangent (an odd function),
the components in $w_k$ and $w_{\Lambda-k}$ contribute with coefficients $c_k$ = $c^*_{\Lambda-k}$ in the reconstruction of the real samples.
In summary, in a decomposition of $\Lambda$ samples, the $\Lambda$ frequency components $\{c_i\}_0^{\Lambda-1}$ are equivalent in pairs,
except for $f_0$, and, when $\Lambda$ is even, for $f_{\Lambda/2}=f_{\text{max}}=\frac{f_s}{2}$.
Both these components are isolated, i.e.\ there is one and only one component at frequency $f_0$ or $f_{\Lambda/2}$ (if $\Lambda$ is even).
In fact, when $k=0$ or $k=\Lambda/2$ the mirror of the frequencies are themselves: $f_{\Lambda/2}=f_{(\Lambda-\Lambda/2) = \Lambda/2}$ and $f_0=f_{(\Lambda-0)=\Lambda}=f_0$.
Furthermore, these two frequencies (zero and Nyquist frequency) do not have a phase offset: their coefficients are strictly real.
Therefore, the number $\tau_\Lambda$ of equivalent coefficient pairs in a decomposition of $\Lambda$ samples is:

\begin{equation}\label{coefsPareados}
	\tau_\Lambda = \frac{\Lambda - \Lambda \% 2}{2} -2 +\Lambda \% 2 = \left \lceil \frac{\Lambda}{2} \right \rceil -2
\end{equation}

This discussion can be summarized in the following equivalences:

\begin{align}
	f_{k}\equiv f_{\Lambda-k}& \;\; , \;\; w_{k}\equiv-w_{\Lambda-k} \label{equivalenciasFreqs} \\
	 a_k = a_{\Lambda -k}& \;\; , \;\;b_k = - b_{\Lambda -k} \label{equavalenciasTermos} \\
	\sqrt{a_k^2 + b_k^2} & = \sqrt{a_{\Lambda - k}^2 + b_{\Lambda -k}^2}\label{equivalenciasModulos} \\
	arctan(b_k, a_k) & = -arctan(b_{\Lambda -k}, a_{\Lambda - k}) \label{equivalenciasFases}
\end{align}

\noindent with
$\forall \; 1 \leq k \leq \tau_\Lambda$, 
$k \in \mathbb{N}$.

To express the general case for components combination in each sample $t_i$,
one can gather the relations for the reconstruction of a real signal (Equation~\ref{moduloEfase}),
for the number of paired coefficients (Equation~\ref{coefsPareados}),
and for the equivalences of modules (Equation~\ref{equivalenciasModulos})
and phases (Equation~\ref{equivalenciasFases}):

\begin{equation}\label{eq:reconsCompleta}
\begin{split}
	t_i = \frac{a_0}{\Lambda} + \frac{ a_{\Lambda/2}}{\Lambda}(1-\Lambda\% 2) + \frac{2}{\Lambda}\sum_{k=1}^{\tau_\Lambda}\sqrt{a_k^2 + b_k^2} \; cos\left[w_k i - arctan(b_k,a_k)\right]
\end{split}
\end{equation}

 \begin{figure*}
     \centering
         \includegraphics[width=.7\textwidth]{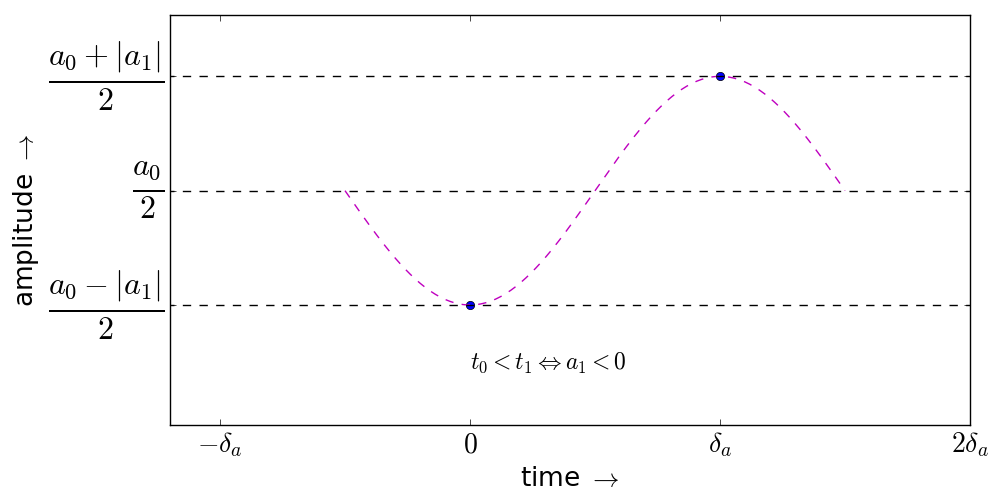}
	 \caption{Oscillation of 2 samples (maximum frequency for any $f_s$). The first coefficient reflects a constant detachment (called \emph{offset}, \emph{bias} or \emph{DC component}) and the second coefficient specifies the oscillation amplitude.}
         \label{fig:amostras2}
 \end{figure*}

Figure~\ref{fig:amostras2} shows two samples and their spectral component. When there is only two samples, the Fourier decomposition has only one pair of coefficients $\{c_k=a_k-j.b_k\}_0^{\Lambda-1=1}$ relative to frequencies $\{f_k\}_0^1=\left\{w_k\frac{f_s}{2\pi}\right\}_0^1=\left\{k\frac{f_s}{\Lambda=2}\right\}_0^1=\left\{0,\frac{f_s}{2}=f_{\text{max}}\right\}$
with energies $e_k=\frac{(c_k)^2}{\Lambda=2}$. The role of amplitudes $a_k$ is clearly observed with $\frac{a_0}{2}$, the fixed offset (also called \emph{bias} or \emph{DC component}), and $\frac{a_1}{2}$ for the oscillation with frequency $f_1=\frac{f_s}{\Lambda=2}$.
This case has special relevance: at least 2 samples are necessary to represent an oscillation and it yields the Nyquist frequency $f_{\text{max}}=\frac{f_s}{2}$, which is the maximum frequency in a sound sampled with $f_s$ samples per second. In fact, any discrete-time signal has this property, not only digitized sound.

\begin{figure*}
     \centering
         \includegraphics[width=.7\textwidth]{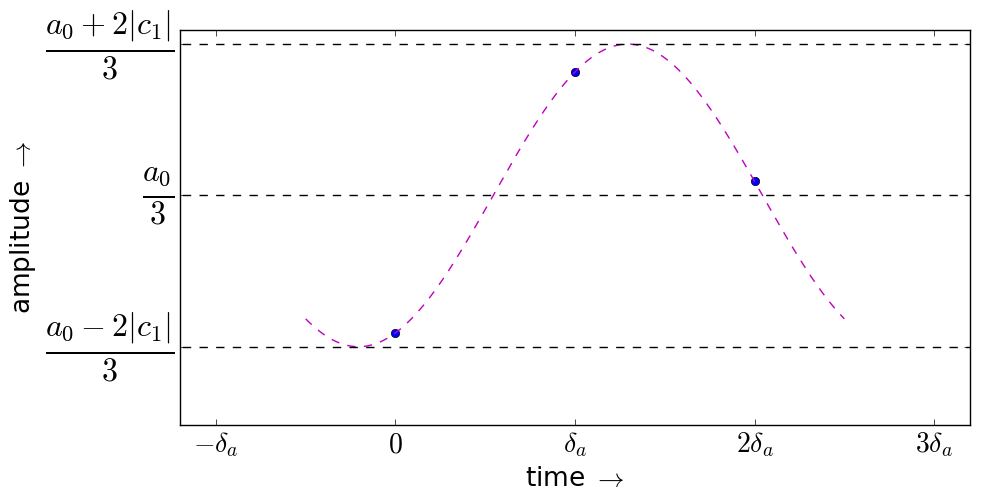}
     \caption{Three fixed samples present only one non-null frequency. $c_1=c_2^*$ and $w_1 \equiv w_2$.}
         \label{fig:amostras3}
 \end{figure*}

All fixed sequences $T$ of only $3$ samples also have just $1$ frequency, since the first harmonic would have $1.5$ samples and exceeds the bottom limit of 2 samples, i.e.\ the frequency of the harmonic would exceed the Nyquist frequency:  $\; \frac{2. f_s}{3} > \frac{f_s}{2}$. 
The coefficients $\{c_k\}_0^{\Lambda-1=2}$ are present in 3 frequency components. One is relative to frequency zero ($c_0$), and the other two ($c_1$ and $c_2$) have the same role for reconstructing a sinusoid with $f=f_s/3$.
This case is illustrated in Figure~\ref{fig:amostras3}.

\begin{figure*}
    \centering
        \includegraphics[width=.7\textwidth]{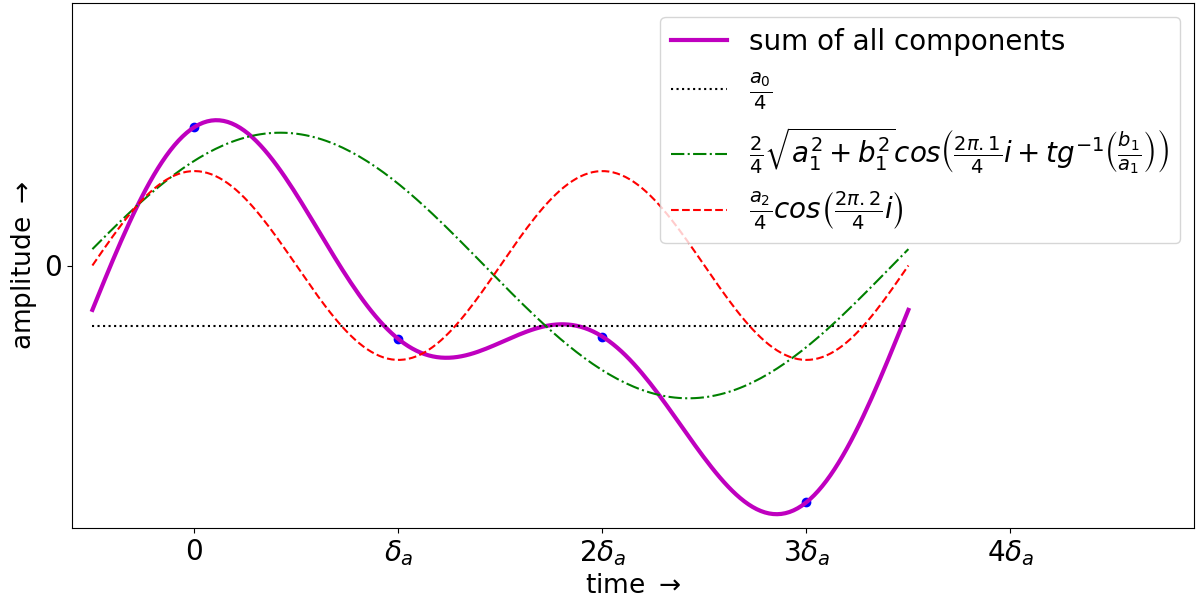}
    \caption{Frequency components for 4 samples.}
        \label{fig:amostras4}
\end{figure*}

\begin{figure*}
    \centering
        \includegraphics[width=.9\textwidth]{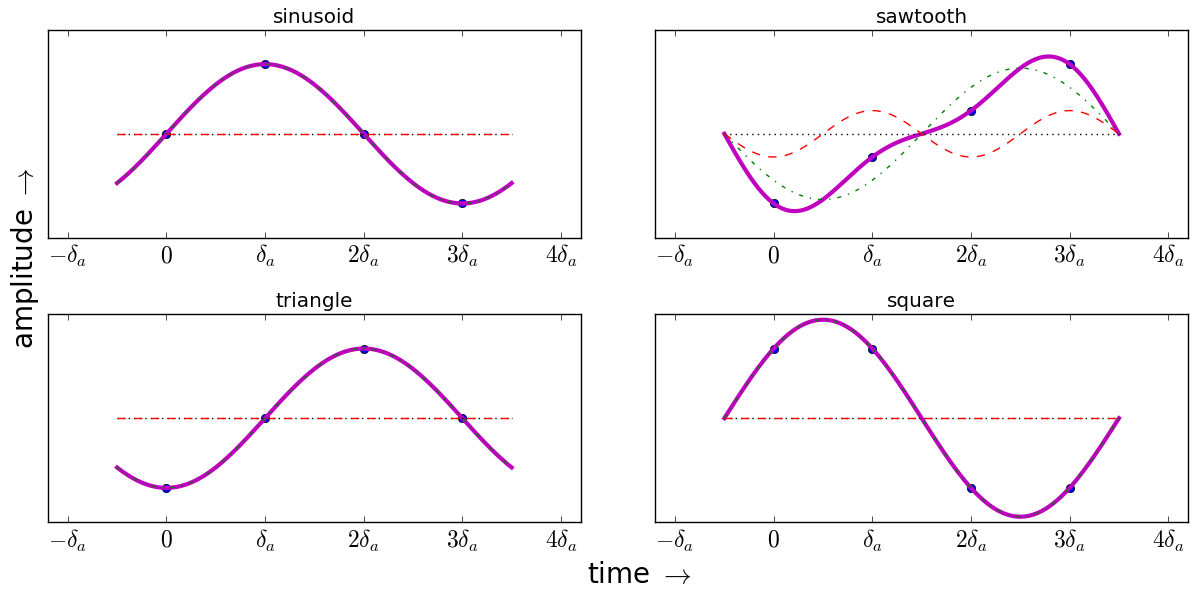}
    \caption{Basic wave forms with 4 samples.}
        \label{fig:formas4}
\end{figure*}

\begin{figure*}[!h!]
    \centering
        \includegraphics[width=.7\textwidth]{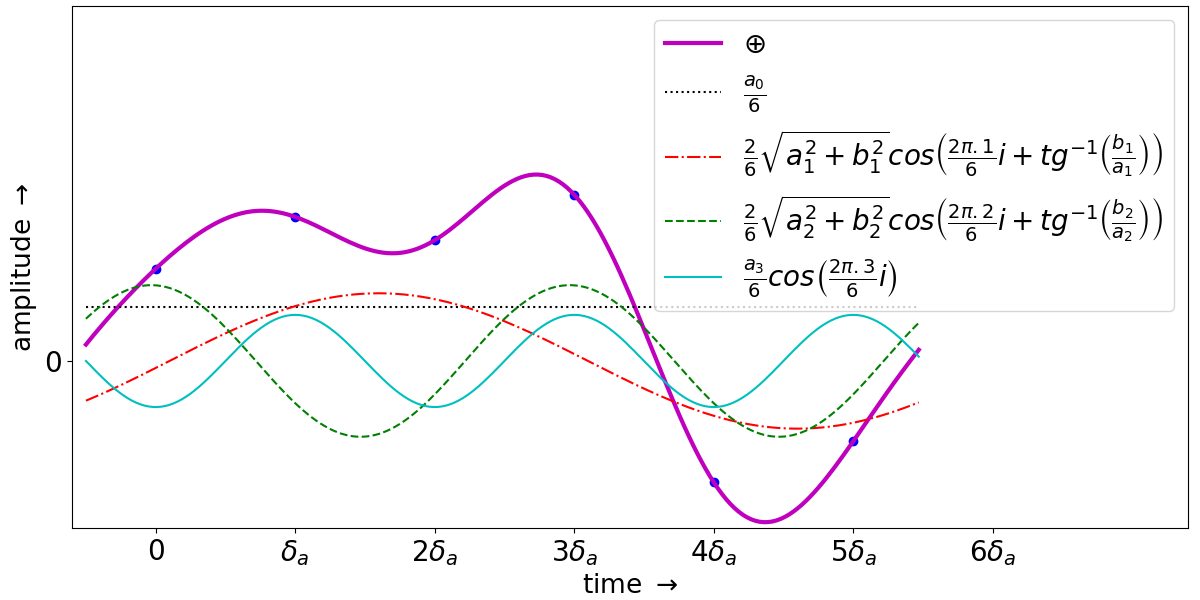}
    \caption{Frequency components for 6 samples: 4 sinusoids, one of them is the \emph{bias} with zero frequency.}
        \label{fig:amostras6}
\end{figure*}

With 4 samples it is possible to represent 1 or 2 frequencies with independence of magnitude and phase.
Figure~\ref{fig:amostras4} depicts contribution of each of the two (possible) components.
The individual components sum to the original waveform and a brief inspection reveals 
the major curvatures resulting from the higher frequency,
while the fixed offset is captured in the component with frequency $f_0=0$.
Figure~\ref{fig:formas4} shows the harmonics for the basic waveforms of 
Equations~\ref{sinusoid},~\ref{sawTooth},~\ref{triangular} and~\ref{square}
in the case of 4 samples.
There is only 1 sinusoid for each waveform, with the exception of the sawtooth, which has even harmonics.

\begin{figure*}[!h!]
    \centering
        \includegraphics[width=.9\textwidth]{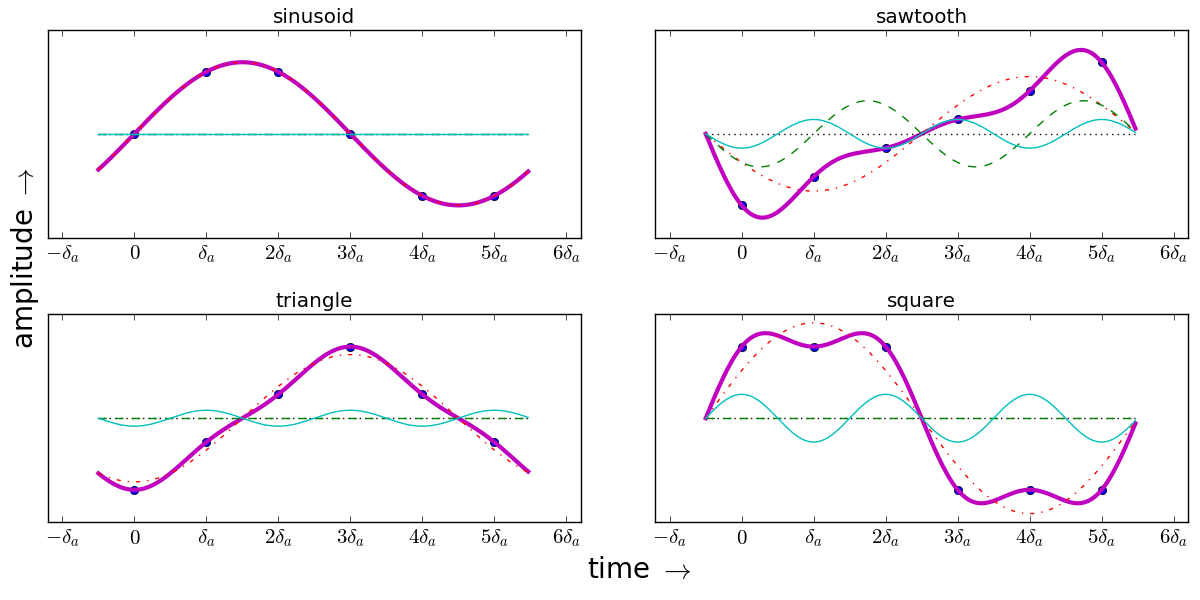}
    \caption{Basic waveforms with 6 samples: triangular and square waveforms have odd harmonics, with different proportions and phases; the sawtooth has even harmonics.}
        \label{fig:formas6}
\end{figure*}

Figure~\ref{fig:amostras6} exposes the sinusoidal components within 6 samples,
while Figure~\ref{fig:formas6} presents the decomposition of the basic wave forms:
square and triangular have the same components but with different proportions, while the sawtooth has an extra component.

\subsection{The basic note}\label{notaBasica}
In a nutshell
a sequence $T$ of sonic samples separated by $\delta_a=1/f_s$ expresses a musical note with
a frequency of $f$ Hertz\footnote{Let $f$ be such that it divides $f_s$.
As mentioned before, this limitation simplifies the exposition for now
and will be overcome in the next section.} and $\Delta$ seconds of duration if,
and only if, it has the periodicity $\lambda_f=f_s/f$ and size $\Lambda=\lfloor f_s . \Delta \rfloor$:

\begin{equation}\label{eq:notaBasica}
T^{f,\; \Delta}=\{t_{i \, \% \lambda_f} \}_0^{\Lambda-1}= \left \{t^f_{i \; \% \left( \frac{f_s}{f} \right) } \right \}_0^{\Lambda-1}
\end{equation}

Such note still does not have timbre: it is necessary to choose a waveform for the samples $t_i$ to have a value.
Any waveform can be used to further specify the note,
where $\lambda_f=\frac{f_s}{f}$ is the number of samples in each period.
Let $L^f \in \{S^f,Q^f,T^f,D^f,R^f \}$
(as given by Equations~\ref{sinusoid},~\ref{sawTooth},~\ref{triangular}
and~\ref{square} and let $R_i^f$ be a sampled waveform)
be the sequence that describes a period of the waveform 
 with duration $\delta_f=1/f$:

\begin{equation}\label{periodoUnico}
L^{f} = \left\{ l_i^f \right\}_0^{\delta_f . f_s -1}=\left\{ l_i^f \right\}_0^{\lambda_f-1}
\end{equation}

Thereafter, the sequence $T$ for a note of duration $\Delta$ and frequency $f$ is:

\begin{equation}\label{eq:notaBasicaTimbre}
T^{f,\; \Delta}=\left\{t_i^f\right\}_0^{\lfloor f_s . \Delta \rfloor -1}=\left \{ l^f_{i\,\%\left(\frac{f_s}{f}\right)} \right \}_0^{\Lambda-1}
\end{equation}

\subsection{Spatialization: localization and reverberation}\label{subsec:spac}
A musical note is always spatialized
(i.e. it is always produced within the ordinary three dimensional physical space)
even though it is not one of its four basic properties in canonical musical theory (duration, loudness, pitch and timbre).
The consideration of this fact is the subject of the spatialization knowledge field and practice\footnote{By
spatialization one might find both:
1) the consideration of cues in sound that derive from the environment,
including the localization of the listener and the sound source;
2) techniques to produce sound through various sources, such as loudspeakers, singers and traditional musical instruments, for musical purposes.
We focus in the first issue although issues of the second are also tackled
and they are obviously intermingled.}.
A note has a source which has a three dimensional position.
This position is the spatial localization of the sound.
It is often (modeled as) a single point but can be a surface or a volume.
The reverberation in the environment in which a sound occurs is another main topic of spatialization.
Both concepts, spatial localization and reverberation,
are widely valued by composers, audiophiles and the music industry~\cite{floEsp}.

\subsubsection{Spatial localization}
It is understood that the perception of sound localization occurs in our nervous system
mainly by three cues: the delay of incoming sound (and its reflections in the surfaces) between both ears,
the difference of sound intensity at each ear and the filtering performed by the human body,
specially its chest, head and ears~\cite{Roederer, hrtf, Heeger}. 

\begin{figure}[h!]
    \centering
        \includegraphics[width=.5\textwidth]{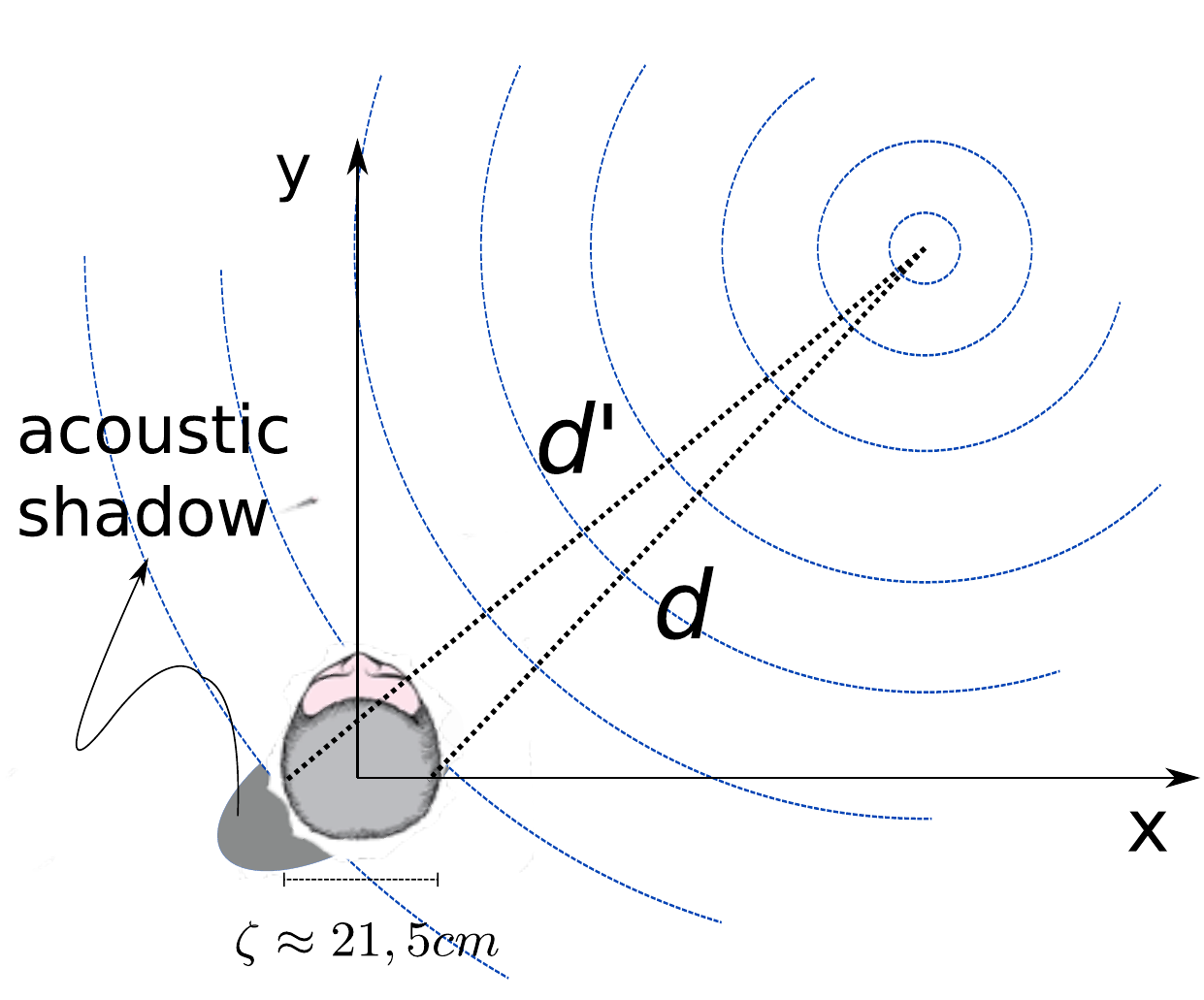}
    \caption{Detection of sound source localization: schema used to calculate the Interaural Time Difference (ITD) and the Interaural Intensity Difference (IID).}
    \label{fig:spac}
\end{figure}

An object placed at $(x,y)$, as in Figure~\ref{fig:spac}, is distant of each ear by:

\begin{equation}\label{eq:distOuvidos}
\begin{split}
d & =\sqrt{\left (x-\frac{\zeta}{2} \right )^2+y^2} \\
d' & =\sqrt{\left (x+\frac{\zeta}{2} \right )^2 + y^2}
\end{split}
\end{equation}

\noindent where $\zeta$ is the distance between the ears, known to be $\zeta \approx 21.5cm$ in an adult human.
The cues for the sonic localization are not easy to calculate, but,
in a very simplified model, useful for musical purposes,
straightforward calculations result in the Interaural Time Difference:

\begin{equation}\label{eq:dti}
ITD=\frac{d'-d}{v_{sound\;at\;air}\approx 343.2 }\quad \text{seconds}
\end{equation}

\noindent and in the Interaural Intensity Difference:

\begin{equation}\label{eq:dii}
IID=20\log_{10}\left (\frac{d}{d'}\right) \quad decibels
\end{equation}

$IID_a=\frac{d}{d'}$ can be used as a multiplicative constant to the right channel of a stereo sound signal
together with ITD~\cite{Heeger}:
\begin{equation}\label{eq:locImpl}
\begin{split}
\Lambda_{ITD} & = \left \lfloor \frac{d'-d}{343,2}  f_s \right \rfloor \\
IID_a & = \frac{d}{d'} \\
\left\{t_{(i+\Lambda_{ITD})}'\right\}_{\Lambda_{ITD}}^{\Lambda+\Lambda_{ITD}-1} & =\left\{IID_a . t_i\right\}_0^{\Lambda-1} \\
\left\{t_i'\right\}_0^{\Lambda_{ITD}-1} & = 0
\end{split}
\end{equation}

\noindent where,
where $\{t_i'\}$ are samples of the wave incident in the left ear, $\{t_i\}$ are samples for the right ear,
and $\Lambda_{ITD}=\lfloor ITD . f_s \rfloor$.
If $\Lambda_{ITD} < 0$, it is necessary to change $t_i$ by $t_i'$
and use $\Lambda_{ITD}'= | \Lambda_{ITD} |$ and $IID_a'=1 / IID_a$.

Spatial localization depends considerably on other cues.
By using only ITD and IID it is possible to specify solely
the horizontal angle (azimuthal) $\theta$ given by:

\begin{equation}\label{eq:angulo}
\theta=\arctan( y, x )
\end{equation}

\noindent with $x,y$ as presented in Figure~\ref{fig:spac}.
Notice that
the same pair of ITD and IID (as defined in Equations~\ref{eq:dti} and~\ref{eq:dii})
is related to all the points in a vertical circle parallel to
the head, i.e. the source can have any horizontal component inside the circle.
Such a circle is called the "cone of confusion".
In general, one can assume that the source is in the same horizontal plane
as the listener and at its front
(because humans are prone to hearing frontal and horizontal sources).
Even in such cases, there are other important cues for sound localization.
Consider the acoustic shadow depicted in Figure~\ref{fig:spac}:
for lateral sources the inference of the azimuthal angle is especially dependent
on the filtering of frequencies by the head, pinna (outer ear) and torso.
Also, low frequencies diffract and arrive to the opposite ear with a greater ITD. 
The complete localization, including height and distance of a sound source,
is given by the Head Related Transfer Function (HRTF).
There are well known open databases of HRTFs, such as CIPIC,
and it is possible to apply such transfer functions in a sonic 
signal by convolution (see Equation~\ref{eq:conv}).
Each human body has its own filtering and there are techniques 
to generate HRTFs to be universally used.~\cite{lazaSPA,CIPIC,floEsp,Heeger,hrtf} 

\subsubsection{Reverberation}
The reverberation results from sound reflections and absorption by the environment (e.g. a room) surface where a sound occurs.
The sound propagates through the air with a speed of $\approx 343.2m/s$ and can be emitted from a source with any directionality pattern. When a sound front encounters a surface there are: 1) inversion of the propagation speed component normal to the surface;  2) energy absorption, especially in high frequencies.
The sonic waves propagate until they reach inaudible levels (and further but then can often be neglected).
As a sonic front reaches the human ear, it can be described as the original sound, with the last reflection point as the source, and the absorption filters of each surface it has reached.
It is possible to simulate reverberations that are impossible in real systems.
For example, it is possible to use asymmetric reflections with relation to the axis perpendicular to the surface,
to model propagation in a space with more than three dimensions,
or consider a listener located in various positions.

There are reverberation models less related to each independent reflection and that explores valuable cues to the auditory system. In fact, reverberation can be modeled with a set of two temporal and two spectral regions~\cite{JOSPhy}:
\begin{itemize}
   \item First period: 'first reflections' are more intense and scattered.
   \item Second period: 'late reverberation' is practically a dense succession of indistinct delays with exponential decay and statistical occurrences.
   \item First band: the bass has some resonance bandwidths relatively spaced.
   \item Second band: mid and treble have a progressive decay and smooth statistical fluctuations.
\end{itemize}

Smith III states that usual concert rooms have a total reverberation time of $\approx 1.9$ seconds, and that the period of first reflections is around $0.1s$.
With these values, there are perceived wave fronts which propagate for $652.08m$ before reaching the ear.
In addition, sound reflections made after propagation for
$34.32m$ have incidences less distinct by hearing.
These first reflections are particularly important to spatial sensation.
The first incidence is the direct sound, described by ITD and IID e.g. as in Equations~\ref{eq:dti} and~\ref{eq:dii}.
Assuming that each one of the first reflections, before reaching the ear,
will propagate at least $3-30m$, depending on the room dimensions,
the separation between the first reflections is $8-90ms$.
Also, it is experimentally verifiable that the number of reflections increases with the square of time.
A discussion about the use of convolutions and filtering to favor the implementation of these phenomena is provided in Section~\ref{subsec:mus2}, particularly in the paragraphs about reverberation.~\cite{JOSPhy}

\subsection{Musical usages}\label{subsec:basMus}
Once the basic note is defined, it is didactically convenient to build musical structures with sequences based on these particles.
The sum of the amplitudes of $N$ sequences with same size $\Lambda$ results in the overlapped spectral contents of each sequence, in a process called mixing:

\begin{equation}\label{eq:mixagem}
\{t_i\}_0^{\Lambda-1}=\left \{ \sum_{k=0}^{N-1}t_{k,i} \right \}_0^{\Lambda-1}
\end{equation}

\begin{figure}[h!]
    {\centering
        \includegraphics[width=.5\columnwidth]{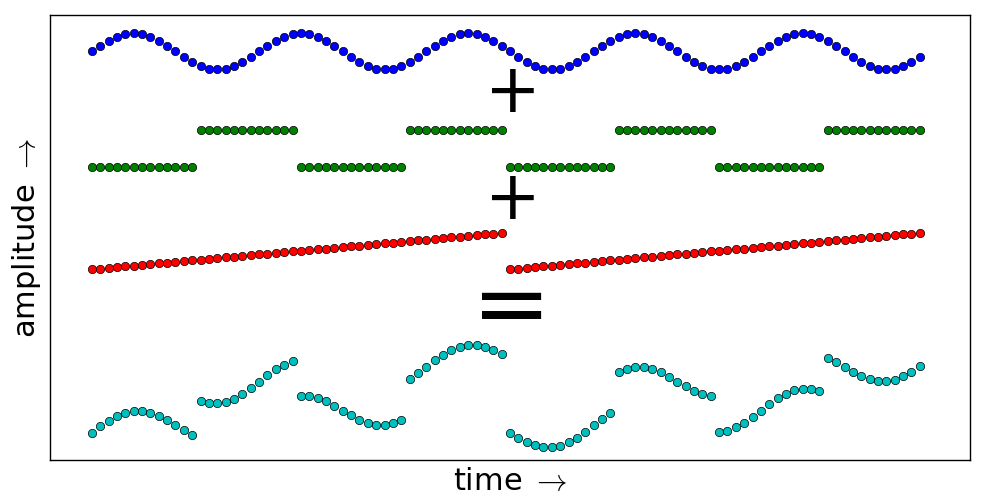}}
    \caption{Mixing of three sonic sequences. The amplitudes are directly summed sample-by-sample.}
        \label{fig:mixagem}
\end{figure}

Figure~\ref{fig:mixagem} illustrates this overlapping process of discretized sound waves, each with 100 samples. If $f_s=44.1kHz$, the frequencies of the sawtooth, square and sine wave are, respectively: $\frac{f_s}{100/2}=882Hz$, $\frac{f_s}{100/4}=1764Hz$ and $\frac{f_s}{100/5}=2205Hz$. The duration of each sequence is very short $\frac{f_s=44.1kHz}{100} \approx 2ms$. One can complete the sequence with zeroes to sum (mix) sequences with different sizes.

The mixed notes are generally separated by the ear according to the physical laws of resonance and by the nervous system~\cite{Roederer}.
This process of mixing musical notes results in musical harmony whose intervals between frequencies and chords of simultaneous notes guide subjective and abstract aspects of music appreciation~\cite{Harmonia} and are addressed in Section~\ref{sec:notesMusic}. 

Sequences can be concatenated in time. If the sequences $\{t_{k,i}\}_0^{\Lambda_k-1}$ represent musical notes,
their concatenation in a unique sequence $T$ is a simple melodic sequence (or melody):

\begin{equation}\label{eq:concatenacao}
\begin{split}
T = \{t_i\}_0^{\sum\Delta_k-1}= & \{t_{l,i}\}_0^{\sum\Delta_k-1}, \;\; \\ l\text{ smallest integer } : & \quad \Lambda_l > i -\sum_{j=0}^{l-1}\Lambda_j
\end{split}
\end{equation}

\begin{figure}[h!]
{    \centering
        \includegraphics[width=.5\columnwidth]{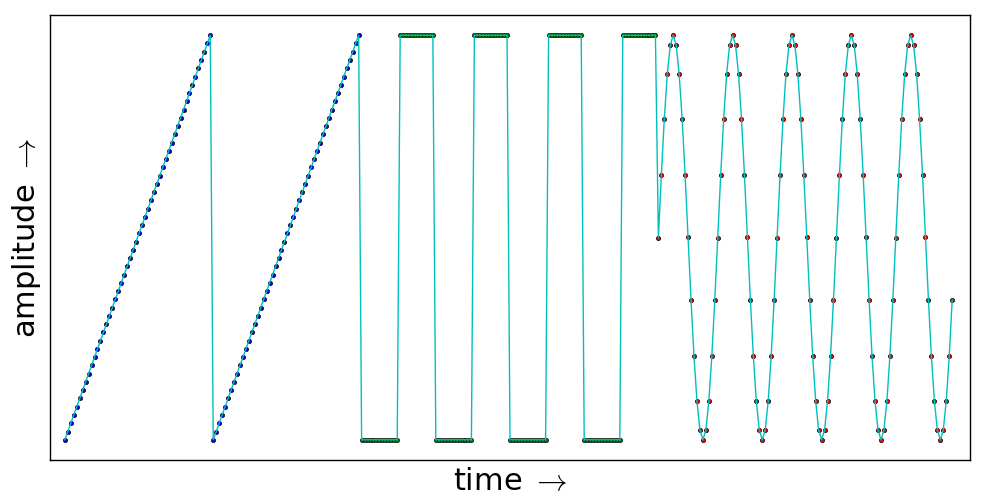}}
    \caption{Concatenation of three sounds.}
        \label{fig:concatenacao}
\end{figure}

This mechanism is illustrated in Figure~\ref{fig:concatenacao} with the same sequences of Figure~\ref{fig:mixagem}. Although the sequences are short for the usual sample rates, it is easy to visually observe the concatenation of sound sequences. In addition, each note has  a duration larger than $100ms$ if $f_s<1kHz$ (but need to oscillate faster to yield audible frequencies).

The musical piece \emph{reduced-fi} explores the temporal juxtaposition of notes, resulting in a homophonic piece.
The vertical principle (mixing) is demonstrated at the \emph{sonic portraits}, static sounds with peculiar spectrum.~\cite{MASSA}

With the basic musical note in discrete-time audio carefully described, the next section develops the temporal evolution of its contents as in \emph{glissandi} and intensity envelopes. Filtering of spectral components and noise generation complements the musical note as a self-contained unit. Section~\ref{notasMusica} is dedicated to the organization of these notes e.g. by using metrics and trajectories, with regards to traditional music theory.

\section{Variation in the basic note}\label{sec:internalVar}\label{sec:varInternas}
The basic digital music note defined in Section~\ref{sec:notaDisc} has the following parameters: duration, pitch, intensity (loudness) and timbre. This is a useful and paradigmatic model, but it does not exhaust all the aspects of a musical note. First of all, characteristics of the note change along the note itself~\cite{Chowning}. For example, a $3s$ piano note has intensity with an abrupt rise at the beginning and a progressive decay, has spectral variations with harmonics decaying and some others emerging along time. These variations are not mandatory, but they are used in sound synthesis for music because they reflect how sounds appear in nature. This is considered so important that there is a rule of thumb: to make a sound that incites interest by itself, arrange internal variations on it~\cite{Roederer}.
To explore all the ways by which variations occur within a note is out of the scope of any work, given the sensibility of the human ear and the complexity of human sound cognition. In this section, primary resources are presented to produce variations in the basic note. It is worthwhile to recall that all the relations in this and other sections are implemented in Python and published in public domain. The musical pieces \emph{ParaMeter transitions}; \emph{Shakes and wiggles}; \emph{Tremolos, vibratos and frequency}; \emph{Little train of impulsive hillbillies}; \emph{Noisy band}; \emph{Bela rugosi}; \emph{Children choir}; and \emph{ADa and SaRa} were made to validate and illustrate concepts of this section. The code that synthesizes these pieces is also part of the toolbox.~\cite{MASSA}
 
\subsection{Lookup table}\label{subsec:lookup}
The \emph{Lookup Table} (LUT) is an array for indexed operations which substitutes continuous and repetitive calculations.
It is used to reduce computational complexity and for employing functions without calculating them directly, e.g. from sampled data or hand picked values.
In music its usage transcends these applications: it simplifies many operations and enables the use a single wave period to synthesize sounds in the whole audible spectrum, with any waveform.

Let $\widetilde{\Lambda}$ be the wave period in samples and $\widetilde{L} = \left\{\, \widetilde{l}_i \,\right\}_0^{\widetilde{\Lambda} -1}$ the sample sequence with the waveform.
A sequence $T^{f,\,\Delta}$ with samples of a sound with frequency $f$ and duration $\Delta$ can be obtained by means of $\widetilde{L}$:

\begin{figure*}[t!]
    \centering
        \includegraphics[width=.7\textwidth]{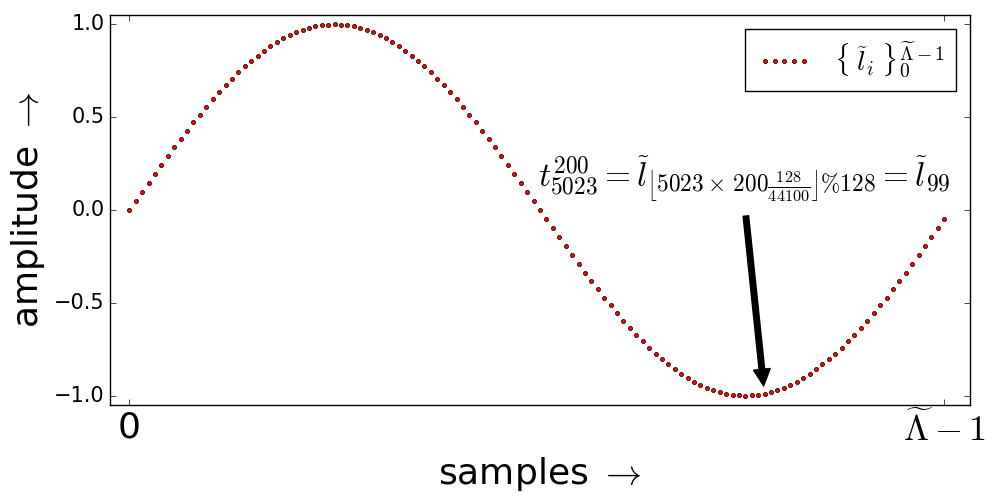}
	\caption{Search (lookup) in a reference table (\emph{lookup table} or LUT) to synthesize sounds at different frequencies using a unique waveform with high resolution.
	Each \emph{i-th} sample $t_i^f$ of a sound with frequency $f$ is related to the samples in the table $\widetilde{L}=\{\widetilde{l}_i\}_0^{\widetilde{\Lambda}-1}$ by $t_i^{f}=\widetilde{l}_{\left\lfloor i f\frac{\widetilde{\Lambda}}{f_s} \right\rfloor \%\,\widetilde{\Lambda}}$ where $f_s$ is the sampling rate.}
        \label{fig:lut}
\end{figure*}

\begin{equation}\label{eq:lut}
\begin{split}
T^{f,\,\Delta}=\left\{t_i^f\right\}_0^{\lfloor \, f_s . \Delta \, \rfloor -1} = \left\{ \, \widetilde{l}_{\gamma_i \% \widetilde{\Lambda} }\, \right\}_{0}^{\Lambda-1}\; , \quad \text{where} \;\; \gamma_i = \left \lfloor i f \frac{ \widetilde{\Lambda}}{f_s} \right \rfloor  
\end{split}
\end{equation}

In other words, with the right LUT indexes ($\gamma_i\%\widetilde{\Lambda}$) it is possible to synthesize sounds at any frequency.
Figure~\ref{fig:lut} illustrates the calculation of a sample $t_i$ from $\left\{\,\widetilde{l}_i\,\right\}$ for $f=200Hz$, $\widetilde{\Lambda}=128$ and adopting the sample rate of $f_s=44.1kHz$. Though this is not a practical configuration (as discussed below), it allows for a graphical visualization of the procedure.

The calculation of the integer $\gamma_i$ introduces noise which decreases as $\widetilde{\Lambda}$ increases.
In order to use this calculation in sound synthesis,
with $f_s=44.1 kHz$, the standard guidelines suggest the use of $\widetilde{\Lambda} = 1024$ samples,
yielding a noise level of $\approx -60dB$.
Larger tables might be used to achieve sounds with a greater quality.
Also, a rounding or interpolation method can be used,
but we advocate the use of a larger table since it does not
introduce relevant computation overhead.~\cite{Geiger}

The expression defining the variable $\gamma_i$ can be understood as $f_s$ being added to $i$ at each second.
If $i$ is divided by the sample frequency, $\frac{i}{f_s}$
is incremented by $1$ at each second. Multiplied by the period, it results in $i \frac{\widetilde{\Lambda}}{f_s}$, which covers the period in one second.
Finally, with frequency $f$ it results in $i f \frac{\widetilde{\Lambda}}{f_s}$ which completes $f$ periods $\widetilde{\Lambda}$ in $1$ second, i.e. the resulting sequence presents the fundamental frequency $f$.

There are important considerations here: it is possible to use practically any frequency $f$.
Limits exist only at low frequencies when the size of table $\widetilde{\Lambda}$ is not sufficient for the sample rate $f_s$.
The lookup procedure is virtually costless and replaces calculations by simple indexed searches 
(what is generally understood as an optimization process).
Unless otherwise stated, this procedure will be used along all the following discussions for every applicable case.
LUTs are broadly used in computational implementations for music,
and are known also as wavetables.
A classical usage of LUTs is known as \emph{Wavetable Synthesis},
which generally consists of many LUTs used together to generate a quasi-periodic music note~\cite{Cook,Wavetable}.

\subsection{Incremental variations of frequency and intensity}\label{subsec:vars}
As stated by the (Weber and) Fechner law~\cite{Weber-Fechner},
human perception holds a logarithmic relation to stimulus.
That is to say, the exponential progression of a stimulus is perceived as linear.
For didactic reasons, and given its use in AM and FM synthesis (Section~\ref{subsec:tvaf}), linear variation is discussed first.

Consider a note with duration $\Delta = \frac{\Lambda}{f_s}$, in which the frequency $f=f_i$ varies linearly from $f_0$ to $f_{\Lambda -1}$. Thus:

\begin{equation}\label{freqLinear}
 F=\{f_i\}_0^{\Lambda-1}=\left\{f_0 + (f_{\Lambda-1}-f_0)\frac{i}{\Lambda-1} \right\}_0^{\Lambda-1}
\end{equation}

\begin{equation}\label{indiceLinear}
\begin{split}
	\Delta_{\gamma_i}=\frac{\widetilde{\Lambda}}{f_s}f_i \quad \Rightarrow \quad \gamma_i & = \left \lfloor \sum_{j=0}^{i} \frac{\widetilde{\Lambda}}{f_s}f_j \right \rfloor \\
	\gamma_i & =  \left \lfloor \sum_{j=0}^{i} \frac{\widetilde{\Lambda}}{f_s} \left [f_0 + (f_{\Lambda-1}-f_0)\frac{j}{\Lambda-1} \right ] \right \rfloor 
\end{split}
\end{equation}

\begin{equation}\label{serieAmostralLin}
 \left\{t_i^{\;\overline{f_0,\, f_{\Lambda-1}}}\right\}_0^{\Lambda-1}=\left\{\,\widetilde{l}_{\gamma_i \% \widetilde{\Lambda}}\,\right\}_0^{\Lambda-1}
\end{equation}

\noindent where $\Delta_{\gamma_i}=f_i\frac{\widetilde{\Lambda}}{f_s}$ is the LUT increment between two samples given the sound frequency of the first sample.
There is a general rule to be noticed here: when a sound has variations in the fundamental frequency,
one should account for them in the LUT indexing.
The resulting indexes can be found by a cumulative sum of each indexing displacement.
The equations for linear pitch transition are:

\begin{equation}\label{freqExponencial}
 F=\{f_i\}_0^{\Lambda-1}=  \left\{f_0 \left ( \frac{f_{\Lambda-1}}{f_0} \right ) ^{\frac{i}{\Lambda -1}} \right\}_0^{\Lambda-1}
\end{equation}

\begin{equation}\label{indiceExponencial}
\begin{split}
 \Delta_{\gamma_i}= \frac{\widetilde{\Lambda}}{f_s}f_i \quad \Rightarrow  \quad \gamma_i= & \left \lfloor \sum_{j=0}^{i} \frac{\widetilde{\Lambda}}{f_s}f_j \right \rfloor \\
	\gamma_i = & \left \lfloor \sum_{j=0}^{i} f_0 \frac{\widetilde{\Lambda}}{f_s} \left ( \frac{f_{\Lambda-1}}{f_0} \right ) ^{\frac{j}{\Lambda -1}} \right \rfloor
\end{split}
\end{equation}

\begin{equation}\label{serieAmostralLog}
 \left\{t_i^{\;\overline{f_0,\,f_{\Lambda-1}}}\right\}_0^{\Lambda-1}=\left\{\,\widetilde{l}_{\gamma_i \% \widetilde{\Lambda}}\,\right\}_0^{\Lambda-1}
\end{equation}

\begin{figure}[h!]
     \centering
         \includegraphics[width=.7\columnwidth]{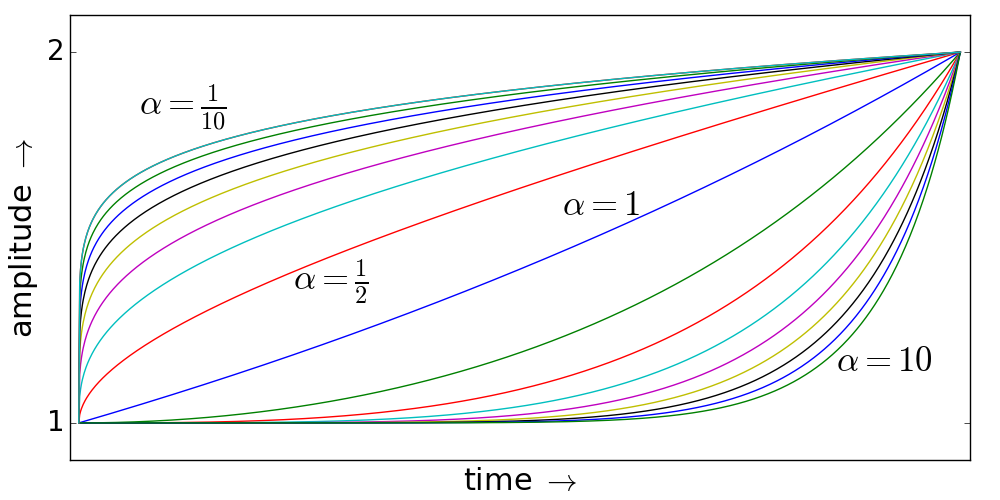}
     \caption{Intensity transitions for different values of $\alpha$ (see Equations~\ref{seqAmp} and~\ref{transAmp}).}
         \label{fig:transicao}
\end{figure}

The term $\frac{i}{\Lambda-1}$ covers the interval $[0,1]$ and it is possible to raise it to a power $\alpha\geq0$ in such a way that the beginning of the transition will be smoother or steeper.
This procedure is especially useful for energy variations with the purpose of changing the loudness\footnote{See Section~\ref{subsec:volume} for considerations about loudness, amplitudes and decibels.}.
Thus, for amplitude variations:
\begin{equation}\label{seqAmp}
\begin{split}
 \{a_i\}_0^{\Lambda-1}= \left \{ a_0 \left ( \frac{a_{\Lambda-1}}{a_0} \right )^{\left ( \frac{i}{\Lambda-1} \right )^\alpha} \right \}_0^{\Lambda-1}= \left \{ \left ( {a_{\Lambda-1}} \right )^{\left ( \frac{i}{\Lambda-1} \right )^\alpha} \right \}_0^{\Lambda-1}
\end{split}
\end{equation}
where $a_0$ is the initial amplitude factor and
$a_{\Lambda-1}$ is an amplitude factor to be reached at the end of the transition.
Applying the loudness transition to a sonic sequence $T$:

\begin{equation}\label{transAmp}
\begin{split}
 T^{'}=T \odot A = \{t_i . a_i\}_0^{\Lambda-1} = \left \{ t_i . (a_{\Lambda-1} )^{\left ( \frac{i}{\Lambda-1} \right )^\alpha} \right \}_0^{\Lambda-1}
\end{split}
\end{equation}

It is often convenient to have $a_0=1$ to start a new sequence with the original amplitude and then progressively change it.
If $\alpha=1$, the amplitude variation follows the exponential progression that is related to the linear variation of loudness. Figure~\ref{fig:transicao} depicts transitions between values 1 and 2 and for different values of $\alpha$, a gain of $\approx 6dB$ as given by Equation~\ref{eq:ampVol}.

Special attention should be dedicated while considering $a=0$.
In Equation~\ref{seqAmp}, $a_0=0$ results in a division by zero and if $a_{\Lambda-1}=0$, there will be a multiplication by zero.
Both cases make the procedure useless, once a ratio of any number in relation to zero is not well defined for our purposes. It is possible to solve this dilemma choosing a number that is small enough like $-80dB\;\Rightarrow a=10^{\frac{-80}{20}}=10^{-4}$ as the minimum loudness for a \emph{fade in} ($a_0=10^{-4}$) or for a \emph{fade out} ($a_{\Lambda-1}=10^{-4}$). A linear fade can be used then to reach zero amplitude, if needed. Another common solution is the use of the quartic polynomial term $x^4$, as it reaches zero without these difficulties and gets reasonably close to the curve with $\alpha=1$ as it departs from zero~\cite{Cook}.

Using Equations~\ref{ampDec} and~\ref{transAmp} specifying a transition of $V_{dB}$ decibels:

\begin{equation}\label{seqAmpDB}
T^{'}=\left\{ t_i 10^{\frac{V_{dB}}{20}\left( \frac{i}{\Lambda-1} \right)^\alpha} \right\}_0^{\Lambda-1}
\end{equation}

\noindent for the general case of amplitude variations following a geometric progression. The greater the value of $\alpha$, the smoother the sound introduction and more intense its end. $\alpha>1$ results in loudness transitions commonly called \emph{slow fade}, while $\alpha<1$ results in \emph{fast fade}~\cite{guillaume}.

For linear amplification -- but not linear perception -- it is sufficient to use an appropriate sequence $\{a_i\}$:

\begin{equation}\label{seqAmpLin}
a_i=a_0 + (a_{\Lambda-1}-a_0)\frac{i}{\Lambda-1}
\end{equation}

The linear transitions will be used for AM and FM synthesis, while exponential transitions are proper tremolos and vibratos, as developed in Section~\ref{subsec:tvaf}. A non-oscillatory exploration of these variations is in the music piece \emph{ParaMeter transitions}~\cite{MASSA}.

\subsection{Application of digital filters}\label{subsec:filtros}
This subsection is limited to a description of sequences processing by convolution and difference equations, and immediate applications, as a thorough discussion of filtering is beyond the scope of this study\footnote{The implementation of filters encompasses an area of recognized complexity, with dedicated literature and software~\cite{Openheim,smith}.}.
With this procedure it is possible to achieve reverberators, equalizers, \emph{delays}, to name a few of a variety of other filters for sound processing used to obtain musical/artistic effects.
Filter employment can be part of the synthesis process or made subsequently as part of processes commonly referred to as ``acoustic/sound treatment''.

\subsubsection{Convolution and finite impulse response (FIR) filters}\label{subsec:conv}
\begin{figure*}
     \centering
         \includegraphics[width=\textwidth]{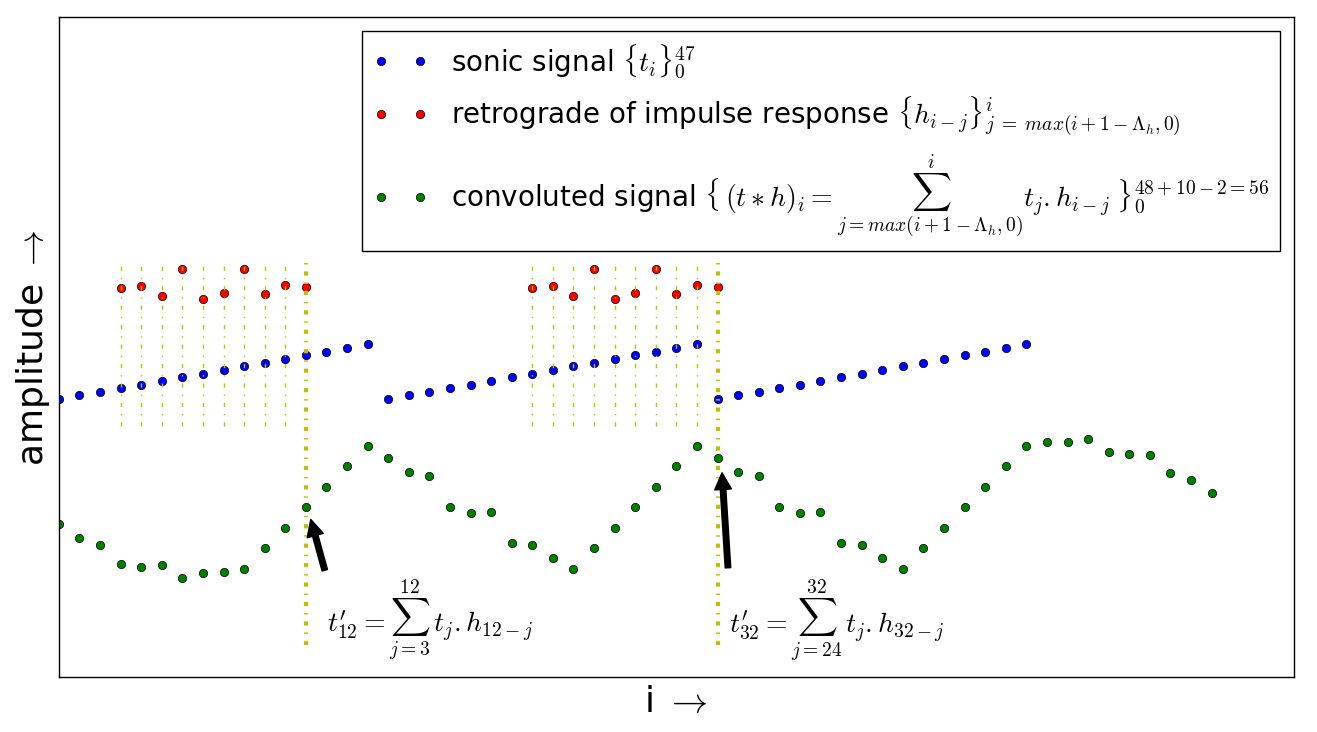}
     \caption{Graphical interpretation of convolution. Each resulting sample is the sum of the previous samples of a signal, with each one multiplied by the retrograde of the other sequence.}
         \label{fig:conv}
\end{figure*}

Filters applied by means of convolution are known by the acronym FIR (Finite Impulse Response) and are characterized by having a finite sample representation.
This sample representation is called `impulse response' $\{h_i\}$.
FIR filters are applied in the time domain by means of convolution of the sound with the respective impulse response of the filter
For the purposes of this work, convolution of $T$ with $H$ is defined as:

\begin{equation}\label{eq:conv}
 \begin{split}
 \left\{t_i'\right\}_0^{\Lambda_t+\Lambda_h-2\; = \;\Lambda_{t\, '}-1} = & \{(T*H)_i\}_0^{\Lambda_{t \, '}-1} = \{(H*T)_i\}_0^{\Lambda_{t \, '}-1} \\
     = & \left \{ \sum_{j=0}^{min(\Lambda_h-1,i)}h_{j} t_{i-j} \right \}_0^{\Lambda_{t\, '}-1} 
     \\ = & \left \{ \sum_{j=max(i+1-\Lambda_h,0)}^{i}t_j h_{i-j} \right \}_0^{\Lambda_{t\, '}-1}
 \end{split}
\end{equation}

\noindent where $t_i=0$ for the samples not given.
In other words, the sound $\{t_i'\}$, resulting from the convolution of $\{t_i\}$, with the impulse response $\{h_i\}$, has each $i$-th sample $t_i$ overwritten by the sum of its last $\Lambda_h$ samples $\{t_{i-j}\}_{j=0}^{\Lambda_h-1}$ multiplied one-by-one by samples of the impulse response $\{h_i\}_0^{\Lambda_h-1}$. This procedure is illustrated in Figure~\ref{fig:conv}, where the impulse response $\{h_i\}$ is in its retrograde form, and $t_{12}'$ and $t_{32}'$ are two samples calculated using the convolution given by $(T*H)_i=t_i'$. The final signal always has the length of $\Lambda_t+\Lambda_h -1=\Lambda_{t'}$.
It is also possible to apply the filter by multiplying the Fourier coefficients of both the sound and the impulse response,
and then performing the inverse Fourier transform~\cite{Openheim}.
This application of the filter in the frequency domain is usually much faster
especially when using a Fast Fourier Transform (FFT) routine.

The impulse response can be provided by physical measurement or by pure synthesis.
An impulse response for a reverberation, for example, can be obtained by recording the sound of the environment when someone triggers a click which resembles an impulse, or obtained by a sinusoidal sweep whose Fourier transform approximates its frequency response.
Both are impulse responses which, properly convoluted with the sound sequence, result in the same sound with a reverberation that resembles the original environment where the measurement was made~\cite{Cook}.

The inverse Fourier transform of an even and real envelope is an impulse response of a FIR filter.
Convoluted with a sound (in the time or frequency domain), it performs the frequency filtering specified by the envelope.
The greater the number of samples, the higher the envelope resolution and the computational complexity, which should often be weighted, for convolution is expensive.

\begin{figure*}
    \centering
        \includegraphics[width=\textwidth]{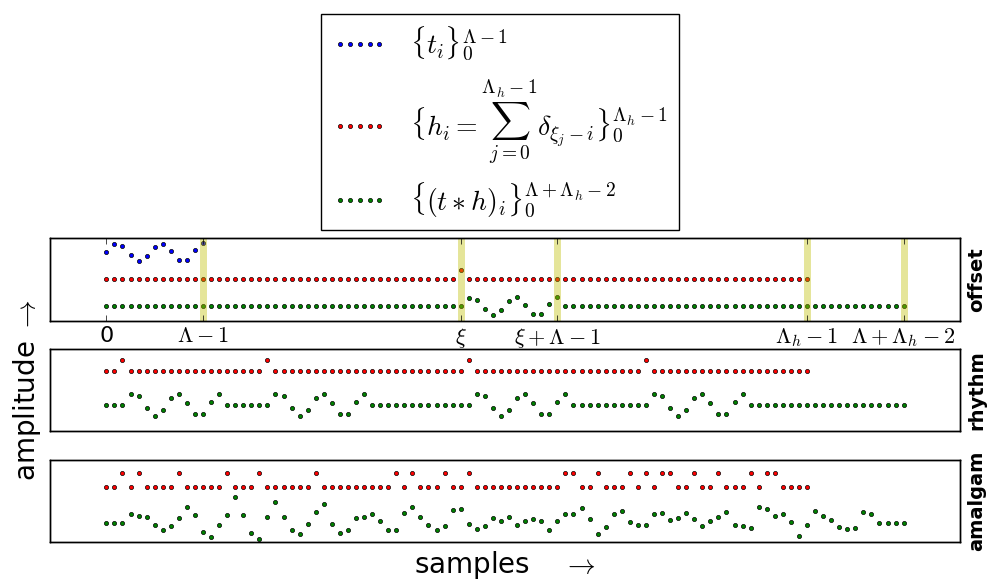}
    \caption{Convolution with different densities of impulses: shifting (a), delay lines (b) and granular synthesis~(c).
    The vertical axis is related to amplitude although one should keep in mind that each subplot has two or three displaced signals.}
        \label{fig:delays}
\end{figure*}

An important property is the time shift caused by convolution with a shifted impulse. Despite being computationally expensive, it is possible to create \emph{delay lines} by means of a convolution with an impulse response that has an impulse for each re-incidence of the sound. Figure~\ref{fig:delays} shows the shift caused by convolution with an impulse. Depending on the density of the impulses, the result is perceived as rhythm (from an impulse for each couple of seconds to about 20 impulses per second) or as pitch (from about 20 impulses per second and higher densities). In the latter case, the process resembles granular synthesis, reverberation and equalization.

\subsubsection{Infinite impulse response (IIR) filters}
This class of filters, known by the acronym IIR, is characterized by having an infinite time representation, i.e.\ the impulse response does not converge to zero. Its application is usually made by the following equation:

\begin{equation}\label{eq:diferencas}
 t_i' = \frac{1}{b_0}\left ( \sum_{j=0}^Ja_j t_{i-j} + \sum_{k=1}^Kb_k t_{i-k}' \right )
\end{equation}

The variables may be normalized: $a_j'=\frac{a_j}{b_0}$ and $b_k'=\frac{b_k}{b_0} \Rightarrow b_0' = 1$.
Equation~\ref{eq:diferencas} is called `difference equation' because the resulting samples 
$\left\{t_i'\right\}$ are given by weighted differences between original samples $\{t_i\}$ 
and previous ones in the resulting signal $\left\{t_{i-k}'\right\}$.

\begin{figure*}
    \centering
        \includegraphics[width=\textwidth]{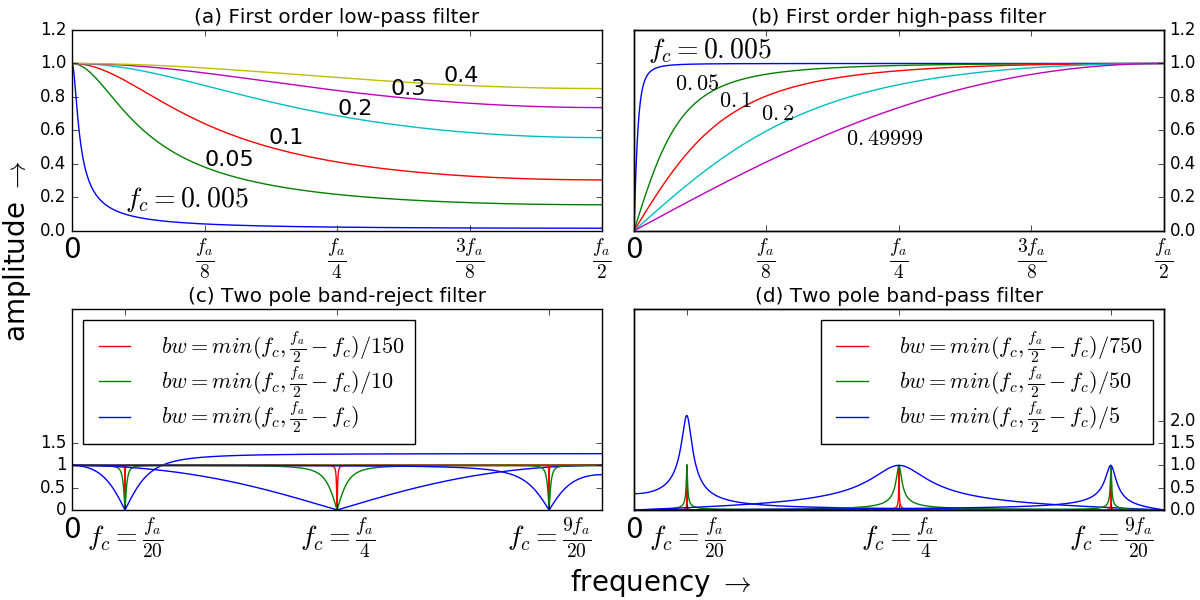}
    \caption{Moduli for the frequency response (a), (b), (c) and (d) for IIR filters of Equations~\ref{eq:passa-baixas},~\ref{eq:passa-altas},~\ref{eq:passa-banda} and~\ref{eq:rejeita-banda} respectively, considering different cutoff frequencies, center frequencies and bandwidth.}
        \label{fig:iir}
\end{figure*}

There are many methods and tools to obtain IIR filters. The text below lists a selection for didactic purposes and as a reference. They are well behaved filters whose characteristics are described in Figure~\ref{fig:iir}. For filters of simple order, the cutoff frequency $f_c$ is where the filter performs an attenuation of $-3dB \approx 0.707 $ of the original amplitude.
For band-pass and band-reject (or 'notch') filters, this attenuation has two specifications: $f_c$ (in this case, the `center frequency') and bandwidth $bw$. In both frequencies $f_c \pm bw$ there is an attenuation of $-3dB \approx 0.707$ of the original amplitude.
There is sound amplification in band-pass and band-reject filters when the cutoff frequency is low and the bandwidth is large enough. In trebles, these filters present only a deviation of the expected profile, extending the envelope to the bass.

It is possible to apply filters successively in order to obtain filters with other frequency responses. Another possibility is to use a biquad 'filter recipe'\footnote{Short for 'biquadratic': its transfer function has two poles and two zeros, i.e. its first direct form consists of two quadratic polynomials in the fraction: $\mathbb{H}(z)=\frac{a_0+a_1.z^{-1}+a_2.x^{-2}}{1- b_1.z^{-1} -b_2 . z^{-2}}$.} or the calculation of Chebichev filter coefficients\footnote{Butterworth and Elliptical filters can be considered as special cases of Chebichev filters~\cite{Openheim,smith}.}.
Both alternatives are explored by~\cite{JOSFM,smith}, and by the collection of filters maintained by the \emph{Music-DSP} community of the Columbia University~\cite{music-dsp,Openheim}.

\begin{enumerate}
  \item Low-pass with a simple pole, module of the frequency response in the upper left corner of Figure~\ref{fig:iir}. The general equation has the cutoff frequency $f_c \in (0,\frac{1}{2})$, fraction of the sample frequency $f_s$ in which an attenuation of $3dB$ occurs. The coefficients $a_0$ and $b_1$ of the IIR filter are given by $x \in [e^{-\pi},1]$:
\begin{equation}\label{eq:passa-baixas}
 \begin{split}
 x & =e^{-2\pi f_c} \\
 a_0 & =  1-x \\
 b_1 & =  x
 \end{split}
\end{equation}
  \item High-pass filter with a simple pole, module of its frequency responses at the upper right corner of Figure~\ref{fig:iir}. The general equation with cutoff frequency $f_c \in (0,\frac{1}{2})$ is calculated by means of $x \in [e^{-\pi},1]$:
\begin{equation}\label{eq:passa-altas}
 \begin{split}
 x & =e^{-2\pi f_c} \\
 a_0 & =  \frac{x+1}{2} \\
 a_1 & =  -\frac{x+1}{2} \\
 b_1 & =  x
 \end{split}
\end{equation}
\item Notch filter.
This filter is parametrized by a center frequency $f_c$ and bandwidth $bw$, both given as fractions of $f_s$, therefore $f,\; bw \in (0,\frac{1}{2})$.
Both frequencies $f_c \pm bw$ have $\approx 0.707$ of the amplitude, i.e.
an attenuation of $3dB$.
The auxiliary variables $K$ and $R$ are:
\begin{equation}\label{eq:varAux}
 \begin{split}
  R & = 1 - 3bw \\
  K & = \frac{1-2R\cos(2\pi f_c) + R^2}{2 - 2 \cos (2 \pi f_c)}
 \end{split}
\end{equation}
The band-pass filter in the lower left corner of Figure~\ref{fig:iir} has the following coefficients:
\begin{equation}\label{eq:passa-banda}
 \begin{split}
 a_0 & =  1 - K \\
 a_1 & =  2(K-R)\cos (2\pi f_c) \\
 a_2 & =  R^2-K \\
 b_1 & =  2R \cos (2\pi f_c) \\
 b_2 & =  -R^2
 \end{split}
\end{equation}
The coefficients of band-reject filter, depicted in the lower right of Figure~\ref{fig:iir}, are:
\begin{equation}\label{eq:rejeita-banda}
 \begin{split}
 a_0 & =  K \\
 a_1 & =  -2K\cos (2\pi f_c) \\
 a_2 & =  K \\
 b_1 & =  2R \cos (2\pi f_c) \\
 b_2 & =  -R^2
\end{split}
\end{equation}
\end{enumerate}

\subsection{Noise}\label{subsec:ruidos}
Sounds without an easily recognizable pitch are generally called noise~\cite{Lacerda}. They are important musical sounds, as noise is present in real notes, e.g. emitted by a violin or a piano. Furthermore, many percussion instruments do not exhibit an unequivocal pitch and their sounds are generally regarded as noise~\cite{Roederer}. In electronic music, including electro-acoustic and dance genres, noise has diverse uses and frequently characterizes the music style~\cite{Cook}. 

The absence of a definite pitch is due to the lack of a perceptible harmonic organization in the sinusoidal components of the sound.
Hence, there are many ways to generate noise. The use of random values to generate the sound sequence $T$ is a trivial method but not outstandingly useful because it tends to produce white noise with little or no variations~\cite{Cook}. Another possibility to generate noise is by using the desired spectrum, from which it is possible to perform the inverse Fourier transform. The spectral distribution should be done with care: if phases of components exhibit prominent correlation, the synthesized sound will concentrate energy in some portions of its duration.

\begin{figure*}
	\hspace*{-.75cm}
         \includegraphics[width=1.\textwidth]{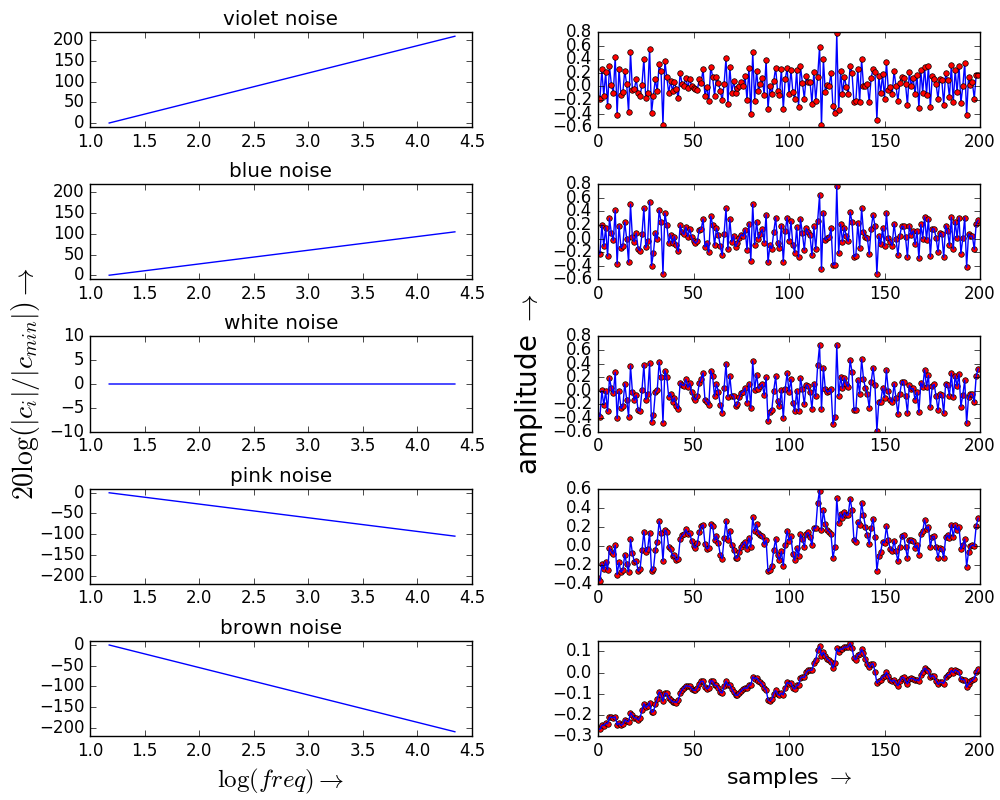}
     \caption{Colors of noise generated by Equations~\ref{eq:branco},~\ref{eq:rosa},~\ref{eq:marrom},~\ref{eq:azul} and~\ref{eq:violeta}: spectrum and example waveforms.}
         \label{fig:ruidos}
\end{figure*}

Some noises with static spectra are listed below. They are called \emph{colored noise} since they are associated with colors for many reasons.
Figure~\ref{fig:ruidos} shows the spectral profile and the corresponding sonic sequence side-by-side. All five noises were generated with the same phase for each component, making it straightforward to observe the contributions of different parts of the spectrum.

\begin{itemize}
 \item The white noise has this name because its energy is distributed equally among all frequencies, such as the white color. It is possible to obtain white noise with the inverse transform of the following coefficients:
\begin{equation}\label{eq:branco}
 \begin{split}
f_{\text{min}} & \approx 15 Hz \\
f_i  &= i \frac{f_s}{\Lambda} \;, \;\; \quad i \;\leq\; \frac{\Lambda}{2},\;\; i\;\in\;\mathbb{N}  \\
c_i & =0\;,\;\; \forall \; i \; : f_i<f_{\text{min}} \\
c_i & =e^{j.x} \;, \; x \; \text{random} \; \in \; [0,2\pi]\;,\;\; \forall \; i \; : f_{\text{min}} \le f_i < f_{\lceil \Lambda/2-1 \rceil}  \\
 c_{\Lambda/2} & = 1 \; \text{, \; if $\Lambda$ even}\\ 
 c_i & = c_{\Lambda - i}^*\;,\;\; \text{for}\;  i \; > \;  \frac{\Lambda}{2}
 \end{split}
\end{equation}

\noindent The minimum frequency $f_{\text{min}}$ is chosen considering that a sound component with frequency below 
		$\approx\; 20Hz$ is usually inaudible.
The exponential $e^{j.x}$ is a way to obtain unitary module and random phase for the value of $c_i$.
		In addition, $c_{\Lambda/2}$ is always real (as discussed in the previous section).

Other noises can be made by a similar procedure.
In the following equations, the same coefficients are used
and weighted using $\alpha_i$.

 \item The pink noise is characterized by a decrease of $3dB$ per octave. This noise is useful for testing electronic devices, being prominent in nature~\cite{Roederer}. 
\begin{equation}\label{eq:rosa}
\begin{split}
	\alpha_i & = \left(10^{-\frac{3}{20}}\right)^{\log _2 \left ( \frac{f_i}{f_{\text{min}}} \right )}  \\
	c_i & =e^{j.x} \alpha_i\;, \; x \; \text{random} \; \in \; [0,2\pi]\;,\;\; \forall \; i \; : f_{\text{min}} \le f_i < f_{\lceil \Lambda/2-1 \rceil}  \\
	c_{\Lambda/2} & = \alpha_{\Lambda/2}\;, \; \text{if $\Lambda$ even} \\ 
\end{split}
\end{equation}

  \item The brown noise (also Brownian noise) received this name after Robert Brown, who described the Brownian movement\footnote{Although its origin is disparate with its color association, this noise became established with this specific name in musical contexts. Anyway, this association can be considered satisfactory once violet, blue, white and pink noises are more strident and associated with more vivid colors~\cite{Cook,guillaume}.}. What characterizes brown noise is the decrease of $6dB$ per octave, with $\alpha_i$ in Equations~\ref{eq:rosa} being:

\begin{equation}\label{eq:marrom}
 \alpha_i=(10^{-\frac{6}{20}})^{\log _2 \left( \frac{f_i}{f_{\text{min}}} \right )}
\end{equation}

 \item In the blue noise there is a gain of $3dB$ per octave in a band limited by the minimum frequency $f_{\text{min}}$ and the maximum frequency $f_{\text{max}}$. Therefore (also based on the Equations~\ref{eq:rosa}):

\begin{equation}\label{eq:azul}
 \begin{split}
 \alpha_i & = (10^{\frac{3}{20}})^{\log _2 \left ( \frac{f_i}{f_{\text{min}}} \right )} \\
 c_i & =0\;,\;\; \forall \; i \; : f_i<f_{\text{min}} \;\; \text{or} \;\; f_i>f_{\text{max}} \\
 \end{split}
\end{equation}

 \item The violet noise is similar to the blue noise, but its gain is $6dB$ per octave:

\begin{equation}\label{eq:violeta}
 \alpha_i = (10^{\frac{6}{20}})^{\log _2 \left ( \frac{f_i}{f_{\text{min}}} \right )}
\end{equation}

 \item The black noise has higher losses than $6dB$ for octave:

\begin{equation}\label{eq:preto}
 \alpha_i=(10^{-\frac{\beta}{20}})^{\log _2 \left( \frac{f_i}{f_{\text{min}}} \right )}\;\;, \quad \beta > 6
\end{equation}

 \item The gray noise is defined as a white noise subject to one of the ISO-audible curves. Those curves are obtained by experiments and are imperative to obtain $\alpha_i$. An implementation of ISO 226, which is the last established revision of these curves, is in the \massa\ toolbox as an auxiliary file~\cite{MASSA}.
\end{itemize}

This subsection discussed only noises with static spectra.
There are also characterizations for noises with a dynamic spectrum along time,
and noises which are fundamentally transient, like clicks and chirps. The former are easily modeled by an impulse relatively isolated, while a chirps is not in fact a noise, but a fast scan of some given frequency band~\cite{Cook}.

\subsection{Tremolo and vibrato, AM and FM}\label{subsec:tvaf}
A vibrato is a periodic variation of pitch and a tremolo is a periodic variation of loudness\footnote{The
jargon may be different in other contexts. For example, in piano music, a tremolo is a vibrato in the classification used here.
The definitions used in this document are usual in contexts regarding music theory and electronic music,
i.e. they are based on a broader literature than the one used for a specific instrument,
practice or musical tradition~\cite{Lacerda,Harmonia}.}.
A vibrato can be achieved by:

\begin{equation}\label{vbrGamma}
 \gamma_i'=\left \lfloor i f' \frac{\widetilde{\Lambda}_M}{f_s} \right \rfloor
\end{equation}

\begin{equation}\label{vbrAux}
 t_i'=\widetilde{m}_{\gamma_i' \;\% \widetilde{\Lambda}_M}
\end{equation}

\begin{equation}\label{vbrF}
 f_i=f \left ( \frac{f + \mu }{f} \right )^{t_i'}=f . 2^{t_i'\frac{\nu}{12}}
\end{equation}

\begin{equation}\label{vbrGamma2}
\begin{split}
	\Delta_{\gamma_i}=\frac{\widetilde{\Lambda}}{f_s}f_i \quad \Rightarrow \quad \gamma_i & = \left \lfloor \sum_{j=0}^{i} \frac{\widetilde{\Lambda}}{f_s}f_j \right \rfloor \\ 
	& = \left \lfloor \sum_{j=0}^{i} \frac{\widetilde{\Lambda}}{f_s}f \left ( \frac{f + \mu }{f} \right )^{t_j'}  \right \rfloor \\
	& = \left \lfloor \sum_{j=0}^{i} \frac{\widetilde{\Lambda}}{f_s}f . 2^{t_j'\frac{\nu}{12}}  \right \rfloor
\end{split}
\end{equation}

\begin{equation}\label{vbrT}
 T^{f, vbr(f',\,\nu)}=\left\{ t_i^{f,vbr(f',\,\nu)} \right\}_0^{\Lambda-1}=\left\{ \widetilde{l}_{\gamma_i \%\; \widetilde{\Lambda} } \right\}_0^{\Lambda-1}
\end{equation}

\begin{figure}[h!]
     \centering
         \includegraphics[width=\columnwidth]{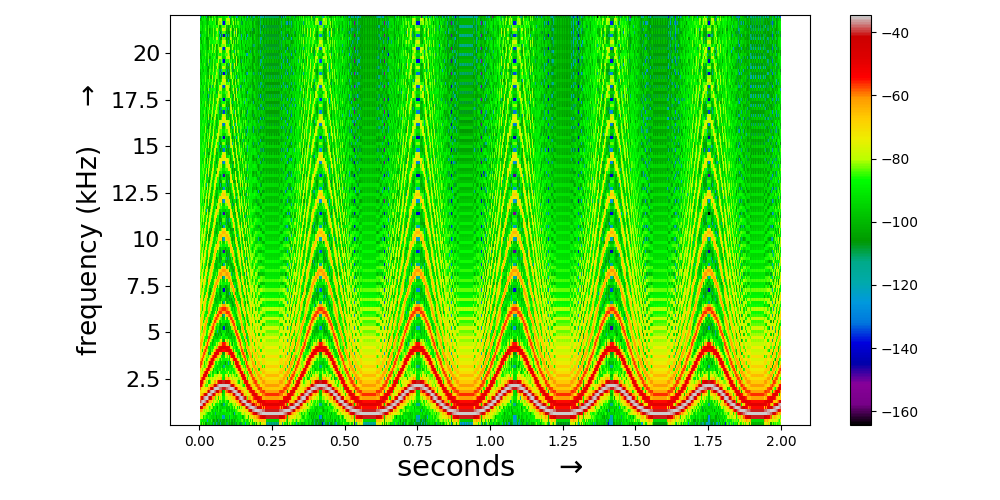}
     \caption{Spectrogram of a sound with a sinusoidal vibrato of $3Hz$ and one octave of depth in a $1000Hz$ sawtooth wave ($f_s=44.1kHz$). The color bar is in decibels.}
         \label{fig:vibrato}
\end{figure}

For the proper realization of the vibrato, it is important to pay attention to both tables and sequences. Table $\widetilde{M}$ with length $\widetilde{\Lambda}_M$ and the sequence of indices $\gamma_i'$ make the sequence $t_i'$ which is the oscillatory pattern in the frequency while table $\widetilde{L}$ with length $\widetilde{\Lambda}$ and the sequence of indices $\gamma_i$ make $t_i$ which is the sound itself. Variables $\mu$ and $\nu$ quantify the vibrato intensity:
\begin{itemize}
    \item $\mu$ is a direct measure of how many Hertz are involved in the upper limit of the oscillation, while
    \item $\nu$ is the direct measure of how many semitones (or half steps) are involved in the oscillation ($2\nu$ is the number of semitones between the upper and lower peaks of the frequency oscillations of the sound $\{t_i\}$).
\end{itemize}

It is convenient to use $\nu=\log_{2}\frac{f+\mu}{f} $ in this case because the maximum frequency increase is not equivalent to the maximum frequency decrease.
The maximum semitone/pitch displacement is the invariant quantity
and is called 'vibrato depth'.
Most often, a vibrato depth is be specified in semitones or cents (one cent $= \frac{1}{100}$ of a semitone).

Figure~\ref{fig:vibrato} is the spectrogram of an artificial vibrato in a note with $1000Hz$,
in which the pitch deviation reaches one octave above and one below.
Practically any waveform can be used to generate a sound and the vibrato oscillatory pattern, with virtually any oscillation frequency and pitch deviation.
Those oscillations with precise waveforms and arbitrary amplitudes are not possible in traditional music instruments, and thus it introduces novelty in the artistic possibilities.

Tremolo is similar: $f'$, $\gamma_i'$ and $t_i'$ remain the same.
The amplitude sequence to be multiplied by the original sequence $t_i$ is:

\begin{equation}\label{trA}
 a_i=10^{\frac{V_{dB}}{20}t_i' } = a_{\text{max}}^{t_i'}
\end{equation}
\noindent and, finally: 

\begin{equation}\label{trT}
\begin{split}
 T^{tr(f')}=\left \{ t_i^{tr(f')} \right \}_0^{\Lambda-1}=\{ t_i . a_i \}_0^{\Lambda-1}= \left \{t_i .10^{t_i' \frac{V_{dB}}{20}}    \right \}_0^{\Lambda-1}=\left\{t_i . a_{\text{max}}^{t_i'} \right\}_0^{\Lambda-1}
\end{split}
\end{equation}

\noindent where $V_{dB}$ is the oscillation depth in decibels and $a_{\text{max}}=10^{\frac{V_{dB}}{20}}$ is the maximum amplitude gain.
The measurement in decibels is suitable because the maximum increase in amplitude is not equivalent to the maximum decrease, while the difference in decibels is preserved.
Notice that the tremolo is applied to a preexisting sound and
thus the characteristics of the tremolo do not need to be accounted for
when synthesizing such sound (if it is synthesized)
in contrast with making a sound with a vibrato.

\begin{figure*}
     \centering
         \includegraphics[width=\textwidth]{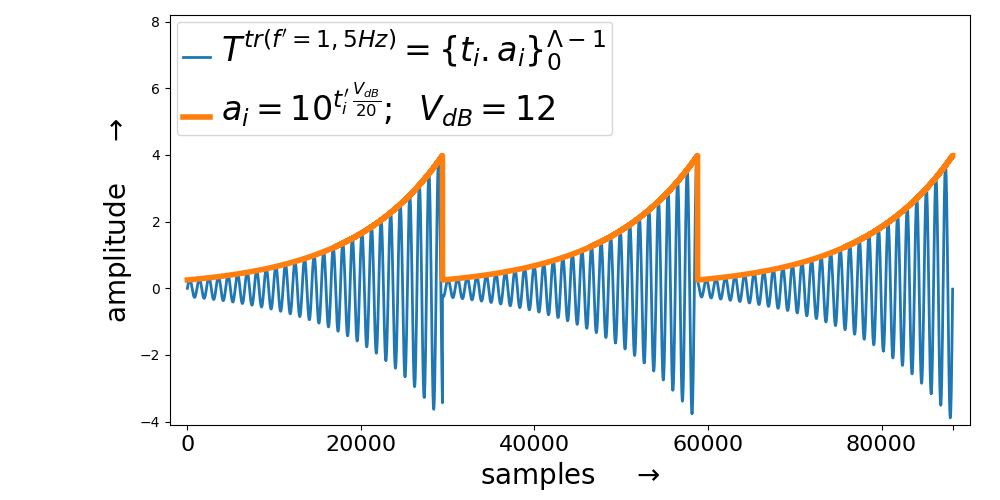}
     \caption{Tremolo with a depth of $V_{dB}=12dB$, with a sawtooth waveform as its oscillatory pattern, with $f'=1.5Hz$ in a sine of $f=40Hz$ ($f_s=44.1kHz$).}
         \label{fig:tremolo}
\end{figure*}

Figure~\ref{fig:tremolo} shows the amplitude of the sequences $\{a_i\}_0^{\Lambda-1}$ and $\{t_i'\}_0^{\Lambda-1}$ for three oscillations of a tremolo with a sawtooth waveform. The curvature is due to the logarithmic progression of the intensity. The tremolo frequency is $1.5Hz$ if $f_s=44.1kHz$ because $\text{duration} = \frac{i_{\text{max}}=82000}{f_s}= 2s \; \Rightarrow \; \frac{3\text{oscillations}}{2s}=1.5$ oscillations per second.

The musical piece \emph{Shakes and wiggles} explores these possibilities given by tremolos and vibratos,
both used in conjunction and independently (tremolos and vibratos occur many times together in a conventional music instrument),
with different frequencies $f'$, depths ($\nu$ and $V_{dB}$),
and progressive variations of parameters.
Aiming at a qualitative appreciation,
the piece also develops a comparison between vibratos and tremolos in logarithmic and linear scales.~\cite{MASSA}

The proximity of $f'$ to $20Hz$ generates roughness in both tremolos and vibratos. This roughness is largely appreciated both in traditional classical music and current electronic music, especially in the \emph{Dubstep} genre. Roughness is also generated by spectral content that produces beating~\cite{Porres,porres2009}. The sequence \emph{Bela Rugosi} explores this roughness threshold with concomitant tremolos and vibratos at the same voice, with different intensities and waveforms.~\cite{MASSA}

As the frequency increases further, these oscillations no longer remain noticeable individually.
In this case, the oscillations become audible as pitch.
Then, $f'$, $\mu$ and the waveform together change the audible spectrum of original sound $T$ in different ways for tremolos and vibratos. They are called AM (\emph{Amplitude Modulation}) and FM (\emph{Frequency Modulation}) synthesis,
respectively. These techniques are well known, with applications in
synthesizers like \emph{Yamaha DX7}, and even with applications outside music, as in telecommunications for data transfer by means of electromagnetic waves (e.g.\ AM and FM radios).

For musical goals, it is possible to understand FM based on the
case of sines and, when other waveforms are employed, to consider the signals by their respective Fourier components (i.e.\ sines as well).
The FM synthesis performed with a sinusoidal vibrato of frequency $f'$ and depth $\mu$ in a sinusoidal sound $T$ with frequency $f$ generates bands centered around $f$ and far from each other by $f'$:

\begin{equation}\label{eq:fmEsp}
\begin{split}
\{t_i'\} & = \left \{ \cos \left [f . 2 \pi \frac{i}{f_s-1} + \mu . sen \left ( f' . 2 \pi \frac{i}{ f_s -1 } \right ) \right ] \right \} = \\
 & = \left \{ \sum_{k=-\infty}^{+\infty} J_k(\mu) \cos \left [ f . 2 \pi \frac{i}{f_s-1} + k . f' . 2 \pi \frac{i}{f_s-1} \right ]  \right \} = \\
 & = \left \{ \sum_{k=-\infty}^{+\infty} J_k(\mu) \cos \left [ (f+k.f') . 2 \pi \frac{i}{f_s-1} \right ]  \right \}
\end{split}
\end{equation}

\noindent where

\begin{equation}\label{eq:Bessel}
J_k(\mu) = \frac{2}{\pi} \int_0^{\frac{\pi}{2}}\left [ cos \left (\overline{k}\;\frac{\pi}{2} + \mu . \sin w \right ) . cos \left ( \overline{k}\;\frac{\pi}{2} + k . w \right ) \right ] dw \;,\; \overline{k} = k \% 2 \;,\; k \in \mathbb{N}
\end{equation}

\noindent is the Bessel function~\cite{BesselCCRMA,JOSFM} and specifies the amplitude of each component in an FM synthesis.

In these equations, the frequency variation introduced by $\{t_i'\}$ does not follow the geometric progression that yields linear pitch variation, but reflects Equation~\ref{freqLinear}.
The result of using Equations~\ref{vbrF} for FM is described in the Appendix D of~\cite{dissertacao}, where the spectral content of the FM synthesis is calculated for oscillations in the logarithmic scale. In fact, the simple and attractive FM behavior is usually observed with linear oscillations, such as in Equation~\ref{eq:fmEsp}, which yield less strident and less noisy sounds.

For the amplitude modulation (AM):

\begin{equation}\label{eq:specAM}
\begin{split}
\{t_i'\}_0^{\Lambda-1} =\{(1+a_i) . t_i\}_0^{\Lambda-1} = \left \{ \left [ 1+M.\sin \left ( f'.2\pi\frac{i}{f_s -1} \right ) \right] .P .\sin \left ( f.2\pi\frac{i}{f_s -1} \right ) \right \}_0^{\Lambda-1} = \\ 
                        =  \left\{P.\sin \left( f.2\pi\frac{i}{f_s -1}  \right ) +  \frac{P.M}{2} \left [ \sin \left( (f-f').2\pi\frac{i}{f_s -1}  \right )  + \sin \left( (f+f').2\pi\frac{i}{f_s -1}  \right ) \right ] \right \}_0^{\Lambda-1}
\end{split}
\end{equation}

The resulting sound is the original one together with the
reproduction of its spectral content below and above with a distance of $f'$.
Again, this is achieved by variations in the linear scale (of the amplitude).
The spectrum of an AM performed with oscillations in the logarithmic amplitude scale is described in Appendix D of~\cite{dissertacao}.
The sequence $T$, with frequency $f$, called `carrier', is modulated by
$f'$, called 'modulator'. In FM and AM jargon, $\mu$ and
$a_{max}=10^{\frac{V_{dB}}{20}}$ are `modulation indexes'.
The following equations are defined for the oscillatory pattern of the modulator sequence $\{t_i'\}$:

\begin{equation}\label{fmGammaAux}
\gamma_i'=\left \lfloor i f' \frac{\widetilde{\Lambda}_M}{f_s} \right \rfloor
\end{equation}

\begin{equation}\label{fmAux}
t_i'=\widetilde{m}_{\gamma_i' \;\% \widetilde{\Lambda}_M}
\end{equation}

In FM, the modulator $\{t_i'\}$ is applied to the carrier $\{t_i\}$ by:

\begin{equation}\label{fmF}
f_i=f + \mu . t_i'
\end{equation}

\begin{equation}\label{fmGamma}
\Delta_{\gamma_i}=f_i\frac{\widetilde{\Lambda}}{f_s} \quad \Rightarrow \quad \gamma_i = \left \lfloor \sum_{j=0}^{i} f_j \frac{\widetilde{\Lambda}}{f_s} \right \rfloor = \left \lfloor \sum_{j=0}^{i} \frac{\widetilde{\Lambda}}{f_s}(f+\mu . t_j') \right\rfloor
\end{equation}

\begin{equation}\label{fmT}
T^{f,\, FM(f',\,\mu)}=\left\{ t_i^{f,\,FM(f',\,\mu)} \right\}_0^{\Lambda-1}=\left\{\,\widetilde{l}_{\gamma_i \%\; \widetilde{\Lambda} } \,\right\}_0^{\Lambda-1}
\end{equation}

\noindent where $\widetilde{l}$ is the waveform period with a length of $\widetilde{\Lambda}$ samples, used for the carrier signal.

To perform AM, the signal $\{t_i\}$ needs to be modulated with $\{t_i'\}$ using the following equations:

\begin{equation}\label{amA}
a_i=1 + \alpha . t_i'
\end{equation}

\begin{equation}\label{amT}
T^{f,\,AM(f',\,\alpha)}=\left\{ t_i^{f,\,AM(f',\,\alpha)} \right\}_0^{\Lambda-1}=\{ t_i . a_i \}_0^{\Lambda-1}= \{t_i . (1 + \alpha . t_i')    \}_0^{\Lambda-1}
\end{equation}

\subsection{Musical usages}\label{subsec:mus2}
At this point the musical possibilities are very wide.
Sonic characteristics, like pitch (given by frequency),
timbre (achieved by waveforms, filters and noise) and loudness (manipulated by intensity)
can be considered in an absolute form or varied during the duration of a sound
or a musical piece.
The following musical usages encompass a collection of possibilities with the purpose of exemplifying types of sound manipulations that result in musical material.
Some of them are discussed more deeply in the next section.

\subsubsection{Relations between characteristics}
An interesting possibility is to establish relations between parameters of
tremolos and vibratos, and of the basic note like frequency. It is possible to have a vibrato frequency proportional
to note pitch, or a tremolo depth inversely proportional to
pitch. Therefore, with Equations~\ref{vbrGamma},~\ref{vbrF} and~\ref{trA}, it is possible to set:

\begin{equation}\label{eq:vinculos}
\begin{split}
f^{vbr} = f^{tr} & = func_a(f) \\
\nu & = func_b(f) \\
V_{dB} & = func_c(f)
\end{split}
\end{equation}

\noindent with $f^{vbr}$ and $f^{tr}$ as $f'$ in the referenced equations.
 $\nu$ and $V_{dB}$ are the respective depth values of vibrato and
tremolo. Functions $func_a$, $func_b$ and $func_c$ are arbitrary and dependent on musical intentions.
The music piece \emph{Bonds} explores such bonds and exhibits variations in the waveforms with the purpose of building a \emph{musical
language} (details in Section~\ref{notasMusica}).~\cite{MASSA}

\subsubsection{Convolution for rhythm and meter}
A musical pulse - such as specified by a BPM
tempo - can be implied by an impulse at the start of each beat: the convolution with an impulse shifts the sound to impulse position, as stated in Section~\ref{subsec:conv}. For example, two impulses equally spaced build a binary division of the
pulse.
Two signals, one with 2 impulses and the other with 3 impulses, both equally spaced in the pulse duration, yield a pulse
maintenance with a rhythm which eases both binary or ternary
divisions.
This is found in many ethnic and traditional musical styles~\cite{Gramani}.
The absolute values of the impulses entail
proportions among the amplitudes of the sonic re-incidences.
The use of convolution with impulses in this context is explored in the music piece \emph{Little train of impulsive hillbillies}.
These procedures also encompass
the creation of `sound amalgams' based on granular synthesis; see Figure~\ref{fig:pulsoSubAgl}.~\cite{MASSA}

\subsubsection{Moving source and receptor, Doppler effect}
According to the discussion in Section~\ref{subsec:spac},
when an audio source (or receptor) is moving,
the IID and ITD are constantly changing and
are ideally updated at each sample of the digital signal 
(if fast computational rendering is not at stake).
As given by basic theory,
the audio source speed $s_s$, with positive values if the source moves away from receptor,
and receptor speed $s_r$, positive when it gets closer to audio source (one might always use $s_r=0$ for musical purposes),
relates the fequency $f$ at the receiver
and the frequency $f_0$ emitted by:

\begin{equation}\label{eq:fDoppler}
    f=\left(\frac{s_{sound}+s_r}{s_{sound}+s_s}\right)f_0
\end{equation}

Using the coordinates as in Figure~\ref{fig:spac},
and Equation~\ref{eq:distOuvidos},
the speed $s_s$ can be found simply by $s_s = f_s(d_{i+1}-d{i})$.
One should also use IID for the intensity progression of the sound,
and ITD to correctly start and end the sonic
sequences related to each ear.
The change in pitch is antisymmetric
upon the crossing of source with receptor:
the same semitones (or fraction of) that are added during the approach are decreased during the departure.
Moreover, the transition is abrupt if source and receptor intersect with zero distance, otherwise, there is a smooth progression.

The musical piece \emph{Doppeleer} explores and exemplifies the musical use of the Doppler effect.~\cite{MASSA}

\subsubsection{Filters and noises}
With the use of filters, the possibilities are even wider.
Convolve a signal to have a reverberated version of it, to remove its noise, to distort or to handle
the audio aesthetically in other ways. For example, sounds originated from an old television or telephone can be simulated with a band-pass filter, allowing only frequencies between $1kHz$ and $3kHz$. By rejecting the frequency of an electric oscillation (usually $50Hz$ or $60Hz$) and the harmonics, one can remove noises caused by audio devices connected to the power supply. A more musical application is to perform filtering in specific bands and to use those bands as an additional parameter to the notes.

Inspired by traditional music instruments, it is possible to apply a
time-dependent filter~\cite{Roederer}. Chains of these filters can perform complex and more accurate filtering routines. The musical piece \emph{Noisy band} explores filters and many kinds and noise synthesis.~\cite{MASSA}

A sound can be altered through different filtering processes and then mixed to create an effect known as \emph{chorus}. Based on what happens in a choir of singers, the sound is synthesized using small and potentially arbitrary modifications of parameters like center frequency, presence (or absence) of vibrato or tremolo and its characteristics, equalization, loudness, etc. As a final result, those versions of the original sound are mixed together (see Equation~\ref{eq:mixagem}).
The musical piece \emph{Children choir} implements a very simple chorus
and applies it to structures described in the next section.~\cite{MASSA}

\subsubsection{Reverberation}\label{subsubsec:reverb}
Using the same terms of Section~\ref{subsec:spac}, the
late reverberation can be achieved by a convolution with a section of pink, brown or black noise, with an exponential decay of amplitude along time. Delay lines can be added as a prefix to the noise
with the decay, and this accounts for both time parts of the reverberation: the early reflections and the late reverberation. Quality can be improved by varying the geometric trajectory and filtering by each surface where the wavefront reflected before reaching the ear in the first $100-200ms$ (mainly with a LP). The colored noise can be gradually introduced with a \emph{fade-in}: the initial moment given by direct
incidence of sound (i.e.\ without any reflection and given by ITD and IID), reaching its maximum at the beginning of the 'late
reverberation', when the geometric incidences loose their relevance to the statistical properties of the decaying noise.
As an example, consider $\Delta_1$ as the duration of the first reverberation section and 
$\Delta_R$ as the complete duration of the reverberation ($\Lambda_1=\Delta_1 f_s$, $\Lambda_R=\Delta_R
f_s$). Let $p_i$ be the probability of a sound to be repeated in the
$i$-th sample.
Following
Section~\ref{subsec:spac}, the sequence $R^1$ with the amplitudes of the
impulse response of the first period can be described as:

\begin{equation}\label{eq:p1rev}
\begin{split}
	R^1  =\left\{r_i^1\right\}_0^{\Lambda_1-1}\;,\; \text{where}\;\;
	r_i^1  =\left\{
        \begin{array}{l l}
            10^{\frac{V_{dB}}{20}\frac{i}{\Lambda_R-1}}\;  & \text{with probability}\quad p_i=\left(\frac{i}{\Lambda_1}\right)^2 \\
                                     0 \; & \text{with probability}\quad 1-p_i \\
        \end{array} \right.
\end{split}
\end{equation}

\noindent where $V_{dB}$ is the total decay in decibels, typically $-80dB$ or $-120dB$.
The sequence $R^2$ with the samples of the impulse response of the second period can be
obtained from a brown noise $N^b$ (or by a pink noise $N^p$) with an exponential amplitude decay of the waveform:

\begin{equation}\label{eq:p2rev}
    R^2=\left\{r_i^2\right\}_{\Lambda_1}^{\Lambda_R-1}=\left\{10^{\frac{V_{dB}}{20}\frac{i}{\Lambda_R-1}}\,.\,r_i^b\right\}_{\Lambda_1}^{\Lambda_R-1}
\end{equation}

Finally:

\begin{equation}\label{eq:rev}
	R=\left\{r_i\right\}_0^{\Lambda_R-1}\;,\; \text{where} \; r_i=\left\{
        \begin{array}{l l}
            r_i^1\;  & \text{if }\quad 0\leq i<\Lambda_1-1 \\
                                     r_i^2 \; & \text{se}\quad \Lambda_1 \leq i < \Lambda_R-1 \\
        \end{array} \right.
\end{equation}

A sound with an artificial reverberation can be achieved by a simple convolution of $R$ (called reverberation impulse response) with the sound sequence $T$, as described in Section~\ref{subsec:filtros}. Reverberation is well known for causing great interest in listeners and to provide sonorities that are more enjoyable. Furthermore, modifications in the reverberation consist in a common technique (almost a \textit{clich\'{e}}) to surprise and attract the listener. The musical piece \emph{Re-verb} explores reverberations in various settings.~\cite{MASSA}

\subsubsection{ADSR envelopes}
The variation of loudness along the duration of a sound is crucial to our timbre perception.
The intensity envelope known as ADSR (\emph{Attack-Decay-Sustain-Release}) has many implementations in both hardware and software synthesizers. A pioneering implementation can be found in the Hammond Novachord synthesizer of 1938 and some variants are mentioned below~\cite{ADSR}. The canonical ADSR envelope is characterized by 4 parameters: attack duration (time at which the sound reaches its maximum amplitude), decay duration (follows the attack immediately), level of sustained intensity (in which the intensity remains stable after the decay) and release duration (after sustained section, this is the duration needed for amplitude to reach zero or final value).
Note that the sustain duration is not specified because it is the difference between the total duration and the sum of the attack, decay and release durations.

\begin{figure}[htp!]
    \centering
        \includegraphics[width=\columnwidth]{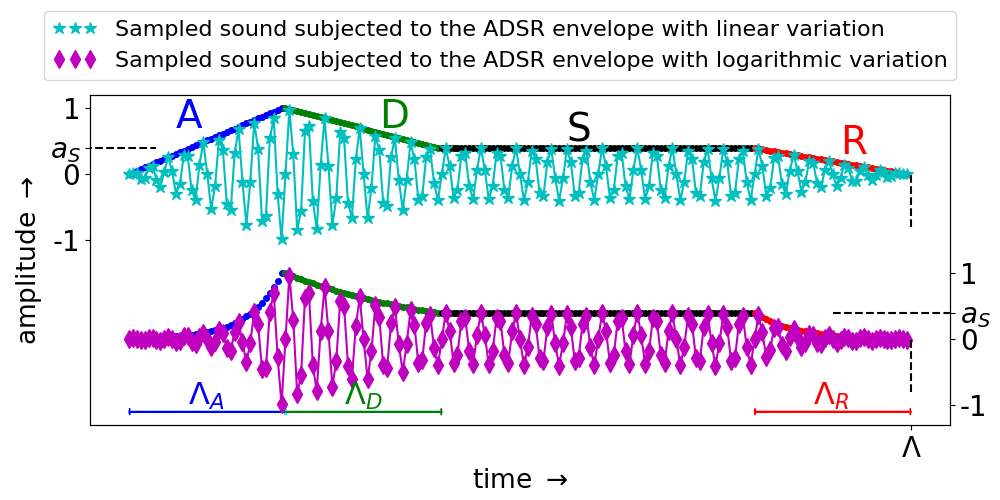}
    \caption{An ADSR envelope (\emph{Attack, Decay, Sustain, Release}) applied to an arbitrary sound sequence. The linear variation of the amplitude is above, in blue. Below the amplitude variation is exponential.}
        \label{fig:adsr}
\end{figure}

The ADSR envelope with durations $\Delta_A$, $\Delta_D$ and $\Delta_R$, with total duration $\Delta$ and sustain level $a_S$, given as the fraction of the maximum amplitude, to be applied to any sound sequence $T=\{t_i\}$ (ideally also with duration $\Delta$), can be expressed as:

\begin{equation}\label{eq:adsr}
\begin{split}
\{a_i\}_0^{\Lambda_A-1}  = & \left\{\xi\left(\frac{1}{\xi}\right)^{\frac{i}{\Lambda_A-1}}\right\}_0^{\Lambda_A-1} \quad \text{ or }\\ = & \left\{\frac{i}{\Lambda_A-1}\right\}_0^{\Lambda_A}\\
\{a_i\}_{\Lambda_A}^{\Lambda_A+\Lambda_D-1} = & \left\{a_S^{\frac{i-\Lambda_A}{\Lambda_D-1}}  \right\}_{\Lambda_A}^{\Lambda_A+\Lambda_D-1} \quad \text{ or } \\ = &  \left\{1-(1-a_S)\frac{i-\Lambda_A}{\Lambda_D-1}\right\}_{\Lambda_A}^{\Lambda_A+\Lambda_D-1}\\
\{ a_i \}_{\Lambda_A+\Lambda_D}^{\Lambda-\Lambda_R-1} = & \left\{ a_S \right\}_{\Lambda_A+\Lambda_D}^{\Lambda-\Lambda_R-1} \\
\{ a_i \}_{\Lambda-\Lambda_R}^{\Lambda-1}  = & \left\{ a_S\left(\frac{\xi}{a_S} \right)^{\frac{i-(\Lambda-\Lambda_R)}{\Lambda_R-1}} \right\}_{\Lambda-\Lambda_R}^{\Lambda-1} \quad \text{ or } \\ = &  \left\{ a_S - a_S\frac{i+\Lambda_R-\Lambda}{\Lambda_R-1}\right\}_{\Lambda-\Lambda_R}^{\Lambda-1} 
\end{split}
\end{equation}

\noindent with $\Lambda_X=\lfloor \Delta_X . f_s \rfloor\;\;\forall\;\; X \; \in
(A,D,R\;)$ and $\xi$ being a small value that provides a satisfactory \emph{fade in} and \emph{fade out}, e.g.\ $\xi=10^{\frac{-80}{20}}=10^{-4}\;$.
The lower the $\xi$, the slower the \emph{fade}, similar to the $\alpha$ illustrated in Figure~\ref{fig:transicao}.
One might also use a linear or quartic ($x**4$) fade at the beginning of the attack and the end of the release
sections to reach zero amplitude (exponential fades never reach zero).
Schematically, Figure~\ref{fig:adsr} shows the ADSR envelope in a classical implementation that supports many variations.
For example, between attack and decay it is possible to add an extra section where the maximum amplitude remains for more than a peak.
Another common example is the use of more elaborated outlines of attack or decay.
The music piece \emph{ADa and SaRa} explores many configurations of the ADSR envelope.~\cite{MASSA}

\begin{equation}\label{eq:adsrApl}
\left\{t_i^{ADSR}\right\}_0^{\Lambda-1} =\{t_i . a_i\}_0^{\Lambda-1}
\end{equation}

\section{Organization of notes in music}\label{notasMusica} \label{sec:notesMusic}
Let $S=\left\{  s_j=T_i^j=\{t_i^j\}_{i=0}^{\Lambda_j-1} \right\}_{j=0}^{H-1}$ be a the sequence of $H$ musical events $s_j$.
Consider $S$ as a `musical structure'.
This section is dedicated to techniques that make $S$ interesting and enjoyable for hearing.
More specifically, what follows is a summary of academic music composition theory and praxis.
This section does not benefit from equations that dictate the amplitude of each sample as deeply as the previous sections.
Even so, we understand that this content is very useful for synthesizing music and is not trivially integrated to Sections~\ref{sec:discNote}
and~\ref{sec:varInternas}.
The concepts are given algorithmic implementations in \mass~\cite{massListings} and can be further formalized~\cite{topos},
although at a cost of prompt intelligibility which we chose to avoid in this exposition.

The elements of $S$ can be overlapped by mixing them together, as in
Equation~\ref{eq:mixagem} and Figure~\ref{fig:mixagem}, for building intervals and chords. This reflects the `vertical thought' in music. On the other hand, the concatenation of events
in $S$, as in Equation~\ref{eq:concatenacao} and in Figure~\ref{fig:concatenacao},
yields melodic sequences and rhythms, which are associated with the `horizontal thought'.
The fundamental frequency $f$ and the starting moment
(attack) are generally considered the most important characteristics of the elements $s_j$.
These observations are convenient to describe and create music constituted by pitches and by temporal metrics and rhythms.
We will start by considering such aspects of musical organization as they are more
traditional in music theory and are usually easier to understand.

\subsection{Tuning, intervals, scales and chords}\label{subsec:afinacao}
\subsubsection{Tuning}
Doubling the frequency is equivalent to ascending one octave ($f=2f_0$).
The octave division in twelve pitches is the canon for classical western music.
Its usage has also been observed
outside western tradition, e.g. in ceremonial/religious and ethnic contexts~\cite{Wisnick}.
The intervals between the pitches need not to be equivalent,
as will become clear in the next paragraphs, but, roughly,
the factor given by $\varepsilon=2^{\frac{1}{12}}$ defines a semitone,
i.e. if $f=2^{\frac{1}{12}}f_0$, there is a semitone between
$f_0$ and $f$.
This entails a note grid along the spectrum in which, given a frequency $f$,
any other fundamental frequency $f'$ is related to $f$ by $f'= \varepsilon^i f$ where $i$ is an integer.
Twelve successive semitones yield an
octave.
Notice that equivalences of semitones and octaves are not absolute:
2 pitches related by an octave ($f_2 = 2f_1$) are different at least because one is higher than the other,
but are equivalent in the sense that they have similar uses and might be added or substituted
in a sound without introducing much novelty or change;
semitones might not be perceived as equivalent "distances"
(this is dependent on context and listener),
but are equivalent e.g. for transposing melodies, harmonies
and other pitch-related structures.

The absolute accuracy of $\varepsilon=2^\frac{1}{12}$ is usual in computational implementations.
Performances with real musical instruments, however, often present semitones that are not exactly $2^{\frac{1}{12}}$
because the pitches yield by such grid do not match the harmonics.
The fixed interval $\varepsilon=2^{\frac{1}{12}}$ characterizes an equally tempered tuning but there are other tunings. The first formalizations of tunings (that the scientific tradition has reported) date from around two thousand years before the advent of the equal temperament~\cite{Roederer}.
Two emblematic tunings are:
\begin{itemize}
    \item The {\bf just intonation}, defined by association of intervals with ratios of low-order integers, as found in the harmonic series. E.g. the white piano keys from C to C are achieved by the ratios of frequency: 1, 9/8, 5/4
    4/3, 3/2, 5/3, 15/8, 2/1. The semitone 16/15 is also often considered.
		There are many ways to perform the division of the 12 notes in the just intonation.
    \item The {\bf Pythagorean tuning}, based on the interval 3/2 (perfect fifth). The 'white piano keys' become: 1, 9/8, 81/64, 4/3, 3/2, 27/16, 243/128, 2/1. Also often used are the 'minor second' 256/243, the 'minor third' 32/27, the 'augmented fourth' 729/512, the 'diminished fifth' 1024/729, the 'minor sixth' 128/81 and the 'minor seventh' 16/9. 
\end{itemize}

In order to account for micro-tonality\footnote{Micro-tonality is the use of intervals smaller
than one semitone and has ornamental and structuring functionalities in music. The division of the octave in $12$ notes has physical grounds but is still a \emph{convention} adopted by western classical music. Other tunings are incident, e.g. a traditional Thai music style uses an octave division in seven notes equally spaced ($\varepsilon=2^{\frac{1}{7}}$),
which allows intervals quite different than those found when $\varepsilon=2^{\frac{1}{12}}$~\cite{Wisnick}.}, non-integer values can be used as factors of $\varepsilon=2^{\frac{1}{12}}$ between frequencies, or one can maintain the usage of integer values and change $\varepsilon$. For example, a tuning that approximates the harmonic series
is proposed with the equal division of the octave in $53$ notes:
$\varepsilon=2^{\frac{1}{53}}$.~\cite{microtonalidade}
Note that if $S=\{s_i\}$ is a pitch sequence related by means of $\varepsilon=2^{1/\eta}$, the sequence $S'$ with the same notes, but related by $\varepsilon'=2^{1/\eta'}$, is 
$S'=\left\{s_i'\right\}=\left\{
s_i \frac{\eta'}{\eta}\right\}$ because:

\begin{equation}\label{eq:micro}
\begin{split}
    F   & = \{f_i\}\\
    S   & = \{s_i\} \Rightarrow f_i = f 2^{s_i/\eta}\\
    S'  & = \{s'_i\} \Rightarrow f_i = f 2^{s'_i/\eta'}\\
    f_i & =  f 2^{s_i/\eta} = f 2^{s'_i/\eta'} \Rightarrow s'_i = s_i\frac{\eta'}{\eta} 
\end{split}
\end{equation}

The music piece \emph{Micro
tone} exemplifies the use of microtonal features.

\subsubsection{Intervals}\label{subsec:intervalos}
Using the ratio $\varepsilon=2^{\frac{1}{12}}$ between note frequencies (i.e.\ one semitone)
the intervals in the twelve tone system can be represented by integers. Table~\ref{eq:intervalos} summarizes the intervals: traditional notation, qualifications of consonance and
dissonances, and number of semitones.

\begin{table*}[htp!]
\centering
    \caption{Musical intervals: traditional notation, basic classification for dissonances and consonances, and number of semitones. Unison, fifth and octave are the perfect (P) consonances. Major (M) and minor (m) thirds and sixths are the imperfect consonances. Minor seconds and major sevenths are the harsh (also strong or sharp) dissonances. Major seconds and minor sevenths are the mild (also weak) dissonances. Perfect fourth is a special case, as it is a perfect consonance when considered as an inversion of the perfect fifth and a dissonance or an imperfect consonance otherwise. Another special case is the tritone (A4 or aug4, d5 or dim5, tri or TT). This interval is consonant in some cultures.
	For tonal music, the tritone indicates a dominant (chord, function or harmonic field, see Section~\ref{subsec:harmonia}) and seeks urgent resolution into a third or sixth. Due to this instability it is considered a dissonant interval.}
\begin{tabular}{ c | c | c }\hline
    \multicolumn{3}{c}{\bf consonances}  \\\hline
   & traditional notation & number of semitones \\
   perfect: & P1, P5, P8 & 0, 7, 12 \\
 imperfect: & m3, M3, m6, M6 & 3, 4, 8, 9 \\\hline\hline
    \multicolumn{3}{c}{\bf dissonances} \\\hline
 & traditional notation & number of semitones \\
 strong: & m2, M7 & 1, 11 \\
 weak: & M2, m7 & 2, 10 \\\hline\hline
    \multicolumn{3}{c}{\bf special cases} \\\hline
 & traditional notation & number of semitones \\
 consonance or dissonance: & P4 & 5 \\
 dissonance in Western tradition: & tritone, aug4, dim5 & 6 \\\hline
\end{tabular}\label{eq:intervalos}
\end{table*}

The nomenclature, based on conveniences for tonal music and practical aspects of manipulating notes, can be specified
as follows~\cite{Roederer,Wisnick}:

\begin{itemize}
    \item Intervals are inspected first by the number of steps between notes. The simple intervals (intervals which are at most an octave wide) are: first (unison), second, third, fourth, fifth, sixth, seventh and eighth (octave).
		Each of these intervals are related to one step less the their names suggest: a third is an interval with two steps.
		As can be noticed in Table~\ref{eq:intervalos}, one step is not one semitone.
		A step, in this sense, is yield by two consecutive notes in a musical scale.
		A scale for now can be regarded as any arbitrary monotonic sequence of pitches and will be discussed in the next section.
            \item The intervals are represented by numeric digits, e.g. 1, 3, 5 are a unison, a third and a fifth, respectively\footnote{Integers might also be used to express the number of semitones in an interval.}.
	\item An interval wider than an octave (e.g. ninth, tenth, eleventh) is called a 'compound interval' and is classified in terms of the simple interval between the same notes but in the same octave. Their notation can be achieved by adding a multiple of
                7 to the simple interval: P11 is an octave plus a forth ($7 + P4 = P11$), M9 is an octave plus a major second ($7 + M2 = M9$),
                m16 is two octaves and a minor second ( $2\times 7 + m2 = m16$).
	\item Quality of each interval: perfect consonances --
                i.e.\ unison, fourth, fifth and octave -- are 'perfect'. The imperfect consonances -- i.e.\ thirds and sixths -- and dissonances -- i.e.\ seconds and sevenths -- can be major and minor. The tritone is an exception to this rule because it is a dissonant interval and cannot be major or minor.
	\item The perfect fourth can be a perfect consonance or a dissonance according to the context and theoretical background. As a general rule, it can be considered a consonance except when it is followed by a third or a fifth by the movement of the notes by seconds.
	\item The tritone is a dissonance in Western music because
		it is typical of the ``dominant'' chord (see Section~\ref{subsec:harmonia}) and represents (or yields) instability.
                Some cultures consider the interval a consonance and use it as a stable interval.
	\item A major interval decreased by one semitone results in a minor interval. A minor interval increased by one semitone results in a major interval.
	\item A perfect interval (P1, P4, P5, or P8), or a major interval (M2, M3, M6 or M7), increased
by one semitone results in an augmented interval (e.g.\
aug3 has five semitones). The augmented forth
is also called tritone (aug4, tri, or TT).
	\item A perfect interval or a minor interval (m2, m3, m6 or m7), decreased by one semitone results in a diminished interval. The
diminished fifth is also called tritone (dim5, tri, or TT).

	\item An augmented interval increased by one semitone results in a `doubly-augmented' interval; a diminished interval decreased by one semitone results in a `doubly-diminished' interval.
	\item Notes played simultaneously yield a harmonic interval.
	\item Notes played as a sequence in time yield a
                melodic interval.  When the lowest note comes first there is an ascending interval, while a descending interval is observed when the highest note comes first. 
	\item A simple interval is inverted if the lower pitch is raised one octave, or if
                the upper pitch is lowered one octave. The sum of an interval and its inversion is
                9 (e.g.\ m7 is inverted to M2: $m7+M2=9$). An inverted major
                interval results in a minor interval and vice-versa. An
                inverted augmented interval results in a diminished interval
                and vice-versa (inverting a doubly-augmented results in a
                doubly-diminished and vice-versa, etc).
                An inverted perfect interval is a perfect interval as well.
\end{itemize}

The augmented/diminished intervals and the doubly-augmented/doubly-diminished intervals have the same number of semitones of other intervals (e.g. minor, major or perfect) and are consequences of the tonal system.
Scale notes are in fact different pitches, with specific uses and functions. Henceforth, in a \textit{C flat} major scale, the tonic -- first degree -- is \textit{C flat}, not \textit{B}, and the leading tone -- seventh degree -- is \textit{B flat}, not \textit{A sharp} or \textit{C double flat}.
To grasp what this entails for intervals, let the second degree (second note) of a scale to be one semitone from the first degree. Consider also the leading tone (i.e. the seventh degree at one ascending semitone from the first degree). There is a diminished third between the seventh and second scale degrees~\cite{Lacerda}.
Notice that the dim3 is only two semitones wide, as is the major second (or e.g. an doubly-augmented unisson!).

This description summarizes the traditional theory of musical intervals~\cite{Lacerda}.
The music piece \emph{Intervals} explores these intervals in both independent and interrelated ways~\cite{MASSA}.

\subsubsection{Scales}\label{subsec:escalas}
A scale is an ordered set of pitches. Strictly speaking, any (ordered) set of pitches can be considered a scale. The complexity about musical scales lean mostly on tradition, i.e. on the scales and their uses which result from practice throughout history. Usually, scales repeat at each octave. The ascending sequence with all notes from the octave division in 12 equal intervals ($\varepsilon=2^{\frac{1}{12}}$) is known as the chromatic scale within the equal temperament. There are 5 perfectly symmetric divisions of the octave within the chromatic scale. These divisions are often regarded as scales themselves owing to the easy and peculiar uses they entail.

Let $e_i$ be integers indexed by $i$ such that
$f=\varepsilon^{e_i} f_0$,
where $f_0$ is any fixed frequency.
The symmetric scales mentioned above can be expressed as:
\begin{equation}\label{escSim}
\begin{aligned}
	\text{chromatic}    & = E^c    & = \{e_i^c\}_0^{11}   & =  \{0,1,2,3,4,5,6,7,8,9,10,11\} & = \{i\}_0^{11}\\
	\text{whole tones}  & = E^{wt} & = \{e_i^{wt}\}_0^{5} & = \{0,2,4,6,8,10\}               & = \{2i\}_0^{5} \\
	\text{minor thirds} & = E^{mt} & = \{e_i^{mt}\}_0^{3} & = \{0,3,6,9\}                    & = \{3i\}_0^3 \\
	\text{major thirds} & = E^{Mt} & = \{e_i^{Mt}\}_0^{2} & = \{0,4,8\}                      & = \{4i\}_0^2\\
	\text{tritones}     & = E^{tt} & = \{e_i^{tt}\}_0^{1} & = \{ 0, 6 \}                     & = \{6i\}_0^1
\end{aligned}
\end{equation}

For example, the third note of the whole tone scale with $f_0=200Hz$ is $f_3=\varepsilon^{e_2^{wt}}.
f_0 = 2^{\frac{4}{12}} .
200 \approxeq 251.98
Hz$.
These `scales', or patterns, generate stable structures by their internal symmetries and can be repeated in a sustained way which is musically effective.
Section~\ref{estCic} discusses other aspects of symmetries in music.
The musical piece \emph{Crystals} uses each one of these scales, in both melodic and harmonic counterpart.

The \emph{diatonic scales} are scales with seven notes in which the consecutive intervals include five whole tones\footnote{A (whole) tone is an interval that is two semitones wide: $f'=2^{\frac{2}{12}}f$.} and two semitones (in each octave). There are seven of them:

\begin{equation}\label{eq:escalas}
\begin{aligned}
	\text{aeolian}    & = \text{natural minor scale}  & = & \\
			  & = E^m = \{e_i^m\}_0^6       & = & \{0,2,3,5,7,8,10\} \\
	\text{locrian}    & = E^{lo} = \{e_i^{lo}\}_0^6 & = & \{0,1,3,5,6,8,10\} \\ 
	\text{ionian}     & = \text{major scale}          & = &  \\
			  & = E^M = \{e_i^M\}_0^6       & = & \{0,2,4,5,7,9,11\} \\
	\text{dorian}     & = E^{d} = \{e_i^{d}\}_0^6   & = & \{0,2,3,5,7,9,10\} \\
	\text{phrygian}   & = E^{p} = \{e_i^{p}\}_0^6   & = & \{0,1,3,5,7,8,10\} \\
	\text{lydian}     & = E^{l}=\{e_i^{l}\}_0^6     & = & \{0,2,4,6,7,9,11\} \\
	\text{mixolydian} & = E^{mi} = \{e_i^{mi}\}_0^6 & = & \{0,2,4,5,7,9,10\}
\end{aligned}
\end{equation}

\noindent They have only major, minor and perfect intervals.
The unique exception is the tritone found as an augmented fourth or a diminished fifth.
Diatonic scales follow a circular pattern of successive intervals \textit{tone, tone, semitone, tone, tone, tone, semitone}. Thus, it is possible to write:

\begin{equation}\label{eq:relacaoDia}
\begin{split}
\{d_i\} & =\{2,2,1,2,2,2,1\} \\
e_0 & =0 \\
e_i & =d_{(i+\kappa)\%7}+e_{i-1} \quad for \;\;  i > 0
\end{split}
\end{equation}

\noindent with $\kappa \in \mathbb{N}$. For each mode there is only one $\kappa \in [0,6]$ for which $\{e_i\}$ matches. 
For example, a brief inspection reveals that
$e_i^{l}=d_{(i+2)\%7}+e_{i-1}^{l}$. Thus, $\kappa=2$ for the lydian scale.

The minor scale have two additional forms, named harmonic and melodic:

\begin{equation}\label{eq:escalasMenores}
\begin{split}
\text{natural minor}&  = E^m = \{e_i^m\}_0^6 = \\
                                           &  = \{0,2,3,5,7,8,10\} \\
\text{harmonic minor}                      &  = E^{mh} = \{e_i^{mh}\}_0^6 = \\
                                           &  = \{0,2,3,5,7,8,11\} \\
\text{melodic minor}                       &  = E^{mm} = \{e_i^{mm}\}_0^{14} = \\
                                           &  = \{0,2,3,5,7,9,11,12,10,8,7,5,3,2,0\} \\
\end{split}
\end{equation}

The different ascending and descending contours of the melodic minor scale is required in tonal music contexts.
The minor scale has one whole tone between the seventh and eigth (or first) degrees but
the separation by one semitone is critical to the polarization of the first degree.
This is not necessary in the descending trajectory, and therefore the scale recovers the standard form.
The harmonic scale presents the modified seventh degree but does not avoid the augmented second between the sixth and seventh degrees; it does not consider the melodic trajectory and thus does not need to avoid the aug2~\cite{Harmonia}.

Although it is not a traditional scale, the harmonic series is often used as such:
\begin{equation}\label{eq:serieHarmonica}
\begin{split}
H = & \{h_i\}_0^{19}= \\
    =  & \{ 0,12,19+0.02,  24,28-0.14, 31+0.2, 34-0.31, \\
                     & 36, 38+0.04,40-0.14, 42-0.49, 43+0.02, \\
                     & 44+0.41, 46-0.31, 47-0.12, \\
                     & 48, 49+0.05, 50+0.04, 51-0.02, 52-0.14   \}
\end{split}
\end{equation}

In this scale, the frequency of the $i$th note $h_i$ is the frequency of $i$th harmonic $f_i=\varepsilon^{h_i} f_0$ from the spectrum generated by $f_0$. Natural sounds have such frequencies (as discussed in Section~\ref{sec:discNote}) usually with deviations from the expected values and with noise.

Many other scales can be expressed using the framework exposed in this section, e.g. the pentatonic scales and the modes of limited transposition of Messiaen~\cite{Messiaen}.

One last observation: the words \emph{scale} and \emph{mode} are often used as synonyms both in the literature and in colloquial discussions.
The word \emph{mode} can also be used to mean two other things:
\begin{itemize}
	\item an unordered set of pitches (i.e. an unordered scale).
	\item A scale used in the context of modal harmony, in the sense presented in Section~\ref{subsec:harmonia}.
\end{itemize}

\subsubsection{Chords}\label{subsec:acordes}
A musical chord is implied by the simultaneous occurrence of three or more notes. Chords are often based on triads, especially in tonal music. Triads are built by two successive thirds
within 3 notes: root, third and fifth. If the lower note of a chord is the root, the chord is in the root position, otherwise it is an inverted chord. A closed position is any in which no chord note fits between two consecutive
notes~\cite{Lacerda}, any non-closed position is an open position. In closed and fundamental positions,
and with the fundamental denoted by $0$, triads can be expressed as:
\begin{equation}\label{triades}
\begin{split}
\text{major triad} = A^M= \{a_i^M\}_0^2=\{0,4,7\} \\ 
\text{minor triad} = A^m = \{a_i^m\}_0^2=\{0,3,7\} \\
\text{diminished triad} = A^d = \{a_i^d\}_0^2=\{0,3,6\} \\
\text{augmented triad} = A^a = \{a_i^a\}_0^2=\{0,4,8\}
\end{split}
\end{equation}

It is commonplace to consider another successive third: it is sufficient to include $10$ as the highest note to achieve a tetrad with a minor seventh, or include $11$ in order to achieve a tetrad with a major
seventh. Inversions and open positions can be obtained with the 
addition of $\pm 12$ to the selected component. Incomplete triadic chords, with extra notes ('dirty' chords), and non-triadic
are also common.
These are often interpreted as the result of further extending the succession of thirds.
E.g. $\{0,2,4,7\}$ will often be understood as a major chord with a major ninth (a major ninth has 14 semitones and $14-12 = 2$).

For general guidance:
\begin{itemize}
        \item A fifth confirms the root (fundamental).
		There are theoretical discussions about why this happens, and the most usual arguments are that the fifth is the first (non-octave) harmonic of a note and that the harmonics of the fifth are in the harmonics of the fundamental.
		Important here is to grasp the fact that musical theory and practice assures that the fifth establishes the fundamental as the root of a chord.
        \item Major or minor thirds from the root entails major or minor chord qualities.
        \item Every tritone, especially if built between a major third and a minor seventh, tends to resolve into a third or a sixth.
        \item Note duplication is avoided. If duplication is needed, the preference is, in descending order: the root, fifth, third and seventh.
        \item Note omission is avoided in the triad. If needed, the fifth is first considered for omission, then third and then the fundamental.
        \item It is possible to build chords with notes different from triads, particularly if they obey a recurrent logic or sequence that justifies these different notes.
        \item Chords built by successive intervals different from thirds -- such as fourths and seconds -- are recurrent in compositions of advanced tonalism or experimental character.
        \item The repetition of chord successions (or of characteristics they hold) fixes a trajectory and makes it possible to 
introduce exotic arrangements without implying in musical incoherence.
\end{itemize}

\subsection{Atonal and tonal harmonies, harmonic expansion and modulation}\label{subsec:harmonia}
Omission of basic tonal structures is the key to achieving modal and atonal harmonies. In the absence of minimal tonal organization,
harmony is (usually) considered modal if the notes match some diatonic scale (see Equations~\ref{eq:escalas}) or if there is only a small number of notes. If basic tonal progressions are absent and notes do not match any diatonic scale and are sufficiently diverse and dissonant (between themselves) to avoid reduction of the notes by polarization\footnote{By polarization we mean having some notes that are way more important than others and to which the other notes are ornaments or subordinates.}, the harmony is atonal. In this classification, the modal harmony is not tonal or atonal and is reduced to the incidence of notes within a (most often diatonic) scale and to the absence of tonal structures. Following this conceptualization, one observes that atonal harmony is hard to be realized and, indeed, no matter how dissonant and diverse
a set of notes is, tonal harmonies arise very easily if not avoided~\cite{harmEXT}.

\subsubsection{Atonal harmony}\label{sec:atonal}
In fact, atonal music techniques avoid that a direct relation of the notes with modes and tonality be established. Manifesting such atonal structures is of such difficulty that the dodecafonism emerged. The purpose of dodecafonism is to use a set of notes (ideally 12 notes), and to perform each note, one by one, in the same
order. In this context, the tonic becomes difficult to be established. Nevertheless, the western listener automatically searches for tonal elements in music and obstinately finds them by unexpected and tortuous paths. The use of dissonant intervals (especially tritones) without resolution reinforces the absence of tonality. In this context, while creating a musical
piece, it is possible to:
\begin{itemize}
     \item Repeat pitches. By considering immediate repetition as an extension of the previous incidence, the use of the same pitch in sequence does not add relevant information.
     \item To play adjacent pitches (e.g. of a dodecafonic progression) at the same time, making harmonic intervals and chords.
     \item Use durations and pauses with freedom, respecting notes order.
     \item Vary note sequences by temporal magnification and translation or pitch transposition and by pitch sequence inversion, retrograde and retrograde inversion. See Sections~\ref{subsec:motivos}
     and~\ref{subsec:usosmusicais3} for what these terms mean.
     \item Make variations in orchestration, articulation, spatialization, among other possibilities in presenting the same notes.
\end{itemize}

The atonal harmony can be observed, paradigmatically, within these presented conditions (which is a simple dodecaphonic model). Most of what was written by great dodecafonic composers,
e.g. Alban Berg and even Schoenberg, had the purpose of mixing tonal and atonal techniques.
Most frequently, atonal music is not strictly dodecafonic, but "serial",
i.e. they use the same kind of techniques based in (arbitrary) sequences (called the series or row)
of pitches and other sonic characteristics.

\subsubsection{Tonal harmony}
In the XX century, music with emphasis on sonorities/timbres, and rhythm, extended the concepts of tonality and harmony. Even so, tonal harmony is very often in artistic movements and commercial venues. In addition, dodecafonism itself is sometimes considered of tonal nature because it was conceived to deny tonal characteristics of polarization.
In tonal or modal music, chords -- like the ones listed in
Equations~\ref{triades} -- built with the root at each degree of a scale  -- such as listed in Equations~\ref{eq:escalas} --  form the pillars of harmony.
Tonal (and modal) harmony deals with chord formation and progressions.
Even a monophonic melody entails harmonic fields, making it possible to perceive the chord progression even in unaccompanied melodies.

In the traditional tonal music, a scale has its tonic (first
degree) on any note, and can be major (with the same notes of the Ionian mode) or minor (same notes of the Eolian mode, the 'natural minor', which has both harmonic and melodic versions, as in Equations~\ref{eq:escalasMenores}). The
scale is the base for triads, each with its root in a degree: $\hat{1},\hat{2},\hat{3},\hat{4},\hat{5},\hat{6},\hat{7}$.
To build triads, the third and the fifth notes above the root are considered together with the root (or fundamental).
$\hat{1},\hat{3},\hat{5}$ is the first degree chord,
 built on top of the scale's first degree and central for tonal music. The chords of the fifth degree $\hat{5},\hat{7},\hat{2}$ ($\hat{7}$ sharp when in a minor scale) and of the forth degree $\hat{4},\hat{6},\hat{1}$ are also important.
 The triads build on the other degrees are less important then these and are usually understood in relation to them. The `traditional harmony' comprises conventions and stylistic techniques to create progressions with such chords~\cite{Harmonia}. 

The `functional harmony' ascribes functions to the three main chords and describes their use by means of these functions. The chord built on top of the first degree is the \textbf{tonic} chord (\textit{T} or \textit{t} for a major or minor tonic, respectively) and its function (role) consists on maintaining a center, usually referred to as a ``ground'' for the music. The chord built on the fifth degree is the \textbf{dominant} (\textit{D}, the dominant is always major) and its function is to lean for the tonic
(the dominant chord asks for a conclusion and this conclusion is the tonic). Thus, the dominant chord guides the music to the tonic. The triad built under the fourth degree is the \textbf{subdominant} (\textit{S} or \textit{s} for a major or minor
subdominant, respectively) and its function is to deviate the music from the tonic.
The tonal discourse aims at confirming the tonic using the tonic-dominant-tonic progression
which is expanded by using other chords in various ways.

The remaining triads are associated to these three most important chords. In the major scale, the associated relative (relative tonic \textit{Tr}, relative
subdominant \textit{Sr} and relative dominant \textit{Dr}) is the triad built a third below, and the associated counter-relative (counter-relative tonic \textit{Ta}, counter-relative subdominant \textit{Sa} and the counter-relative dominant \textit{Da}) is the triad built in a third above. In the minor scale the same happens, but the triad a third below is called counter-relative (tA, sA) and the triad a third above is called relative (tR,
sR). The precise functions and musical effects of these chords are
controversial but are basically the same as the chords they are associated to.
Table~\ref{tab:harmonia} shows relations between the triads built at each degree of the major scale.

\begin{table}[htp!]
\centering
\caption{Summary of tonal harmonic functions on the major scale.
Tonic is the musical center, the dominant leans to the tonic and the subdominant moves the music away from
the tonic. The three chords can, in principle, be replaced by their
respective relative or counter-relative.}
\begin{tabular}{l | c | r}
relative & main chord of the function & counter-relative \\\hline\hline
$\hat{6},\hat{1},\hat{3}$ & tonic:       $\hat{1},\hat{3},\hat{5}$ & $\hat{3}, \hat{5},      \hat{7}$ \\
$\hat{3},\hat{5},\hat{7}$ & dominant:    $\hat{5},\hat{7},\hat{2}$ & [ $\hat{7},\hat{2},\hat{4}\#$ ] \\
$\hat{2},\hat{4},\hat{6}$ & subdominant: $\hat{4},\hat{6},\hat{1}$ & $\hat{6},\hat{1},       \hat{3}$
\end{tabular}
\label{tab:harmonia}
\end{table}

The dominant counter-relative should form a minor chord. It explains the change in the forth degree by a semitone above $\hat{4}\#$. The diminished chord
$\hat{7},\hat{2},\hat{4}$, is generally considered a `dominant seventh chord with the root omitted'~\cite{Koellheuteur}.
In the minor mode, there is a change in $\hat{7}$ by an ascending semitone to achieve a separation between $\hat{7}$ and $\hat{1}$ of a semitone. This is important for the dominant function (which should lean to the tonic). In this way, the dominant is always major, for both major and minor scales and, therefore, in a minor scale the relative dominant remains a third below, and the counter-relative remains a third above.

\subsubsection{Tonal expansion: individual functions and chromatic mediants}
Each chord can be stressed and developed by performing their individual dominant or subdominant, which are the triads based
on a fifth above or a fifth below, respectively. These individual dominants and subdominants,
in the same way, have also subdominants and dominants of their own. Given a tonality, any chord can occur, no matter
how distant it is from the most basic chords and from the notes of the scale.
The unique (theoretical) condition is that the occurrence presents a coherent trajectory of dominants and subdominants (or their relatives and counter-relatives) to the original tonality.

There are four mediants for each chord,
they are a third apart from the original chord and are simple triads,
as the relatives and counter-relatives,
but retain the major/minor quality of the reference chord.
The
'chromatic mediants'
are the upper mediant, formed with the root at the third of the original chord;
and the lower mediant, formed by the fifth at the third of the original chord.
If two chromatic alterations exist, i.e.\ two notes are altered by one semitone,
it is a 'doubly-chromatic mediant'.
Again, there are two forms: 
the upper form, with a third in the fifth of the original triad; and the lower form, with a third in the root
of the original triad.
This relation between chords is considered of advanced tonalism, sometimes even considered as an expansion and dissolution of tonalism,
with strong and impressive effects although they are simple, consonant major/minor triads.
Chromatic mediants are used since the end of Romanticism by Wagner, Lizt, Richard Strauss, among others~\cite{Harmonia,Salzer}.

\subsubsection{Modulation}
Modulation is the change of key (tonic, or tonal center) in music, being characterized by start and end keys, and transition artifacts.
Keys are always (thought of as) related by fifths and their relatives and counter-relatives. Some ways to perform modulation include:

\begin{itemize}
    \item Transposing the discourse to a new key, without any preparation. It is a common Baroque procedure although incident in other periods as well. Sometimes it is 
    called phrasal modulation or unprepared modulation.
    \item Careful use of an individual dominant, and perhaps also the individual
    subdominant, to confirm change in key and harmonic field.
    \item Use of chromatic alterations to reach a chord in the new key by starting from a chord in the previous key. Called chromatic modulation.
    \item Featuring a unique note, possibly repeated or suspended with no accompaniment, common to start and end keys, it constitutes a peculiar way
    to introduce the new harmonic field.
    \item Changing the function, without changing the notes, of a chord.
        This procedure is called enharmony.
    \item Maintaining the tonal center and changing the key quality from major to minor
    (or vice-versa) is a `parallel modulation'. Keys with same tonic but
		different (major/minor) qualities are known as homonyms.
\end{itemize}

The dominant has great importance and is a natural pivot in modulations,
a fact that leads to the circle of fifths~\cite{Harmonia,Salzer,Koellheuteur,Harmony}.
Other inventive ways to modulate are possible, to point but one common example, the minor thirds tetrad ($E_i^{tm}$ in Equations~\ref{escSim}) can be sustained to bridge between tonalities, with the facility that both its tritones can be resolved in a number of ways.
The music piece \emph{Acorde cedo} explores these chord relations~\cite{MASSA}.

\subsection{Counterpoint}\label{subsec:contraponto}
Counterpoint is a set of techniques for the conduction of simultaneous melodic lines, or ``voices''.
The bibliography covers systematic ways to conduct voices, leading to scholastic genres like canons, inventions and fugues~\cite{Fux,SchoenbergContra}. It is possible to
summarize the rules of scholastic counterpoint, and it is known that Beethoven --
among others -- also outlined such a digest of counterpoint.

\begin{figure}[h!]
    \centering
        \includegraphics[width=.8\columnwidth]{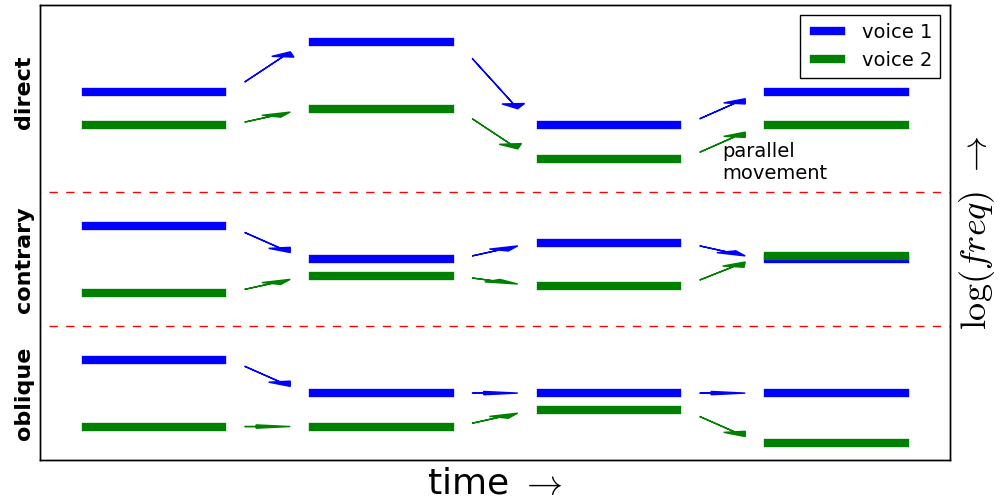}
    \caption{Different motions of counterpoint aiming to preserve independence
        between voices. There are 3 types of motion: direct, contrary and
        oblique. The parallel motion is a type of direct motion.
	A voice is often called a 'melodic line' or a melody.}
        \label{fig:movContraponto}
\end{figure}

The main purpose of (scholastic) counterpoint is to conduct voices in a way that they sound independent.
In order to do that, the relative motion of voices (in pairs) is crucial and
categorized as: direct, oblique and contrary, as depicted in Figure~\ref{fig:movContraponto}.
The parallel motion is a direct motion in which the starting and final intervals are the same.
The golden rule here is to take care of the direct motions, avoiding them
when ending in a perfect consonance.
The parallel motion should occur only between
imperfect consonances and no more than three consecutive times.
Dissonances can be forbidden or used only when followed and preceded by consonances of neighbor
degrees, i.e.\ adjacent notes in a scale.
The motions that lead to a
neighbor note in the scale sound coherent and are prioritized.
When having 3 or more voices, the melodic
relevance lies mainly in the highest and then in the lowest of the voices~\cite{Fux,Tragtenberg,SchoenbergContra}.

These rules were used in the musical piece \emph{Count point}~\cite{MASSA}.

\subsection{Rhythm}\label{subsec:ritmo}
\begin{table*}[htp!]
\caption{Durations heard as rhythm, as pitch and transition.}
\begin{tabular}{  l | r r r r   r r r    r r r || r r  } \hline
& \multicolumn{10}{c}{\bf perception of durations as rhythm} & \multicolumn{2}{c}{}  \\
	duration (s) & \multirow{2}{*}{\bf ...}     & {\bf 32,}     & {\bf 16,}   & {\bf 8,}  & {\bf 4,}   & {\bf 2,}   & {\bf 1,}   & {\bf 1/2,} & {\bf 1/4,} & {\bf 1/8,} & \multirow{2}{*}{ ... transition } & \\
frequency (Hz) & & {\color{gray} 1/32,}   & {\color{gray} 1/16,} & {\color{gray} 1/8,} & {\color{gray} 1/4,} & {\color{gray} 1/2,} &  {\color{gray} 1,}  & {\color{gray} 2,}   & {\color{gray} 4,}   & {\color{gray} 8,} & & \\
& \multicolumn{10}{c}{ - } & \multicolumn{2}{c}{} \\ \hline
\end{tabular}
\vspace{.2cm}

	\begin{tabular}{  l | p{1.7cm}  || r r || r } \hline
& \multicolumn{1}{c}{} & \multicolumn{2}{c}{transition} &  \\
duration (s) & \multirow{2}{*}{\bf rhythm ...} & $\frac{1}{16}=62.5ms$ , & $\frac{1}{20}=50ms$ & \multirow{2}{*}{\bf ... pitch} \\
frequency (Hz) & & 16, & 20 &  \\
& \multicolumn{1}{c}{} & \multicolumn{2}{c}{transition} & \\ \hline
\end{tabular}
\hspace{3.36cm}
\vspace{.2cm}

	\begin{tabular}{  l | p{1.7cm} || r r r r r r }\hline
& \multicolumn{1}{c}{} &  \\
	duration (s) & \multirow{2}{*}{transition ...} & {\color{gray} 1/40} & {\color{gray} 1/80  } & {\color{gray} 1/160 } & {\color{gray} 1/320 } & {\color{gray} 1/640 } & \multirow{2}{*}{\bf ... } \\
frequency (Hz) & & {\bf 40}   & {\bf 80}   & {\bf 160}   & {\bf 320}   & {\bf 640}   & \\
& \multicolumn{1}{c}{} & \multicolumn{6}{c}{\bf perception of durations as pitch} \\ \hline
\end{tabular}
\hspace{2.70cm}
\label{tab:duracoes}
\end{table*}

Rhythmic notion is dependent on events separated by durations~\cite{Lacerda}.
Such events can be heard individually if their onsets are spaced by at least $50-63ms$.
For the temporal separation between them to be perceived as a duration,
the period should even a bit larger, around $100ms$~\cite{microsound}.
It is possible to summarize the durations heard as rhythm or pitch
as in Table~\ref{tab:duracoes}~\cite{Alfaix, microsound}.

The transition span in Table~\ref{tab:duracoes} is minimized because the limits
are not well defined. In fact, the duration where someone begins to perceive a
fundamental frequency, or a separation between occurrences, depends on the
listener and sonic characteristics~\cite{microsound,Roederer}.
The rhythmic metric is commonly based on a basic duration called pulse, which typically is between $0.25$ and $1.5s$ ($240$
and $40 BPM$, respectively\footnote{BPM stands for Beats Per Minute and is just a frequency measure like Herz, but is the number of incidences per minute instead of second. BPM is often used as a measure of musical tempo and of heart rate.}). In music education and cognitive studies, it is common to associate this range of frequencies with the durations of the heart beat, movements of respiration and steps of a walking or running person~\cite{Lacerda,Roederer}.

\begin{figure*}
    \centering
        \includegraphics[width=.9\textwidth]{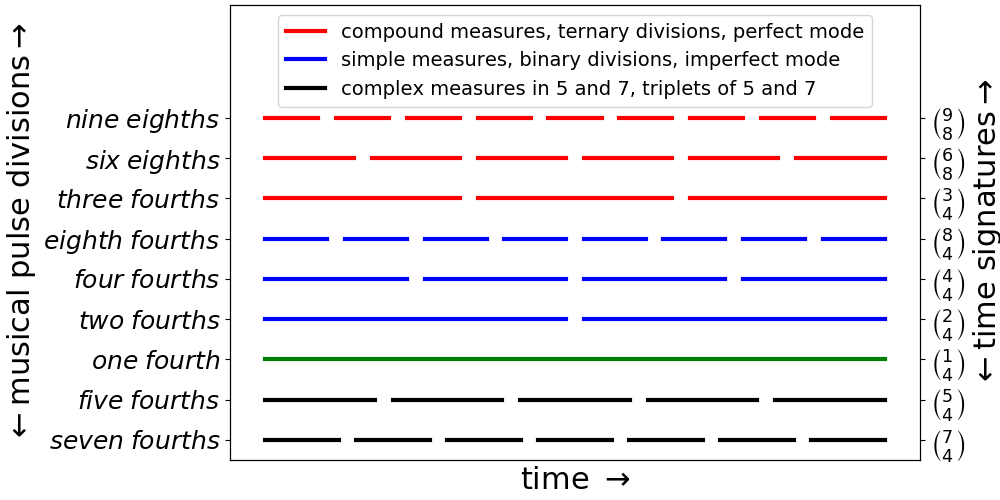}
    \caption{Divisions and groupings of the musical pulse for establishing a metric. Divisions of the quarter note, regarded as the
        pulse, is presented on the left. The time signature yielded by
        groupings of the music pulse is presented on the right.}
        \label{fig:pulsoSubAgl}
\end{figure*}

The pulse is subdivided into equal parts and is also repeated in sequence. These relations (division and concatenation) usually follow relations of small
integers. By far, the most often musical pulse divisions (and their sequential groupings), in written and ethnic
music, are: 2, 4 and 8; 3, 6 (two groups of 3 or 3 groups of 2), 9 and 12 (three and 4 groups of 3). At last, the prime numbers 5 and 7, completing
1-9 and 12. Other metrics are less common, like division or grouping in 13, 17, etc, and are mainly used in experimental music or classical music of the XX and XXI centuries. No matter how complex they seem, metrics are almost always compositions and decompositions of 1-9 equal parts~\cite{Gramani,Roederer}.
This is illustrated in Figure~\ref{fig:pulsoSubAgl}.

Binary divisions are frequent in dance rhythms and celebrations, and are called ``imperfect''.
Ternary relations are typical of
ritualistic and sacred music and are called ``perfect''.
Strong units (accents) fall in the `head' of the units (the first subdivision) and are called downbeats. In binary divisions (2, 4
and 8), strong units alternate with weak units
(e.g.\ division in 4 is: strong, weak, average strong, weak). In ternary divisions
(3, 6 and 9) two weak units succeed the downbeat (e.g.\ division in 3 is:
strong, weak, weak). Division in 6 is considered compound but can also
occur as a binary division. Binary division units which suffer a ternary division yields two units divided into three units each: strong (subdivided in strong,
weak, weak) and weak (also subdivided in strong, weak, weak). Another way to perform
the division in 6 is with a ternary division whose units subdivide as binary,
resulting in: a strong unit (subdivided in strong and weak) and two weak units
(subdivided in strong and weak each).

An accent in the weak beat is a `backbeat', whereas a note starting on a weak beat and persisting across a strong beat is a `syncope'.
These are often found in ethnic and popular music and was used with parsimony in classical music before the XX century.

Notes can occur inside and outside of these divisions of the \emph{'musical metric'}. In most well-behaved cases, notes occur exactly on these divisions, with greater incidence on strong beats.
In extreme cases, rhythmic metric cannot be perceived~\cite{Roederer}. 
Noteworthy is that (usually small or progressive) variations along the temporal grid are crucial for musical interpretation
styles~\cite{Cook}.

Let the pulse be the grouping level $j=0$, the first pulse subdivision be level $j=-1$,
the first pulse agglomeration be level $j=1$ and so on. Accordingly, let $P_i^j$ be the $i$-th unit at grouping level $j$: $P^0_{10}$ is the tenth pulse, $P^{1}_3$ is the third grouped unit (possibly the third measure),
$P^{-1}_2$ is the second part of pulse subdivision. The limits of $j$ are of special interest: pulse divisions are durations perceivable as rhythm; furthermore, the pulses sum, at its maximum, a music or a cohesive set of musical pieces. In other words, a duration given
by $P^{min(j)}_i$, $\forall \; i$, should be greater than $50 ms$ and the durations
summed together $\sum_{\forall i}P^{\text{max}(j)}_i$ should be less than a few
minutes or, at most, a few hours.
These limits might be extrapolated in extreme cases and with aesthetic goals.

Each level $j$ has some parts $i$. When $i$ has three different
values (or multiple of three) there is a perfect (i.e. ternary or compound) relation. When $i$ has only
two, four or eight possible values, than there is an imperfect relation (i.e. binary or simple),
as shown in Figure~\ref{fig:pulsoSubAgl}. Any unit can be
specified as:

\begin{equation}\label{eq:rhythmicUnit}
P^{ \{ j_k \} }_{ \{ i_{k} \}}
\end{equation}

\noindent where $j_k$ is the grouping level and $i_k$ is the unit itself.

As an example, consider $P^{-1,0,1}_{3,2,2}$ as the third subdivision $P^{-1}_3$ of the
second pulse $P^0_2$ and of the second pulse group $P^1_2$ (possibly second measure).
Each unit $P_i^j$ can be associated with a sequence of temporal samples $T$ that constitutes e.g. a
note.
In practice, there is an underlying reference duration, usually associated with the pulse,
e.g. $d_r=1$ second, and the durations of each segment are specified by:
\begin{itemize}
	\item a `temporal notation': where each entry is a relative duration to
		be multiplied by the reference duration. E.g. $durs=\{1,0.5,4\}$
		is mapped to $\{d_id_r\} = \{1d_r, 0.5d_r, 4d_r\}$. Or:
	\item a `frequential notation': where each entry is how many the
		entry that fits a same duration. E.g. $durs=\{4, 2, 16\}$
		is mapped to $\{d_r/d_i\} = \{d_r/4, d_r/2, d_r/16\}$.
		This notation might be less intuitive but it is more tightly
		related to traditional music theory, where e.g. the duration related
		to the number 4 is twice the duration related to the number 8.
\end{itemize}
See the function \texttt{rhythymToDurations} in file \texttt{src/aux/functions.py}
for an implementation of both notations that allows the specification of tuplets
(the use of arbitrary divisions of a reference duration).
The music piece \emph{Poli Hit Mia} uses different metrics.~\cite{MASSA}

\subsection{Repetition and variation: motifs and larger units}\label{subsec:motivos}
Given the basic musical structures, both frequential (chords and scales) and rhythmic (simple, compound and complex beat divisions and agglomerations), it is
natural to present these structures in a coherent and meaningful way~\cite{Boulez}.
The concept of an arc is essential in this context:
by departing from a context and returning, an arc is made.
One important, and maybe trivial, case is the arc from and to the absence of a unit: from the beginning to the end.
The audition of melodic and harmonic lines is permeated by
arcs due to the cognitive nature of the musical hearing: as the mind divides an excerpt, and groups excerpts, each of the units yields an arc.
Accordingly, the note can be considered the smallest (relevant) arc, and each motif and melody as an arc as well.
Each beat and subdivision, each measure and musical
section, constitutes an arc. Music in which the arcs do not present consistency with one another can be understood as music with no coherence. Coherence impression
comes, mostly, from the skilled handling of arcs in a music piece.

Musical arcs are abstract structures and amenable to basic operations. A spectral arc, like a chord, can be inverted, magnified and permuted, to mention just a few possibilities. Temporal arcs, like a melody, a motif, a measure or a note, are also
prone to variations. Let
$S=\left\{s_j=T^j=\{t_i^{j}\}_0^{\Lambda_j-1}\right\}_0^{H-1}$ be a sequence
of $H$ musical events $s_j$, each event with its $\Lambda_j$ samples $t_i^j$
(refer to the beginning of this Section~\ref{notasMusica} if needed). Bellow is a list of basic techniques
for variation.
\begin{itemize}
        \item Temporal translation is a displacement
    $\delta$ of a specific material to another instant $\Gamma'=\Gamma + \delta$
    of the music. It is a variation that changes temporal localization in
    a music:
    $\left\{s_j'\right\}=\left\{s_j^{\Gamma'}\right\}=\left\{s_j^{\Gamma+\delta}\right\}$
    where $\Gamma$ is the duration between the beginning of the piece (or another reference)
        and the first event $s_0$ of the original structure $S$, and
    $\delta$ is the time offset of the displacement.

    \item Temporal expansion or contraction is a change in duration of each
    arc by a factor $\mu\,:\; s_j'^{\Delta}=s_j^{\mu_j . \Delta}$. Possibly,
    $\mu_j=\mu$ is constant.

    \item Temporal reversion consists on generating a sequence with elements
    in the reverse order of the original sequence $S$, thus: $S'=\left\{s_j'\right\}_0^{H-1}=\left\{s_{(H-j-1)}\right\}_0^{H-1}$.

    \item Pitch translation, or transposition, is a displacement $\tau$ of the pitches.
        It is a variation that changes pitch
        localization:
        $\left\{s_j'\right\}=\left\{s_j^{\Xi'}\right\}=\left\{s_j^{\Xi+\tau}\right\}$
        where $\Xi$ is a reference value, such as the pitch of a section $S$ or of the first event $s_0$.
        If $\tau$ is given in semitones, the transposition displaces a
        frequency $f$ to $\tau_f=f2^{\frac{\tau}{12}}$.
        and the pitch $\Xi_i$ to $\Xi'_i=\Xi_i +12
		\log_2\left(\frac{f'_i}{f_i}\right)$.\footnote{In the MIDI protocol, $\Xi_{f}=55Hz$ when pitch $\Xi=33$
	(an \textit{A1} note). Another good MIDI reference is $\Xi_{f}=440Hz$ and
		$\Xi=69$ (\textit{A4}). The difference ($\Xi_1 - \Xi_2$) is in semitones.
        $\Xi$ is not a measure in semitones: $\Xi=1$ is not a semitone, it is a note with an audible frequency as rhythm, with
		less than 9 occurrences each second (see Table~\ref{tab:duracoes}).}

        \item Interval inversion is either: 1) the inversion of note pitch order, within the octave equivalence,
            such as described in Section~\ref{subsec:intervalos};
            or 2) the inversion of interval orientation. In the former case, the number of semitones
        is preserved in the ``strict inversion'': $f'_i = 2^{-e} f_i$ where $e$ is a positive constant;
        the inversion is said tonal if the distances are
        considered in terms of the diatonic scale $E_k$:
        $f'_i = f.2^{\left(\frac{12-e_{\left(7-j_e\right)}}{12}\right)}$
        where $j_e$ is the index in $E$ (as in Equation~\ref{eq:relacaoDia}).

        \item Rotation of musical elements is the translation of all elements
        a number of positions ahead or behind, with the care to fill empty positions
        with events which are out of the slots. Thus, a
        rotation of $\tilde{n}$ positions is $s'_n=s_{(n+\tilde{n})\%H}$. If
        $\tilde{n}<0$, it is sufficient to use $\tilde{n}'=H-\tilde{n}$. It is
        usual to associate $\tilde{n}>0$ (events advance) with the clockwise rotation and
        $\tilde{n}<0$ (elements delay) with the anti-clockwise rotation.
        Additional information about rotations is given in Section~\ref{estCic}.

        \item The insertion and removal of material in $S$ can be
    ornamental or structural: $S'=\{s_j'\}=\{s_j \text{ if condition A,
    otherwise } r_j\}$, for any music material $r_j$, including silence.
    Elements can be inserted at the beginning, like a prefix
    for $S$; at the end, as a suffix; or in the middle, splitting $S$ into both
    a prefix and a suffix. Both materials can be mixed in a variety of ways.

    \item Changes in articulation, orchestration and spatialization, or
    $s_j'=s_j^{*_j}$, where $*_j$ is the new characteristic incorporated by 
    element $s_j'$.
    
    \item Accompaniment.
        Musical material presented when $S$ occurs can be modified to yield a variation.
\end{itemize}

From these processes, many others are derived, such as the inverted retrograde, the temporal contraction with an external suffix, etc.
Variations are often thought about in the terms above but are also often very loose,
such as an arbitrary shuffle of the notes in a melody which the composer or performer finds interesting.
As a result, a whole process of mental and neurological activity is unleashed for relating the arcs, responsible for feelings, memories and imaginations, typical of a diligent musical listening.
This cortical activity is critical to
musical therapy, known by its utility in cases of depression and neurological injury.
Also, it is known that regions of the human brain responsible for sonic processing are also used for other activities,
such as for performing verbal discourse and mathematics.~\cite{Sacks,Roederer}

Paradigmatic structures guide the creation of new musical material.
One of the most established structures is the tension/relaxation dipole.
Other traditional dipoles include tonic/dominant, repetition/variation,
consonance/dissonance, coherence/rupture, symmetry/asymmetry,
equivalence/difference, arrival/departure, near/far, and stationary/moving.
All these dipoles are often thought of as parallel or even as equivalent.
Ternary constructions tend to relate to the circle and to unification. The
`transcendental' ternary communion, `modus perfectus', opposes to the `passionate'
dichotomic, `modus imperfectus'.
For a scholastic discussion on the composition of motives,
phrases, melodies, themes and musical form (such as rondo, ternary, theme and variations) see~\cite{Schoenberg}.

\subsection{Directional structures}\label{subsec:dir}
\begin{figure}
    \centering
        \includegraphics[width=.8\columnwidth]{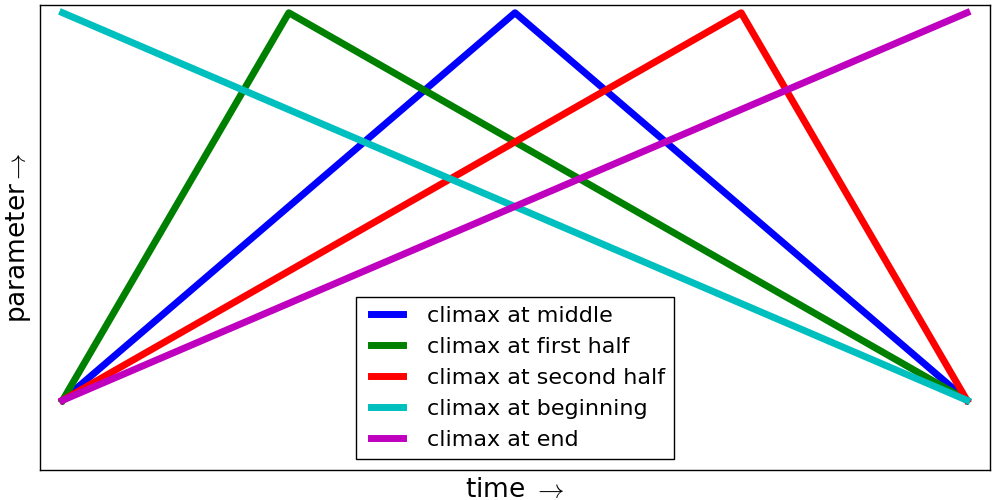}
        \caption{Canonical distinctions of musical climax in a given melody and
        other arcs. The possibilities considered are: climax at the beginning, at the first half, in the middle, in the second half and 
        at the end. The x and y-axis parameters can be non-existent and yield only a reference structure.}
        \label{fig:climax}
\end{figure}

The arcs can be decomposed in two sections: the first reaches the apex and the second returns from apex to start region.
This apex is called climax by traditional music theory. It is
usual to distinguish between arcs whose climax is at the beginning, middle, end,
or the first or second half of the duration. These structures are
shown in Figure~\ref{fig:climax}. The varying parameter can be non-existent, a case in which
the arc consists only of a reference structure, a case which resembles a note without the fundamental frequency.~\cite{Schoenberg}

Consider the sequence $S=\{s_i\}_0^{H-1}$ with a monotonic variation of a characteristic.
The sequence
$R=\{r_i\}_0^{2H -2}=\left\{s_{(H-1-|H-1-i|)}\right\}_0^{2H-2}$
presents perfect specular symmetry, i.e.\ the second half is the
mirrored version of the first. In musical terms, the climax is
in the middle of the sequence. It is possible to modify this
by using sequences with different sizes. All the mathematics of
sequences, already well established and taught routinely in calculus courses, can be used to generate these arcs~\cite{Guidorizzo,Schoenberg}.
Theoretically, when applied to any characteristic of musical events,
these sequences produce arcs, since they imply a deviation and return of an initial context (parametrization, state, etc).
Henceforth, it is possible for a given sequence to have
numerous distinct arcs, with different sizes and climax. 
This is an interesting and useful resort, and the correlation of arcs yields coherence~\cite{Salzer}.

In practice, and historically, there is special incidence and use of the golden ratio.
The Fibonacci sequence might be generalized as follows in order for any two numbers to be used
and approximate the golden ratio.
Given any two numbers $x_0$
and $x_1$, define the elements of the sequence $\{x_n\}$ as: $x_n=x_{n-1}+x_{n-2}$.
The greater $n$ is, the more $\frac{x_{n}}{x_{n_1}}$ approaches the golden ratio
($1.61803398875...$). The sequence converges fast even with discrepant
initial values.
E.g. let $x_0=1$, $x_1=100$ and $y_n=\frac{x_n}{x_{n+1}}$, the error for the first values with
respect to the golden ratio is, approximately, $\{ e_n \}
=\left\{100\frac{y_n}{1.61803398875}-100 \right\}_1^{10}=\{6080.33, -37.57, 23,
-7.14, 2.937, -1.09, 0.42, -0.1601, 0.06125, -0.02338\}$. The Fibonacci sequence
presents the same error progression, but starts at the second step of a more discrepant initial setting
($\frac{1}{1}\approx\frac{100+1 = 101}{100}$).
One might benefit from the On-Line Encyclopedia of Integer Sequences (OEIS~\cite{oeis})
for exploring various sequences.

The musical piece \emph{Dirracional} uses arcs into directional structures.~\cite{MASSA}

\subsection{Cyclic structures}\label{estCic}
The philosophical understanding that human thought is founded on the recognition of similarities and differences (e.g. as perceived in stimuli), places symmetries
at the core of cognition~\cite{Deleuze}.
Mathematically, it is commonplace to express symmetries as algebraic groups, and a finite group is always isomorphic to a permutation
group (by Cayley's theorem).
In a way, this states that permutations can express any symmetry in a
finite system~\cite{gruposFascination}.
Also, any permutation set can be used as a generator of algebraic groups~\cite{permMusic}.
In music, permutations are ubiquitous in scholastic techniques,
 which confirms their central role.
The successive application of permutations generates cyclic arcs~\cite{change,Zamacois,permMusic} and
e.g. these two academic documents report on the generation of musical structures using permutation groups~\cite{figgusOriginal, figgusEspacializacao}.
The properties defining a group $G$ are:

\begin{equation}\label{eq:groups}
\begin{split}
\forall \;\; p_1,p_2 \in G \Rightarrow  \quad   & p_1 \bullet p_2  = p_3 \in G \\ 
     & \text{(closure property)} \\
\forall \;\; p_1,p_2,p_3 \in G \Rightarrow \quad & (p_1\bullet p_2)\bullet p_3  = p_1\bullet (p_2\bullet p_3) \\
     & \text{(associativity property)} \\
\exists \;\; e \in G :                  \quad    & p \bullet e  = e \bullet p \;,\;\;\; \forall\; p \in G  \\ 
     &  \text{(existence of the identity element)} \\
\forall \;\; p \in G, \;\exists\; p^{-1} :\quad  &  p\bullet p^{-1}i =p^{-1}\bullet p = e \\
     &  \text{(existence of the inverse element)}
\end{split}
\end{equation}

From the first property follows that two permutations act as one permutation. In fact, it is possible to apply a
permutation $p_1$ and another permutation $p_2$, and, comparing both initial and final orderings, observe another permutation $p_3$.
Every element $p$ operated with itself a sufficient number of times $n$ reaches the identity element $p^n=e$ (otherwise the group generated by $p$ would be infinite).
The order $n$ of an element $p$ is
the lowest $n\,:\;p^n=e$.
Thus, a finite
permutation $p$, successively applied, reaches the initial ordering of its
elements, and yields a cycle. This cycle, if used for parameters of notes or other musical structures,
yields a cyclic arc.

These arcs can be established by using one or a set of permutations.
As a historical
example, the \emph{change ringing} tradition conceives music through
bells played one after another and then played again, but in a different
order. This process is repeated until it reaches the initial ordering. The sequence of
different orderings is a \emph{peal}. Table~\ref{tab:change}
presents a traditional \emph{peal}, named ``Plain Change''~\cite{change}, for 3 bells (\textcolor{red}{1}, \textcolor{blue}{2} and \textcolor{green}{3}), which explores
all possible orderings. Each line indicates one bell ordering to be
played. Permutations occur between each line. In this case, the musical structure
consists of permutations that entail a cyclic behavior.

\begin{table}[htp!]
\centering
\caption{Change Ringing: a traditional \emph{peal} for 3 bells. Permutations
occur between each line. Each line is a bell ordering and each ordering is played at a time.} 
\begin{tabular}{l c r}
\textcolor{red}{1} & \textcolor{blue}{2} & \textcolor{green}{3} \\
\textcolor{blue}{2} & \textcolor{red}{1} & \textcolor{green}{3} \\
\textcolor{blue}{2} & \textcolor{green}{3} & \textcolor{red}{1} \\
\textcolor{green}{3} & \textcolor{blue}{2} & \textcolor{red}{1} \\
\textcolor{green}{3} & \textcolor{red}{1} & \textcolor{blue}{2} \\
\textcolor{red}{1} & \textcolor{green}{3} & \textcolor{blue}{2} \\
\textcolor{red}{1} & \textcolor{blue}{2} & \textcolor{green}{3}
\end{tabular}
\label{tab:change}
\end{table}

The use of permutations in music can be summarized in the following way:
let $S=\{s_i\}$ be sequence of musical events $s_i$ (e.g.\ notes), and $p$ a
permutation. $S'=p(S_i)$ comprises the same elements of $S$ but in a
different order. Permutations have two notations: cyclic and
natural. The natural notation basically indicates the original indexes in the order that results from
the permutation. Thus, given the original ordering of the sequence by its indexes $[0\;1\;2\;3\;4\;5\;...]$, the permutation is noted by the sequence of indexes it
produces (e.g. $[1\;3\;7\;0\;...]$).
In the cyclic notation, a permutation is expressed
by swaps of elements and its successors.
E.g. $(1,2,5)(3,4)$ in cyclic notation is equivalent to $[0,2,5,4,3,1]$ in natural notation.

In the auralization of a permutation, it is not necessary to permute elements of $S$,
but only some characteristic. Thus, if $p$ is a permutation and $S$ is a sequence of basic notes as in the end of Section~\ref{notaBasica}, the
sequence $S'=p^f(S)=\left\{s_i^{p(f)}\right\}$ consists of the same
musical notes, following the same order and maintaining the same characteristics, but with the
fundamental frequencies permuted according to $p$.

Two subtleties of this procedure should be commented upon. 
First, a permutation $p$ is not restricted to involve all elements of $S$, i.e.\ it can operate in a subset of $S$.
Second, not all elements $s_i$ need to be executed at each access to $S$.
To exemplify, let $S$ be a sequence of music notes $s_i$. 
If $i$ goes from $0$ to $n$, and
$n>4$, at each sequence of $4$ notes it is possible to execute e.g. only the first $4$
notes.
The other notes of $S$ can occur in other events where permutations 
allocate such notes to the first four events.
The execution of disjoint sets of $S$
is the same as modifying the permutation and executing the first $n$ notes.

In summary, to each permutation $p$, we have to determine:
1) note characteristics where it operates (frequency, duration, \emph{fades},
intensity, timbre, etc); and
2) the period of incidence (how many times $S$ is used before a permutation is
applied).

The PPEPPS/FIGGS and the \emph{Tr\^es Trios} present respectively a computational implementation
and an instrumental musical piece that use
permutations to achieve
musical structures~\cite{MASSA,figgusOriginal,figgusEspacializacao,figgus,3Trios}.

\subsection{Serialism and post-serial techniques}
Recapitulating concepts from Sections~\ref{sec:atonal} and~\ref{subsec:motivos},
sequences of characteristics can be predefined and
used throughout a musical piece.
These sequences can be of intensities, timbre, durations,
density of events, etc.
Sequences can be used very strictly or loosely,
such as by skipping some elements.
The sequences can be of different sizes, yielding arcs
until the initial condition is reached again (i.e. cycles).
One paradigmatic case is the ``total serialism'' where all
the musical characteristics are serialized.
Although the use of sequences is inherent to music (e.g. scales, metric pulses),
their use with greater emphasis than tonal (or modal) elements
in western music, as an artistic trend, took place only in the
first half of the twentieth century and is called ``serialism''.
Post-serial techniques are numerous, but here is a description of
important concepts:
\begin{itemize}
	\item Spectralism:
		consists on use the (Fourier) spectrum of a sound or the harmonic series for musical composition,
		such as
		to obtain harmonies, sequences of pitches or a temporal evolution of the overall spectrum.
		For example, the most prominent frequencies can be used as pitches,
		real notes can be used to mimic an original spectrum
		(e.g. use piano notes to mimic the spectrum of a spoken sentence) or portions of the spectrum made to vary.~\cite{grisey}
	\item Spectromorphology:
		can be considered a spectral music (spectralism) theoretical framework~\cite{smalley,schaeffer}
		that examines the relation between sound spectra and their temporal evolution.
		The theory poses e.g. different onsets, continuations and terminations; characteristics of (sonic) ``motion'';
		and spectral density.
	\item Stochastic music:
		the use of random variables to describe musical elements are extensively considered in stochastic music~\cite{formalized}.
		In summary, one can use probability distributions for the synthesis of basic sounds and for obtaining larger scale musical structures.
		Changes in these distributions or in other characteristics yield the discourse.
	\item Textures:
		sounds can be assembled in terms of a ``sonic texture''.
		A sonic texture is often thought about very abstractly
		as a sonic counterpart of visual texture.
		Parameters that can be used: range between highest and lowest note,
		density of notes, durations of notes, motives, number of voices, etc.
		If the sounds are small enough (typically $<$ $100ms$) the process can be though of in terms of \emph{granular synthesis}~\cite{microsound}.
\end{itemize}

\subsection{Musical idiom?}
In numerous studies and aesthetic endeavors, there are models, discussions and exploitation of a `musical language'.
Some of them are linguistic theories applied to music
and some discern different `musical
idioms'~\cite{Lerdahl, Harmonia, Salzer,Alfaix}. Simply put, a musical idiom or language
is the result of chosen materials together with variation techniques and
relations established between elements along a music piece. In these matters,
dichotomies are prominent, as explained in Section~\ref{subsec:motivos}:
repetition and variation, relaxation and tension, stability and instability, consonance and dissonance, etc.
A thorough discussion of what can be considered a musical language is out of the scope of this article, but this brief consideration of the subject is useful as a convergence of all the previous content.

\subsection{Musical usages}\label{subsec:usosmusicais3}
The basic note was defined and characterized in quantitative terms in Section~\ref{sec:notaDisc}.
Next, the internal note
composition was addressed within both internal transitions and elementary sonic treatment
(Section~\ref{sec:varInternas}). Finally, this section aims at organizing these notes in music. The numerous resources and consequent infinitude
of praxis possibilities is typical and highly relevant for artistic contexts~\cite{Harmonia,Webern}.

There are studies and further developments for each of the presented resources.
For example, it is possible to obtain `dirty' triadic harmonies (with notes out of the triad) by superposition of perfect fourths.
Another interesting example is the superimposition of rhythms in different metrics, constituting what is
called \emph{polyrhythm}. The music piece \emph{Poli-hit my}~\cite{MASSA} explores these simultaneous metrics by impulse trains convolved with notes.

Microtonal scales are important for 20th
century music~\cite{microtonalidade} and yielded diverse remarkable results throughout history, e.g.
fourths of a tone ($\epsilon=2^{\frac{1}{24}}$) are often used in some genres of Indian and Arabic music.
The musical sequence \emph{Micro Tone}~\cite{MASSA} explores these possibilities,
including microtonal melodies and harmonies
with many pitches in a very reduced frequency bandwidth.

As in Section~\ref{subsec:mus2}, relations between
parameters are powerful to achieve musical pieces.
The number of permuted
notes can vary during the music, a relationship between permutations and the piece
duration. Harmonies can be made from triads (Equations~\ref{triades}) with duplicated
notes at each octave and more numerous duplication when the depth and frequency of
vibratos are lower (Equations~\ref{vbrGamma},~\ref{vbrAux},~\ref{vbrF},~\ref{vbrGamma2},~\ref{vbrT}).
Incontestably, the possibilities are very wide, which is made evident by the numerous musical pieces and styles.

The symmetries at octave divisions (Equation~\ref{escSim}) and the
symmetries presented as permutations (Table~\ref{tab:change} and
Equations~\ref{eq:groups}) can be used together. In the music piece \emph{3 trios},
this association is performed in a systematic way in order to achieve a specific style.
This is an instrumental piece, not included as a source code but available online~\cite{3Trios}.

\emph{PPEPPS} (Pure Python EP: Project Solvent) is an EP (Extended Play) synthesized using
resources presented in this document. With minimal parametrization, the scripts
generate complete musical pieces, allowing easy composition of sets of
music.~\cite{figgus} A simple script of a few lines specifies music delivered as 16 bit
44.1kHz PCM files (WAVE). This facility and technological
arrangement creates aesthetic possibilities for both sharing and education.

\section{Conclusions and further developments}
\label{cap:conclusao}
In our understanding,
this article is effective in relating musical elements to digital audio.
We aimed at achieving a concise presentation of the subject because it involves many knowledge fields, and therefore can very easily blast into thousands of pages.
Some readers might benefit from the text alone, but the \emph{scripts} in the \massa\ toolbox, where all the equations and concepts are directly and simply implemented as software (in Python),
are very helpful for one to achieve elaborated implementations and deeper understandings.
The scripts include routines that render musical pieces to illustrate the concepts in practical contexts.
This is valuable since art (music) can involve many non-trivial processes and is often deeply glamorized, which results in a nearly unmanageable terrain for a newcomer.
Moreover, this didactic report and the supplied open source scripts should facilitate the use of the framework.
One Supporting Information document~\cite{massListings} holds listings of sections, equations, figures, tables, scripts and other documents.
Another Supporting Information document~\cite{massCode} holds a PDF presentation of the code related to each section
because many readers might not find it easy to browse source code files.

The possibilities provided by this exposition pour from both the organization of knowledge and the ability to achieve sounds which are extremely true to the models.
For example, one can produce noises with an arbitrary resolution of the spectrum and a musical note can be synthesized with the parameters (e.g. of a vibrato) updated sample-by-sample.
Furthermore, software for synthesis and processing of sounds for musical purposes
by standard restricts the bit depth to 16 or 24.
This is achievable in this framework but by standard Python uses more bits per floating point number.
These ``higher fidelity'' characteristics can be crucial
e.g. for psychoacoustic experiments or to generate high quality musical sounds or pieces.
Simply put, it is compelling
for many scientific and artistic purposes.
The didactic potential of the framework is evident when noticed that:
\begin{itemize}
	\item the integrals and derivatives, ubiquitous in continuous signal processing,
are all replaced, in discrete signals, by summations,
which are more intuitive and does not require fluency in calculus.
	\item The equations and concepts are implemented in a simple and straightforward manner as software which can be easily assembled and inspected.
\end{itemize}
\noindent In fact, this framework was used in a number of contexts, including courses, software implementations and for making music~\cite{vimeoLM,vivace,dissertacao}.
Such detailed analytical descriptions, together with the computational implementations, have not been covered before in the literature, as far as the authors know, such as testified in the literature review (Appendix G of~\cite{dissertacao}), where books, articles and open software are related to this framework.

The free software license, and online availability of the content,
facilitate collaborations and the generation of sub-products in a co-authorship fashion,
new implementations and development of musical pieces.
The scripts can be divided in three groups: implementation of all the equations and topics of music theory; routines for rendering musical pieces that illustrate the concepts; scripts that render the figures of this article and the article itself.

This framework favored the formation of interest groups in topics such as musical creativity and computer music.
In particular, the project \url{labMacambira.sourceforge.net} groups Brazilian and foreign co-workers
in diverse areas that range from digital direct democracy and georeferencing to art and education.
This was only possible because of the usefulness of audiovisual abilities in many contexts, in particular because of the knowledge and mastery condensed in the \massa\ framework.\footnote{There are more than 700 videos, written documents,
 original software applications and contributions in well-known external software (such as Firefox, Scilab, LibreOffice, GEM/Puredata, to name just a few)~\cite{siteLM,wikiLM,vimeoLM}.
Some of these efforts are available online~\cite{dissertacao}.
It is evident that all these contributions are a consequence of more that just \massa, but it is also evident to the authors that \massa\ had a primary role in converging interests and attracting collaborators.}

Future work might include application of these results in artificial intelligence
for the generation of attractive artistic materials. Some psychoacoustic effects were detected,
which need validation and should be reported, specially with~\cite{quadrosSonoros}.\footnote{The sonic portraits where sent to a public mailing list~\cite{metaLista} and the fifth piece was reported by some individuals to induce a state in which noises from the own tongue, teeth and jaw of the individual echoed for some seconds (the GMANE archives with the descriptions of the effect by listeners in the public email list is unfortunately offline at the moment.} Other foreseen advances
are: a JavaScript version of the toolbox, better hypermedia deliverables of this framework, user guides
for different goals (e.g. musical composition, psychophysic experiments, sound synthesis, education), creation
of more musical pieces, open experiments to be studied with EEG recordings, a linked data representation
of the knowledge in \massa\ through SKOS and OWL to tackle the issues exposed in Section~\ref{sec:disc},
data sonification routines, and further analytical specification of musical
elements in the discrete-time representation of sound as feedback is received from the community.

\begin{acks}
    This work was supported by Capes, CNPq and FAPESP (project 17/05838-3).
\end{acks}

\bibliographystyle{ACM-Reference-Format}
\bibliography{./article2}

\end{document}